\documentclass[final,authoryear,5p,times,twocolumn]{elsarticle}

\usepackage{lineno,hyperref}
\modulolinenumbers[5]

\makeatletter
\journal{Journal of Financial Stability}
\def\ps@pprintTitle{%
	\let\@oddhead\@empty
	\let\@evenhead\@empty
	\def\@oddfoot{}%
	\let\@evenfoot\@oddfoot}
\makeatother

\usepackage{numcompress}\biboptions{authoryear}

\usepackage[english]{babel}
\usepackage{amsmath}
\usepackage{amsfonts}
\usepackage{mathtools}

\usepackage{graphicx}
\usepackage{placeins}
\usepackage{dcolumn}% Align table columns on decimal point
\usepackage{bm}% bold math
\usepackage{parskip}
\usepackage{booktabs}
\usepackage{color,soul}

  \usepackage{algorithm}
\usepackage{algpseudocode}

\begin{document}
	
	\begin{frontmatter}
		\title{Estimating the loss of economic predictability from aggregating \\ firm-level production networks}

		\author[csh,wu]{Christian Diem} \ead{diem@csh.ac.at} 
		\author[cb,csh]{András Borsos} 
		\author[csh]{Tobias Reisch} 
		\author[ceu,csh]{János Kertész}
		\author[csh,cosy,sfi]{Stefan Thurner} 	\ead{stefan.thurner@muv.ac.at}

		\address[csh]{Complexity Science Hub Vienna, Josefst\"adter Stra\ss e 39, A-1080, Austria}
		\address[wu]{Institute for Finance, Banking and Insurance, WU Vienna University of Economics and Business, Welthandelsplatz 1, A-1020, Austria}
		\address[cb]{Financial Systems Analysis, Central Bank of Hungary, Szabadság tér 9, Budapest 1054, Hungary} 
		\address[cosy]{Section for Science of Complex Systems, Medical University of Vienna, Spitalgasse 23, A-1090, Austria} 
		\address[sfi]{Santa Fe Institute, 1399 Hyde Park Road, Santa Fe, NM 87501, USA} 
		\address[ceu]{Department of Network and Data Science, Central European University, Quellenstrasse 51, 1100 Vienna Austria} 
		
		\begin{abstract}
		To estimate the reaction of economies to political interventions or external disturbances, input-output (IO) tables --- constructed by aggregating data into industrial sectors --- are extensively used. However, economic growth, robustness, and resilience crucially depend on the detailed structure of non-aggregated firm-level production networks (FPNs). Due to non-availability of data little is known about how much aggregated sector-based and detailed firm-level-based model-predictions differ. Using a nearly complete nationwide FPN, containing 243,399 Hungarian firms with 1,104,141 supplier-buyer-relations we self-consistently compare production losses on the aggregated industry-level production network (IPN) and the granular FPN. For this we model the propagation of shocks of the same size on both, the IPN and FPN, where the latter captures relevant heterogeneities within industries. In a COVID-19 inspired scenario we model the shock based on detailed firm-level data during the early pandemic. We find that using  IPNs instead of FPNs leads to errors up to 37\% in the estimation of economic losses, demonstrating a natural limitation of industry-level IO-models in predicting economic outcomes. We ascribe the large discrepancy to the significant heterogeneity of firms within industries: we find that firms within one sector only sell 23.5\% to and buy 19.3\% from the same industries on average, emphasizing the strong limitations of industrial sectors for representing the firms they include. Similar error-levels are expected when estimating economic growth, CO2 emissions, and the impact of policy interventions with industry-level IO models. Granular data is key for reasonable predictions of dynamical economic systems. 
		%\bl{244/250 words}
		\end{abstract}		
		
	\end{frontmatter}

\begin{figure*}[ht]
	\centering
	\includegraphics[width= .99\textwidth]{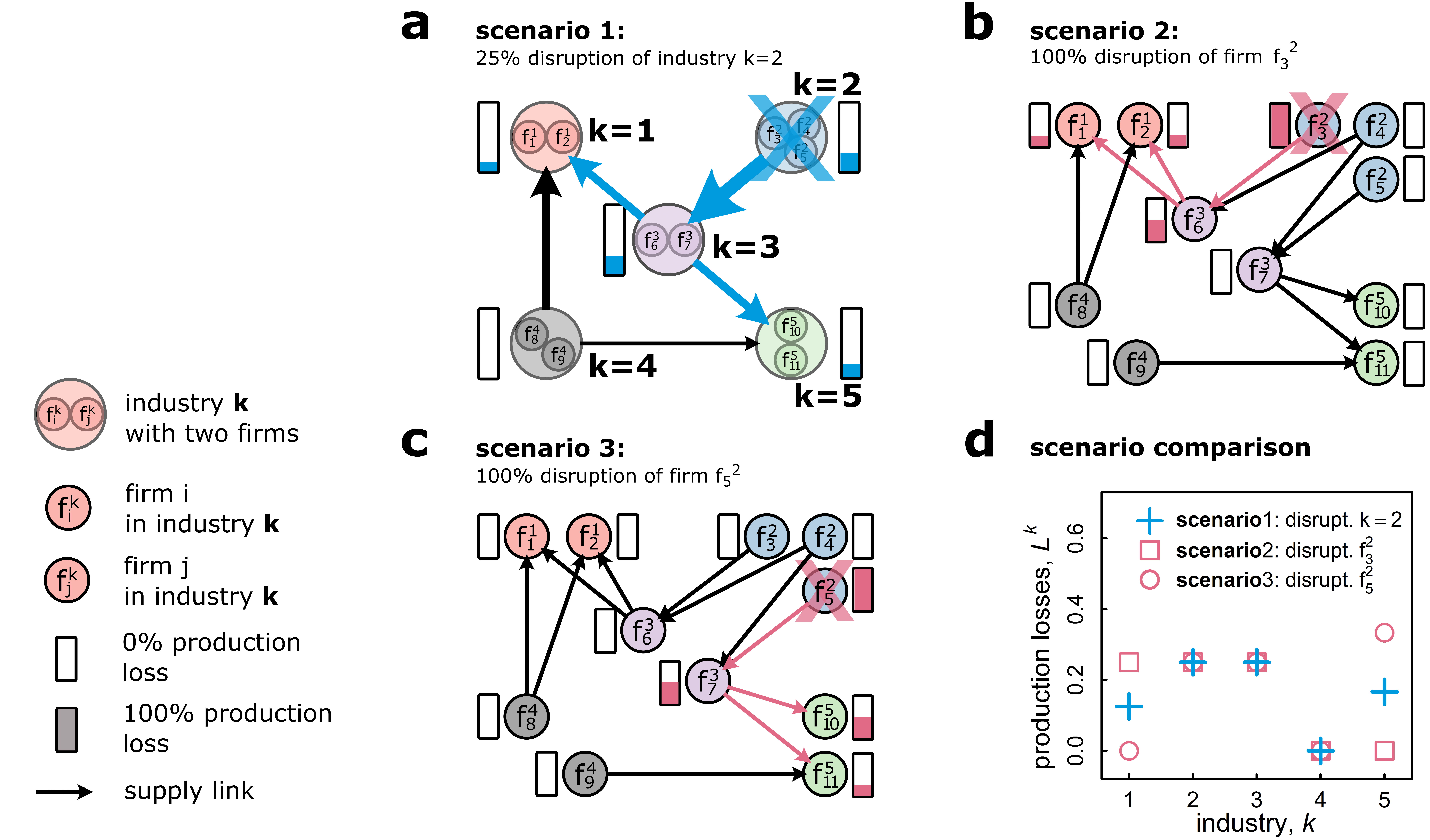} 
	\caption{Schematic demonstration of the origin of errors in production loss estimates when aggregating firm-level production networks (FPN) to the industry-level (IPN). 
	\textbf{a)} Shock propagation on the IPN in response to a \emph{25\% initial disruption of industry 2} (blue X) resulting in a 25\% production loss (blue bar marks 25\% reduction); scenario 1. The sector disruption can originate from various combinations of shocks on the level of the firms $f_3^2, f_4^2, f_5^2$. The initial shock spreads downstream to sector 3 and leads to a 25\% production loss, and further to sectors 1 (12.5\%  loss) and sector 5 (16.7\%  loss).
	\textbf{b)} Shock propagation on the FPN in response to a \emph{100\% disruption of firm 3} (red X, red bar); scenario 2. The disruption propagates downstream (red edge) to firm 6, (50\% production loss), and further to firms $1$ and $2$ (25\%  loss). Other nodes are not affected (0\% loss, empty bars). 
	\textbf{c)}  Shock propagation on the FPN in response to a different \emph{100\% disruption of firm 5} (red X, red bar), resulting in a 0\% production level; scenario 3. The disruption propagates downstream (red edge) to firm 7, (50\% loss) and to firms $10$ (50\% loss) and $11$ (25\% loss). Other nodes are not affected.
	\textbf{d)} Comparison of industry-specific production losses, $L^k$, (y-axis) for industries 1,2,3,4, and 5 (x-axis), in response to the aggregated 25\% disruption of sector 2 (blue `+') and the two 100\% firm-level shocks, of firms 3 and 5 (red squares and circles). Note that both firm-level shocks scenarios lead to the same aggregated 25\% shock to industry 2.
	The production losses of industries 2, 3, and 4 are 0.25, 0.25, and 0, respectively, for all three cascades the symbols `+', circle, and square overlap. However, the output losses of sectors 1 and 5 are remarkably different for the three different shocks (symbols do \emph{not} overlap). FPN-losses are seen to vary from 0 to 0.25 for sector 1, and from 0 to 0.33 for sector 5, whereas the IPN-losses are the same for both firm-level failure scenarios: 0.125 for sector 1, and 0.167 for sector 5. Remarkably, the IPN-loss deviates from the FPN-loss estimates by about 100\%.} 
	\label{fig1}
\end{figure*}

Supplier-buyer relationships between economic agents such as firms and companies --- the production network (PN) --- constitutes the backbone of every economy. The PN is crucial for understanding and predicting central economic processes, including innovation, growth, development, adaptation, and transition.
The network structure of the PN, and its ability to change, fundamentally determine how economies respond to severe crises, policy interventions, supply chain disruptions, or shocks in general. 
The decisions of firms largely determine the way the PN operates, restructures over time, or how it adapts to changes in global markets, their environment, and society. 
These decisions include what and how much to produce, the combination of  material inputs and technology, prices, who to hire, how to finance production and innovation, and --- importantly --- how to react to crises and shocks like, for example, to natural disasters, loss of essential suppliers, wars, pandemics, trade wars, or economic sanctions. The decisions of firms cause the PN to constantly change and hence alter its systemic features such as its efficiency, robustness, or resilience. 

So far the study of economic processes on and the systemic features of PNs has yielded fascinating insights. PNs determine and constrain the paths of future economic growth of regions and countries \citep{hidalgo2007product, neffke2011regions}. The position of industries within the national and international PN is predictive of price trends, changes in productivity, and future economic growth \citep{mcnerney2022production}. The ways economic shocks affect the agents of an economy depend on the PN. Firms that fail might be essential suppliers for other firms, which have to stop their production as a consequence \citep{ivanov2014ripple, yan2015nexus}. Consequently,  production disruptions can cascade, similarly to financial contagion \citep{battiston2012debtrank, glasserman2016contagion, diem2020minimal, thurner2022Handbook}. In this context it is important to mention that PNs can amplify micro-level sector shocks, to cause fluctuations of macro-economic relevance \citep{acemoglu2012network, carvalho2019production, moran2019may}. The COVID-19 pandemic showed that  models utilizing PNs can produce high quality forecasts of the economic effects of lock-downs \citep{pichler2022forecasting}. Tightly connected to shock propagation in PNs is the topic of the resilience of countries and industries with respect to economic shocks \citep{henriet2012firm, contreras2014propagation, klimek2019quantifying, han2019predicting,martin2015notion}. PNs are directly linked with the environment and the climate crisis \citep{willner2018global}; they determine the CO2 emission levels of industries and countries \citep{wiedmann2009emissions, davis2010consumption, wiedmann2015Material}, and in the other direction, natural disasters may lead to direct and in-direct economic damages that need to be quantified \citep{hallegatte2008adaptive, otto2017modeling, colon2021criticality}. Finally, PNs  are an integral part of national accounting of almost every economy \citep{horowitz2006concepts, eurostat2008eurostat, un2018supplyuse}, and serve as essential inputs for growth forecasts, employment projections, and estimates for policy interventions. 

However, the PNs behind these insights are generally accessible only on an aggregated level in the form of industry-level {\em input-output tables} (IOTs) that record how the entire output of one  industry enters as a production input into other industries. For almost a century, IOTs have been used to represent countries' PNs  \citep{leontief1936quantitative, miller2009input}. They are widely available and highly standardised \citep{eurostat2008eurostat, un2018supplyuse}, so that they can be globally connected \citep{dietzenbacher2013construction, yamano2006oecd}, thus, enabling the study of global PNs \citep{timmer2015illustrated, otto2017modeling, klimek2019quantifying}. Typically, the dimensionality of IOTs ranges from 56 industries, e.g., in the world input output database 2016 release \citep{dietzenbacher2013construction, timmer2015illustrated}), to  405  sectors, as in the US-American economy (Bureau of Economic Analysis, benchmark input-output statistics) \citep{horowitz2006concepts}. Industry-level IOTs are a cornerstone of economic research and modelling. However, industry-level production networks (IPNs), such as IOTs, are highly aggregated representations of the economy and can not capture the details of the supply-chain relations between firms. The aim of this paper is to demonstrate that these details (manifesting themselves in significant inhomogeneities) are often essential, and their omission can be a source of considerable errors in economic predictions. 

Studying firm-level production networks (FPNs) has been almost impossible until recently, when large-scale FPNs that include (almost) all firms and (almost) all their supply links have become available for countries such as Japan \citep{fujiwara2010large} (1.1 million firms, 5.5 million links), Belgium \citep{dhyne2015belgian} (0.8 million firms, 17.3 million links), or Hungary \citep{borsos2020unfolding} (0.25 million firms, 1.2 million links); for a review see \citep{bachilieri2022firmlevel}. Subsequently, new methods have been developed to reconstruct FPNs \citep{brintrup2020supply, wichmann2020extracting,  reisch2022monitoring,  ialongo2022reconstructing, kosasih2022machine, MUNGO2023Reconstructing, mungo2023Revealing}. Based on this firm-firm supply network data, novel insights are gained on the effects of shock propagation after natural disasters \citep{inoue2019firm, carvalho2021supply}, on interactions of the financial system with the FPN  \citep{demir2022financial, huremovic2020production, borsos2020shock}, and quantifying systemic risk contributions of individual firms in an economy have become possible \citep{diem2022quantifying}. Further, the importance of indirect exposures of firms to imports and exports through the FPN was shown in \citep{dhyne2021trade}, the origins of firm-size heterogeneity identified \citep{bernard2022origins}, and the question of how price changes (inflation) propagate through the FPN was understood \citep{duprez2018price}. 

Aggregating FPNs containing millions of firms to IPNs consisting of a few dozens of industries leads to a massive loss of information on production processes and to possibly substantial biases, as this was the case even when aggregating (the already aggregated) IOTs \citep{kymn1990aggregation, su2010input, lenzen2011aggregation}. Before we illustrate \emph{two severe problems} that emerge when aggregating firms and their supply relations into IPNs, we specify the necessary notation. 

The IPN consisting of $m$ industries is represented by the weighted directed adjacency matrix, $Z$, where, a link, $Z_{k l}$, denotes the sales of goods or services (price times quantity) from industry $k$ to industry $l$ for a given time period.
Figure \ref{fig1}a shows an example, $Z$, with $m=5$ industry sectors, where, e.g., industry 3 buys inputs needed for its production process from industry 2 and sells its output to sectors 1 and 5. Colors represent the different industries and link weights indicate sales volume. Figure \ref{fig1}b shows the corresponding FPN, $W$, with $n=11$ firms. A link, $W_{ij}$, denotes the sales of firm $i$ to firm $j$ for the same time period. Every firm, $i$, belongs to one of the $m$ industries, specified by the $i^\text{th}$ element of the industry classification vector, $p$, where, $p_i \in \{1,2,...,m \}$. In the example, firm $i$ within industry $k$ ($p_i = k$) is denoted by $f_i^k$, and e.g., firms $f_3^2, f_4^2$, and $f_5^2$ of sector 2 sell to firms $f_6^3$ and $f_7^3$ of sector 3. Due to data constraints we assume that each firm $i$ only produces one product, corresponding to its industry classification, $p_i$, as in \citep{henriet2012firm, inoue2019firm, diem2022quantifying}. We construct the IPN, $Z$, by aggregating all product flows between firms from the respective industries, e.g., $Z_{23} = W_{36} + W_{46} + W_{47} + W_{57}$ and more generally $Z_{kl} = \sum_{i=1}^{n} \sum_{j=1}^{n} W_{ij} \delta_{p_i, k} \delta_{p_j, l}$.\footnote{Official IOTs are constructed differently and are based on surveys and other data sources \citep{eurostat2008eurostat, miller2009input}.} The total number of sales of firm $i$ to all other firms in the FPN, are measured by its {\em out-strength}, $s_i^\text{out} = \sum_{j=1}^{n} W_{ij}$. It is a proxy for firm $i$'s output (amount produced). The {\em in-strength}, $s_i^\text{in} = \sum_{j=1}^{n} W_{ji}$, represents all purchases of $i$ from other firms. 

\paragraph{\textbf{Problem 1: Aggregated industries are not representative}}
Figure \ref{fig1}b demonstrates how aggregation causes the \emph{first problem}. $f_6^3$ and $f_7^3$ of sector $3$ have no \textit{overlap} in their customers' industries; $f_6^3$ sells only to firms in sector 1, and $f_7^3$ sells only to firms in sector 5. Aggregation to the industry-level erases this information and industry 3 sells equally to industry 1 and industry 5; see Fig. \ref{fig1}a. This means that the output vector of industry 3 is \emph{not representative} of the output-vectors of the firms it contains. Similarly, the IPN, $Z$, is not representative of the FPN, $W$.

\paragraph{\textbf{Problem 2: Aggregation mis-estimates economic dynamics}}
The \emph{second problem} is that aggregation leads to a mis-estimation of firm-level economic dynamics. Figure \ref{fig1} illustrates how the mis-estimation of production losses arises by comparing the same production shock propagating on the industry-level network, $Z$, an the firm-level, $W$. We compare three scenarios.
Figure \ref{fig1}a shows scenario 1, a \emph{25\% initial disruption of industry 2} (blue X), at time $t=1$. The production of sector 2 drops by 25\%  (indicated by the bar to the right filled 25\% blue), and the  production level, $h_2(t)$, is $h_2(1) = \phi_2 = 0.75$. This initial shock is specified by the vector of remaining production levels, $\phi = (1, 0.75, 1, 1, 1)$.  Then, the shock spreads downstream (blue edge) to sector 3 at $t=2$ (25\% production loss, $h_3(2)=0.75$), and at $t=3$ to sectors 1 (12.5\% production loss, $h_1(3)=0.875$), and 5 (16.7\% production loss, $h_5(3)=0.833$).\footnote{The index $t$, denotes the ``internal'' time steps of the shock propagation model, \textit{not} calendar time.} The shock propagates, as industries 3, 1, and 5 lack inputs for their production processes. Note that the 25\% disruption of industry 2 could originate from various combinations of individual shocks to firms, $f_3^2, f_4^2$ and $f_5^2$, in industry 2.
Figure \ref{fig1}b shows scenario 2, the \emph{100\% disruption of firm 3, $f_3^2$,} (red X, red bar). The production of firm 3 drops to 0\%, i.e., a total operational failure ($h_{3}(1)=\psi^1_3=0$). The firm-level shock  is specified by the remaining production level vector $\psi^1$, where $\psi^1_3=0$ and  $\psi^1_i=1$, for all $i\neq 3$. The disruption propagates downstream (red edge) to $f_6^3$ (50\% production loss, $h_{6}(2)=0.5$), and further to firms $1$ and $2$ (25\% production loss, $h_{1}(3)=h_{2}(3)=0.75$). Aggregating the production losses of firms yields a loss of 25\% for industries 1, 2 and 3 and a 0\% loss for industries 4 and 5.
Figure \ref{fig1}c shows scenario 3, the propagation of a \emph{100\% disruption of firm 5, $f_5^2$,} (red X, red bar). 
Aggregating the resulting production losses yields a loss of 25\% for industries 2 and 3,  a 0\% loss for industries 1 and 4, and a 33\% loss for industry 5.
Figure \ref{fig1}d compares for each industry, $k$, (x-axis) the industry-specific production loss, $L^k$, (y-axis), across the three scenarios,  25\% shock to sector 2 (blue `+'),  100\% shock to firm 3 (red squares) and firm 5 (red circles). When aggregated both firm-level shocks yield the industry-level shock of 25\% disruption of industry 2 and for industry 3 and 4 the production losses form shock propagation are also the same (0.25, and 0, respectively) --- the symbols `+', circle, and square overlap. However, the output losses of sectors 1 and 2 are vastly different across the three shocks --- `+', circle, square do \emph{not} overlap. The FPN-based losses vary from 0 to 0.25 for sector 1 and from 0 to 0.33 for sector 5, whereas the aggregation-based IPN losses are the same for both firm-level shocks, 0.125 for sector 1 and 0.167 for sector 5. The IPN-based loss mis-estimates the FPN-based losses by 100\%. Other network dynamics such as growth, innovation, or productivity spill overs, --- happening to a large extent at the firm- and not the industry-level --- are potentially affected in similarly drastic ways. 

In this paper we quantify the relevance of these two problems by utilizing a unique data set that allows us to observe almost every firm-level supply chain relation of the entire production network of Hungary, containing 243,399 firms and 1,104,141 links in 2019, see Data and Methods. 
First, we assess how representative industry-level production networks are of real-world firm-level production networks. We do that by quantifying the intra-sector overlaps  of firms' input- and output vectors.  
Second, we quantify the estimation-errors of economy-wide and industry-specific production losses that arise when using industry-level production networks to approximate firm-level shock propagation dynamics. Firm-level labor data with monthly time resolution enables us to realistically estimate the size of the COVID-19 shock for  \textit{individual firms} in the beginning of 2020. Then, we compare the production losses from propagating a realistic COVID-19 shock and 1,000 synthetic shock realizations, either on the firm- or the industry-level production network. We sample the synthetic shocks such that they are of the same size  when aggregated to the industry-level, but affect firms within industries differently. This feature allows us to clearly show the effects of intra-sector heterogeneity in firms' input-output vectors for estimating production losses, while controlling for size and industry effects.

\begin{figure*}[t]
	\centering
	\includegraphics[width= 0.95\textwidth]{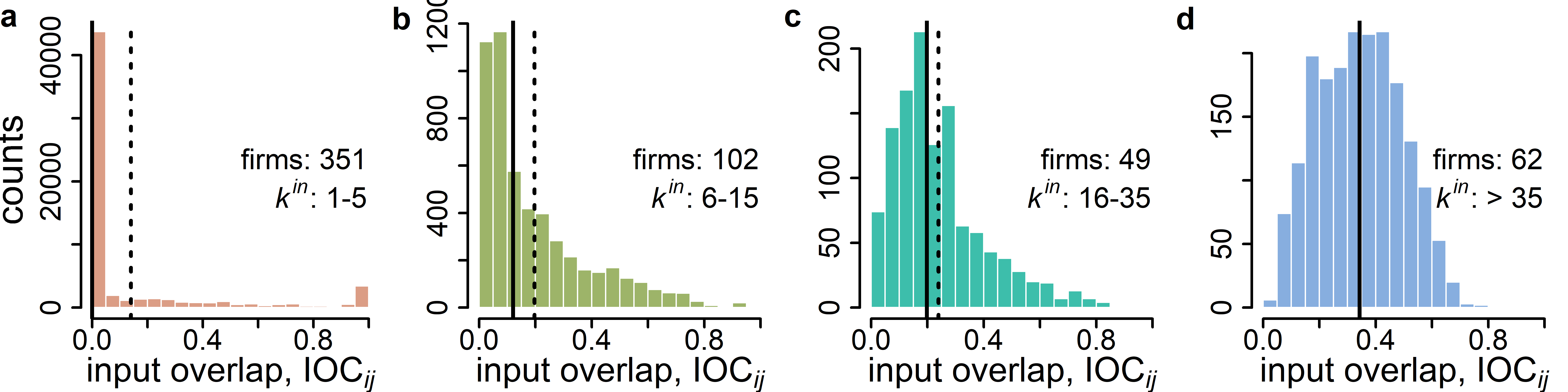} 
	\includegraphics[width= 0.95\textwidth]{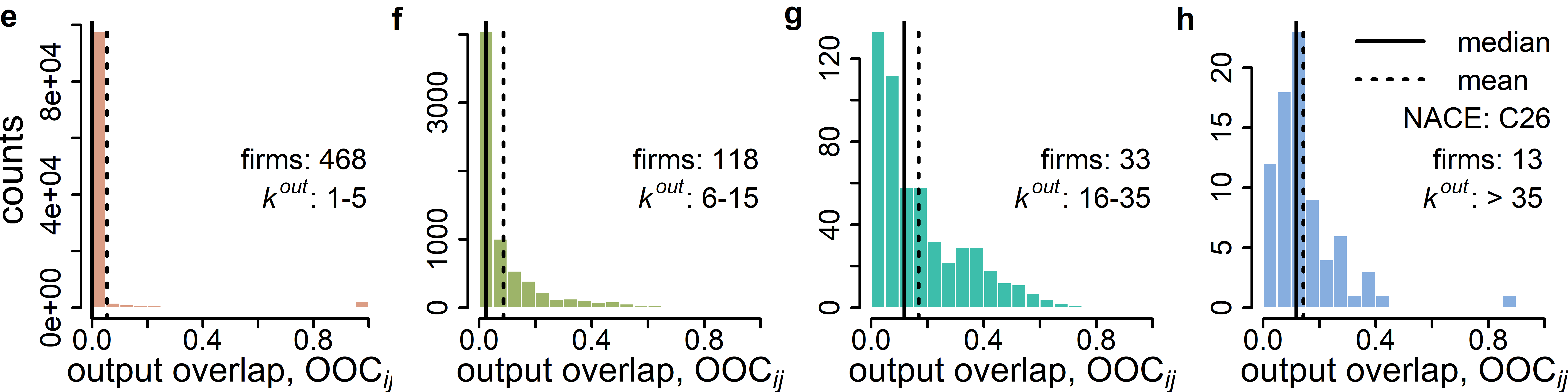} 
	\caption{Pairwise similarity distributions of input and output vectors for firms of the NACE class 26, `manufacture of computer, electronic and optical products'.  
	a-d) show input vector overlap coefficients, IOC$_{ij}$, and e-h) output vector overlap coefficients, OOC$_{ij}$, for four in-degree, $k_i^{in}$, (number of suppliers) and out-degree, $k_i^{out}$ (number of buyers) bins, respectively. Vertical solid lines correspond to median, dashed lines to the average overlap coefficients.
	\textbf{a)} IOC$_{ij}$ for 351 firms with $1 \leq k_i^{in} \leq 5$ suppliers; 
  	\textbf{b)} 102 firms with $6 \leq k_i^{in} \leq 15$suppliers;
 	\textbf{c)} 49 firms with $16 \leq k_i^{in} \leq 35$ suppliers; 
 	\textbf{d)} for 62 firms with 
  	more than $35$ suppliers. It is clearly visible that the similarity of input vectors is low for all numbers of supplier, but increases on average with the number of suppliers.
	\textbf{e)} OOC$_{ij}$ distribution for 468 firms with $1 \leq k_i^{out} \leq 5$ customers; 
 	\textbf{f)} 118 firms with $6 \leq k_i^{out} \leq 15$ customers; 
	\textbf{g)} 33 firms with $16 \leq k_i^{out} \leq 35$ customers; and 
	\textbf{h)} 13 firms with more than $ 35$ customers. 
    The similarity of output vectors is even lower than for input vectors, and also increases on average with the number of buyers. If industry-level aggregation were fully representative for the IO-vectors of firms in NACE C26 in all panels the distributions would correspond to one single bar at an overlap value of 1.
}
	\label{fig2nace2}
\end{figure*}

\begin{figure*}[t]
	\centering
	\includegraphics[width=0.9\textwidth]{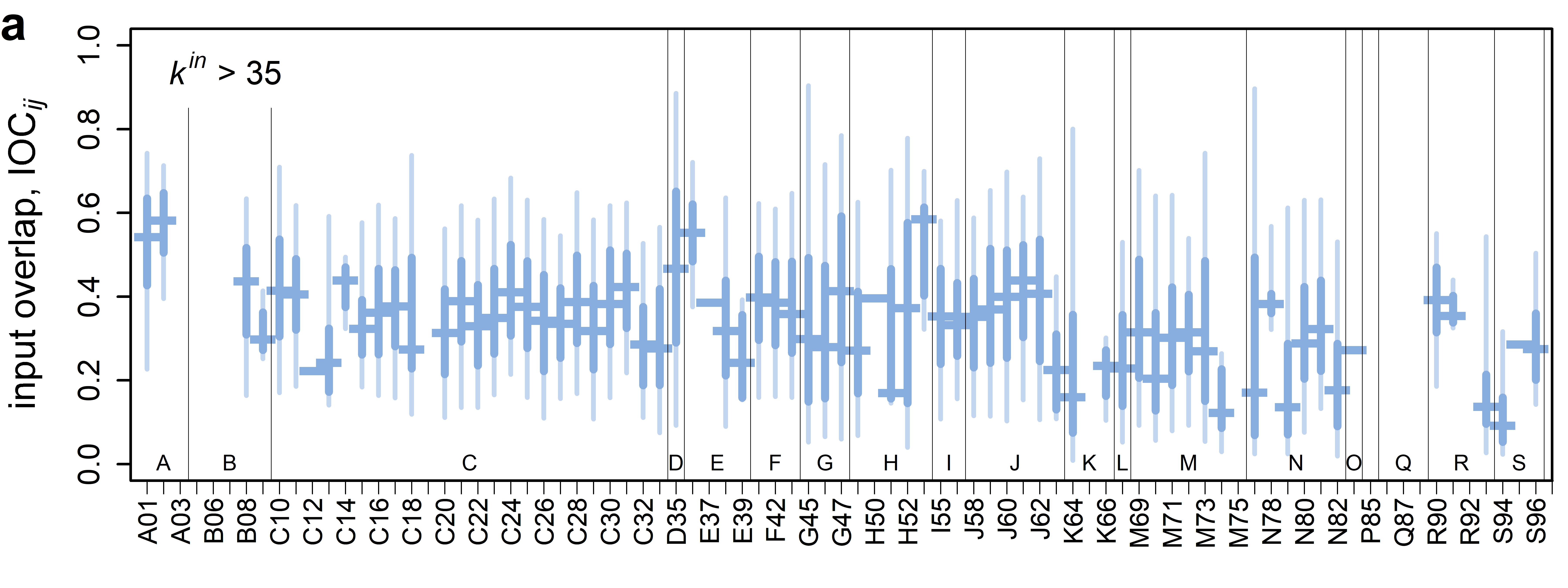} 
	\includegraphics[width=0.9\textwidth]{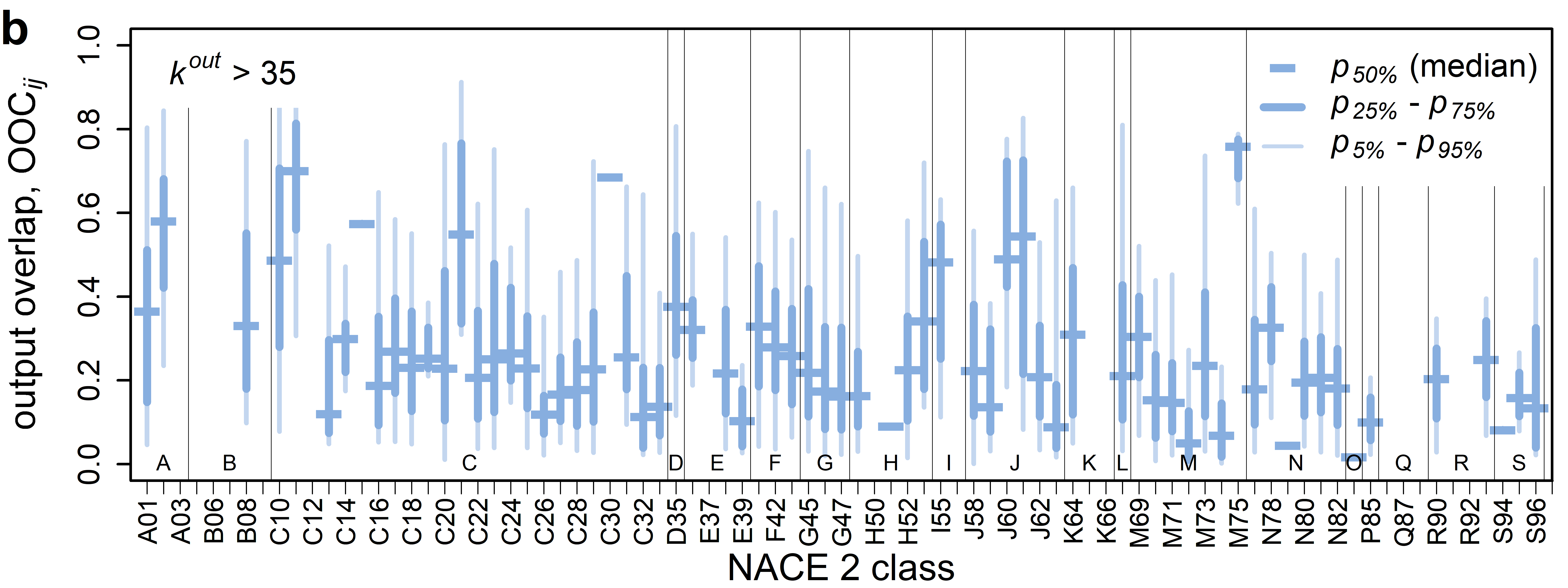} 
	\caption{Pairwise similarity distributions of input- and output-vectors of firms within all NACE2 industries. The overlap coefficient is computed for firms with more than 35 suppliers (a) and buyers (b), respectively. The dark blue horizontal bars in  the boxplots correspond to the median, ($p_{50\%}$), dark blue vertical lines to the inter-quartile range ($p_{25\%}$ -- $p_{75\%}$), and thin light blue vertical lines to error bars ($p_{5\%}$ -- $p_{95\%}$). Thin black vertical lines separate the NACE1 classes. Empty columns indicate  sectors with less than two firms in this degree bin. 
	\textbf{a)} Intra-industry input overlap coefficients, IOC$_{ij}$. The average of the mean (median) input overlaps, across all NACE2 industries is 0.35 (0.33) and the standard deviation of mean (median) input overlaps is 0.084 (0.102). The average standard deviation is 0.156. Relatively low input overlaps are the norm with few outliers such as `agricultural industry' (A1-A2), `water collection, treatment and supply' (E36) and  `transport' (H53).	
	\textbf{b)} Intra-industry output overlap coefficients, OOC$_{ij}$. The average of the mean (median) output overlaps, across all NACE2 industries is 0.282 (0.257) and the standard deviation of mean (median) output overlaps is 0.147 (0.161). Again we see small overlaps. 
    Output overlaps are on average lower than the input overlaps, but there appears to be more variation across industries. 
    If industry-level aggregation were fully representative for the IO-vectors of firms in both panels all distributions would correspond to a single bar at an overlap value of 1. Not a single industry is even close to that value, the highest similarities are found for sectors such as Veterinary activities (M75), Manufacture of beverages (C11), Manufacture of other transport equipment (C30). 
    }
	\label{fig3_nace2main}
\end{figure*}

\section*{Results}

\subsection*{\textbf{Quantifying input and output vector overlaps of firms} }
Large overlaps (firms within sectors are similar) would suggest that aggregation to the industry-level does not lead to large distortions of network dynamics. Small overlaps (firms within sectors are heterogeneous) would lead to potentially large aggregation effects. 
First, we aggregate for every firm its firm-level in- and output vector to the industry-level (NACE2), see \ref{SI_othersimmeas}. Second, for each pair of firms, $i$, and, $j$, within a given NACE2 industry we calculate the  input overlap coefficient (IOC) and the output overlap coefficient (OOC) as, 
\begin{equation} \label{eq_ioc} 
	\text{IOC}_{ij} 
 	=  \sum_{k=1}^{m} \min \left[  \bar{\Pi}^\text{in}_{ik}, \; \bar{\Pi}^\text{in}_{jk} \right] \qquad,  
\end{equation}
\begin{equation} \label{eq_ooc} 
	\text{OOC}_{ij} 
 	=  \sum_{k=1}^{m} \min \left[  \bar{\Pi}^\text{out}_{ik}, \; \bar{\Pi}^\text{out}_{jk} \right] \quad , 
\end{equation}
where $m$ is the number of NACE2 industries (here 86), and $\bar{\Pi}^\text{in}_{i\cdot}$ and $\bar{\Pi}^\text{out}_{i\cdot}$ are the normalized input- and output vectors of firm, $i$, respectively, see \ref{SI_othersimmeas}. $\text{IOC}_{ij}$ specifies the fraction of total inputs, $i$ and $j$ buy from the same industries. It quantifies the common exposure of $i$ and $j$ to supply shocks originating from the same upstream industries and indicates the fraction of a  demand shock that is forwarded  by $i$ and $j$ to the same upstream industries. $\text{OOC}_{ij}$, specifies the fraction total sales,  $i$ and $j$ sell to the same industries. It quantifies the common exposure of $i$ and $j$ to demand shocks originating from the same downstream industries and indicates the fraction of a shock that is forwarded  by $i$ and $j$ to the same downstream industries. For more information, see \ref{SI_details_on_IOC_OOC}.
In Fig. \ref{fig1}b, the relative input vector is $\bar{\Pi}^\text{in}_{10} = (0,0,0.5,0.5,0)$ for firm 10 and $\bar{\Pi}^\text{in}_{11} = (0,0,1,0,0)$ for firm 11, hence, $\text{IOC}_{10, 11} = 0.5$. The propagation of  upstream shocks by 10 and 11 will only overlap by 50\% (sector 3), while 50\% spread to distinct sectors. 

\paragraph*{\textbf{Firms within industries are highly different}}
We show the distribution of the pairwise similarities $\text{IOC}_{ij}$ and $\text{OOC}_{ij}$ for all firms in NACE2 industry C26, `Manufacture of computer, electronic and optical products' in Fig. \ref{fig2nace2}. 
Figure \ref{fig2nace2}a-d show the $\text{IOC}_{ij}$ distributions stratified by their number of suppliers (in-degree, $k_i^{in}$). Figure \ref{fig2nace2}a contains all firms that have 1 to 5 suppliers, Fig. \ref{fig2nace2}b 6 to 15, Fig. \ref{fig2nace2}c 16-35, and Fig. \ref{fig2nace2}d more than 36.  The average similarity of firms' input vectors is small across all four groups for which the median (vertical solid line) and mean (dashed line) overlaps are 0, 0.121, 0.199 343, and 0.141, 0.196, 0.239, 0.343,  respectively. Clearly, the average similarity of input vectors is increasing for firms with more suppliers. The distribution for firms with one to five suppliers (Fig. \ref{fig2nace2}a) is bi-modal, most pairs of firms have either almost no overlap or almost perfect overlap. For firms with a few suppliers (\ref{fig2nace2}b-c) the distributions become unimodal and right skewed, implying that very high similarities appear in the right tail, but are not very frequent. Finally, the distribution of input overlaps for firms with more than 35 suppliers are centered around 0.34 (\ref{fig2nace2}d). 
Figure \ref{fig2nace2}e-h show the distribution of the pairwise output overlap coefficients, OOC$_{ij}$, grouped according to their number of buyers (out-degree, $k_i^{out}$). The bin sizes are the same as before. The average similarity of output vectors is visibly smaller than those of input vectors. The median and mean overlaps for the respective out-degree bins are 0, 0.025, 0.119, 0.119 and 0.054, 0.087, 0.169, 0.143, respectively. The  distributions are more concentrated towards low overlaps and remain right skewed for all out-degree bins. 

\paragraph{\textbf{Similarity of firms is low and varies across industries}}
We now show the summary statistics of the pairwise IOC$_{ij}$ and OOC$_{ij}$ distributions for all NACE2 industries in Fig. \ref{fig3_nace2main}, in particular, the mean, 5\%, 25\%, 50\% (median) 75\%, and 95\% percentiles. Only firms with more than 35 suppliers and buyers are included. The x-axis shows the 86  NACE2 codes present; the y-axis represents the overlap coefficients, each boxplot corresponds to one NACE2 class. Dark blue horizontal bars indicate the median, ($p_{50\%}$), thick dark blue vertical lines indicate the inter-quartile range ($p_{25\%}$ -- $p_{75\%}$), thin light blue vertical lines indicate error bars ($p_{5\%}$ -- $p_{95\%}$), and thin vertical black lines separate NACE1 class affiliations. Empty columns indicate that less than 2 firms exist in the respective sector and degree bin.
Figure \ref{fig3_nace2main}a shows that the low input overlaps of industry C26 are not just an outlier. The mean of the mean (median) input overlaps, IOC$_{ij}$, across NACE2 industries is 0.35 (0.33) and the standard deviation of mean (median) input overlaps is 0.084 (0.102). This indicates that relatively low input overlaps are the norm with few outliers. The highest median IOC$_{ij}$ are found in the `agricultural industry' (A1-A2), `water collection, treatment and supply' (E36) and in the `transport' sectors (H53), whereas the lowest median IOC$_{ij}$ are found in service sectors, such as `other professional', `scientific and technical activities' (M74), `travel agency, and related activities' (N79), `sports activities and amusement and recreation activities' (R93) and `activities of membership organisations'  (S94).  The average standard deviation is 0.156. The standard deviation of standard deviations is small 0.048, and the length of error bars appears to be relatively homogeneous across sectors, suggesting that the variation of pairwise input overlaps, IOC$_{ij}$, is relatively constant across sectors. 
Figure \ref{fig3_nace2main}b shows that output overlaps, OOC$_{ij}$, are on average lower than the input overlaps, but have a higher variation across industries. The mean of the mean (median) output overlaps, OOC$_{ij}$, across all NACE2 industries is 0.282 (0.257) and the standard deviation of mean (median) output overlaps is 0.147 (0.161), indicating  that relatively low output overlaps are the norm with several outliers. For more details, see \ref{SI_further_details_on_IOC_OOC}.

In \ref{SI_other_degree_buckets} we show that for the degree bins 1-5, 6-15, and 16-35 the mean over  mean (median) input overlaps, are 0.132, (0.009) 0.202 (0.148), 0.269 (0.241), respectively; the respective values for output overlaps are slightly lower. As for industry C26, generally input and output vectors of firms within industries become more homogeneous with the number of suppliers and buyers. In \ref{SI_nace4_simmeas} we show the same analysis for NACE4 industries based on NACE4-level input-output vectors and find that the intra-sector variation of input-output vectors is higher than at the NACE2 level. In \ref{SI_jaccard_index}, we show that our results are robust with respect to the choice of the similarity measure. In \ref{SI_section_auto_sim} we show  that the similarity of input and output vectors of firms \textit{over time} is substantially higher than intra-industry similarities. Individual firms show significant similarity from one year to the next, as expected, while the observed low level of intra-industry similarities capture fundamental heterogeneities.

Overall, we clearly see that input and output overlaps of firms within industries are surprisingly low, across industries and across degree bins. The high level of heterogeneity of input-output vectors of firms within industries shows that for most industries sector-level aggregates are practically not representative for the actual firm-level supply chain inter-linkages and very likely will mis-represent dynamic processes occurring on the firm-level network.

\begin{figure}[t]
	\centering
	\includegraphics[width=0.98\columnwidth]{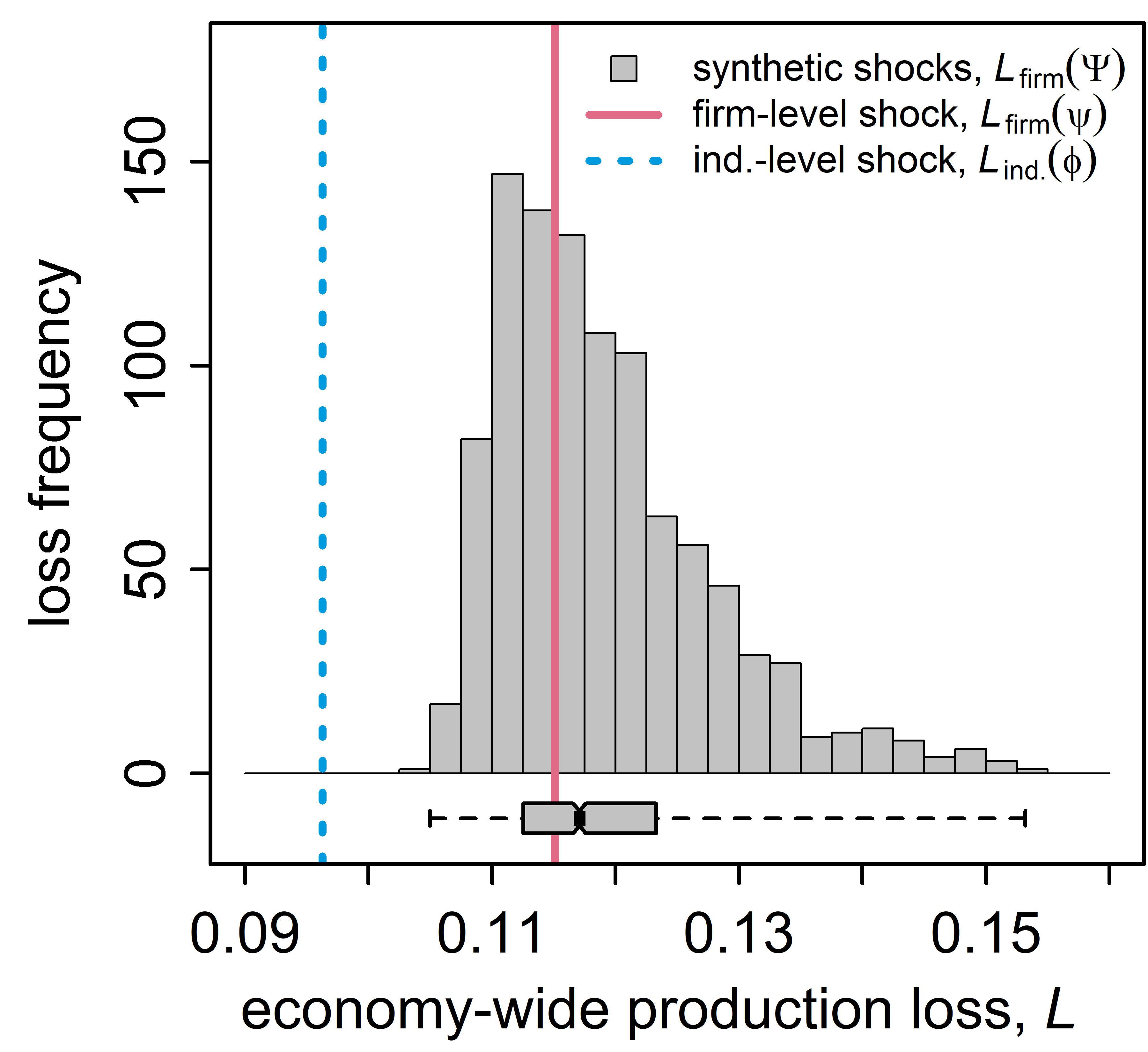} 
	\caption{Economy-wide production losses, $L$, obtained from an empirically calibrated and 1,000 synthetic COVID-19 shocks propagating on the aggregated industry-level production network, IPN, (blue dashed line) and on the firm-level production network, FPN, (red line, histogram). The FPN and IPN correspond to the production network of Hungary in 2019; the firm-level shock, $\psi$, correspond to firms reducing their production level proportional to their reduction in employees between January and May 2020, and are taken from monthly firm-level labor data. The NACE2 level shock, $\phi$, is the aggregation of $\psi$. The 1,000 synthetic shocks, $\Psi$, are sampled such that (when they are aggregated to the NACE2 level) they all have the same size as $\phi$.
The empirically calibrated shock, $\psi$, yields a FPN-based loss, ${L}_{\text{firm}}(\psi)$, of 11.5\% (red line).
The synthetic shocks yield a distribution of FPN-based production losses, ${L}_{\text{firm}}(\Psi)$, ranging from 10.5\% to 15.3\% of national output (histogram). The median is 11.7\% (see boxplot). As a reference, the Hungarian GDP declined by 14.2\% in Q2 2020. Note that for the IPN all realizations, $\Psi$, result in the same production loss, ${L}_{\text{ind.}}(\phi) $, of 9.6\%, by construction. The aggregation to the IPN causes a substantial underestimation of the FPN-based production losses. 
}
\label{fig5_total_losses}
\end{figure}

\subsection*{\textbf{Production loss mis-estimations from aggregating networks}}
We now compare the economy-wide production losses for Hungary caused by a COVID-19 shock propagating once on the firm-level production network (FPN), and once on the industry-level production network (IPN). Based on firms' actual employment reductions, the shock realistically captures how \textit{individual firms} were affected by COVID-19 in the beginning of 2020. The shock is represented by the vector, $\zeta$, where, $\zeta_i$, is the relative reduction of firm $i$'s labor input from January to May 2020, $\zeta_i=\max[0, \: 1-e_i(\text{may})/e_i(\text{jan})]$, and $e_i$ is the number of $i$'s employees in the respective month. The \textit{remaining production capacities} of firms (after the shock) are given by the vector $\psi=1-\zeta$, where, $\psi_i \in [0,1]$, is the remaining fraction of firm $i$'s production, e.g., if $i$ reduced its employees by 20\%, its remaining  capacity is $\psi_i = 0.8$. Aggregating the capacities, $\psi_i$, of all firms $i$ in sector $k$ gives   sector $k$'s remaining production capacity, $\phi_k$. For details on  shock construction and aggregation, see Data and Methods.

Following the COVID-19 shock, we simulate how the adaptation of firms' supply- and demand-levels propagate  \textit{downstream} and \textit{upstream} along the PN, once on the firm-level and once on the industry-level. We employ the simulation model of \citep{diem2022quantifying}, where each firm (industry) is equipped with a generalized Leontief production function, see Data and Methods for details. The simulation stops when the production levels of firms have reached a new stationary state at (model-internal) time, $T$. Every firm $i$ (or sector $k$) has a \textit{final production level}, $h_i(T,\psi) \in [0,1]$ ($h_k(T, \phi)\in [0,1]$), that depends explicitly on the details of the shock $\psi$ ($\phi$). It represents the fraction of the original production, $s_i^\text{out}$, firm $i$ (sector $k$) maintains after the shock has propagated. We define the \textit{FPN-based economy-wide production-loss} as 
\begin{equation} \label{ESRI_net}
	{L}_{\text{firm}}(\psi) = \sum_{i=1}^{n} \frac{s^\text{out}_i }{\sum_{j=1}^n s^\text{out}_j }\big(1-h_i(T,\psi) \big) \quad .
\end{equation}
It is the fraction of the overall revenue in the network (measured in out-strength, $s_i^\text{out}$, see Data and Methods) that is lost due to the shock and the in-direct effects of its propagation. The \textit{IPN-based economy-wide production-loss}, ${L}_{\text{ind.}}(\phi)$, is defined accordingly, see Eq. [\ref{ESRI_net_ind}] in Data and Methods. 

\begin{figure*}[t]
	\centering
	\includegraphics[width= 1\textwidth]{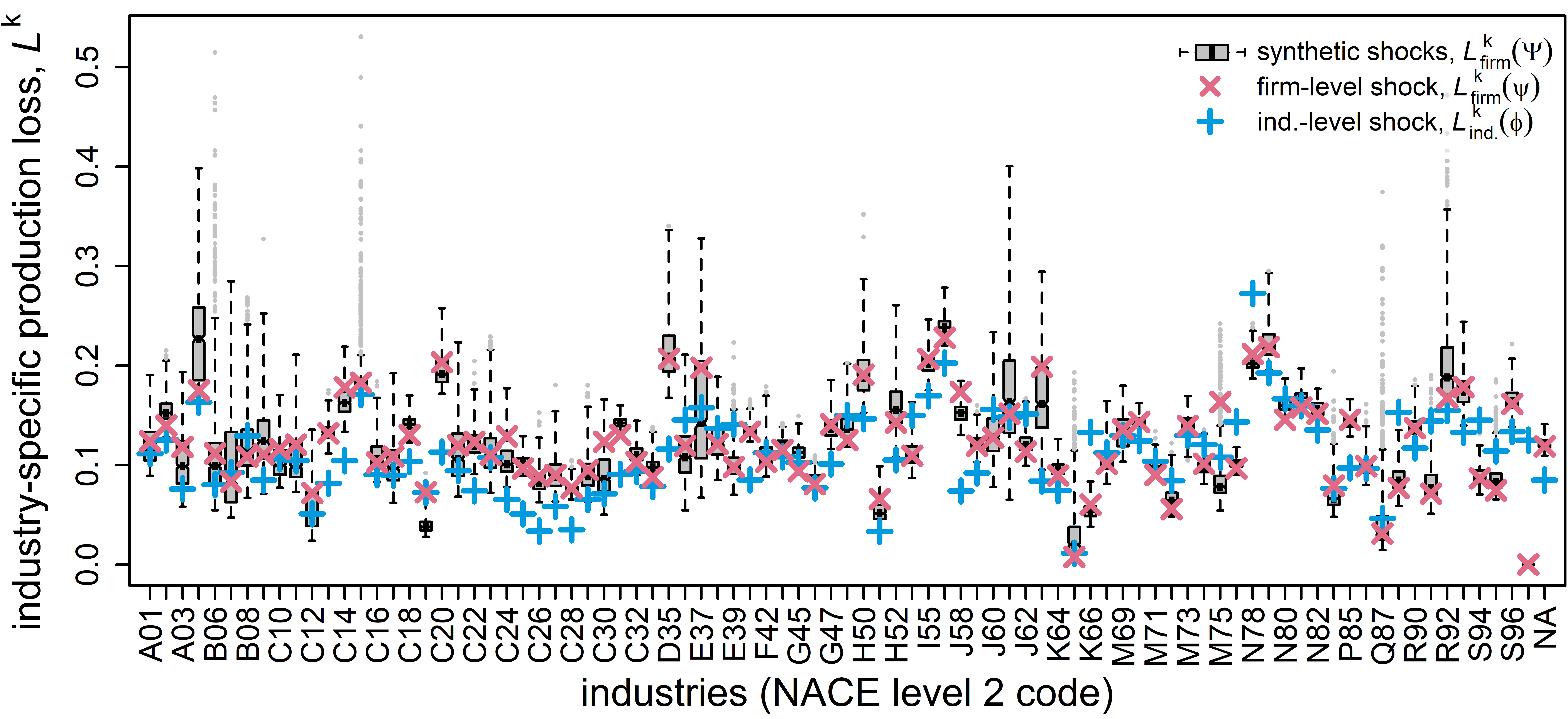} 
	\caption{Comparison of industry-specific production losses, $L^k$, obtained from an empirically calibrated and 1,000 synthetic COVID-19 shocks propagating on the aggregated industry-level production network, IPN, (blue `+'es) and on the firm-level production network, FPN, (red `x'es, boxplots). For most industries the FPN-based production losses, ${L}^k_{\text{firm}}(\Psi)$, (boxplots)  vary strongly across the 1,000 synthetic shocks even though shocks have the same size when aggregated to the industry-level. Shock propagation on the industry-level (blue `+'es) can not capture this variation. IPN-based production-losses typically under-estimate the FPN-based production losses severely.}
	\label{boxplot_emp_shock_may_GL_nace2}
\end{figure*}

Figure \ref{fig5_total_losses} compares the production losses for the two simulations, FPN and IPN. The propagation on the FPN leads to a production loss, ${L}_{\text{firm}}(\psi)$, of 11.5\% (red solid line), while propagation on the IPN yields a loss,  ${L}_{\text{ind.}}(\phi)$, of 9.6\% (blue dashed line). Aggregated industry-level shock propagation substantially underestimates the production losses caused by firm-level shock propagation dynamics, for the COVID-19 shock, $\psi$, by 16.5\%. We quantify the size of mis-estimations if the firm-level shock was slightly different. We sample 1,000 distinct, synthetic realizations of the COVID-19 shock that are of the same size when aggregated to the industry-level, but affect firms within industries differently. For every sector, $k$, we take the empirical distribution, $\zeta_i$, of firms $i$ belonging to that sector. Then, we sample for every company, $i$, of sector $k$ a new value, $\zeta_i^{\rm sample}$,  from this distribution, replace the old $\zeta_i$, and calculate the corresponding remaining production capacity vector, $\psi^{\rm sample}=1-\zeta^{\rm sample}$. In this way we generate the set of 1,000  synthetic capacity vectors, $\Psi=\{\psi^1,\psi^2, \dots, \psi^{1,000} \}$; for the full algorithm, see \ref{SI_shock_construction}. The resulting distribution of FPN-based economy-wide production-losses ${L}_{\text{firm}}(\Psi)$ is shown as histogram and boxplot in Fig. \ref{fig5_total_losses}. The losses vary strongly from 10.5\% to 15.3\% of economy-wide production, i.e. losses can vary by a factor of up to 1.46 for different initial shocks of the same size. The actual Hungarian GDP declined by 14.2\% in Q2 2020 \citep{OECD2023},  showing that the losses obtained by our computations are within perfectly realistic bounds; a gross output estimate is not available for comparison. 

Note that the 1,000 synthetic shocks propagating on the IPN always lead to the same economy-wide production-loss (9.6\%, blue dashed line) because all firm-level shocks, $\Psi$, impact the industry-level production capacities by exactly the same amount, $\phi$. On average the IPN based production losses underestimate FPN-based losses by 2.3\% of the economy-wide production. In relative terms losses are on average underestimated by 
 18.7\%. For 10\% of the shocks the underestimation is even larger than 26.3\% and the maximum underestimation is 37.1\%.
This tail of large losses is clearly visible in the histogram and is caused by shocks affecting systemically relevant firms stronger \citep{diem2022quantifying}. The median and mean of the 1000 losses, ${L}_{\text{firm}}(\Psi)$, are 11.7\% and 11.9\%, respectively, and lie close to the FPN-based production loss, $L_{\text{firm}}(\psi) = 11.5\%$, based on the original COVID-19 shock, $\psi$ (red line).

\paragraph*{\textbf{Mis-estimating industry-specific production losses}}
We now compare the IPN- and FPN-based production losses for every NACE2 industry separately. We define the \textit{FPN-based industry-specific production loss} of industry, $k$, in response to the COVID-19 shock, $\psi$,  as
\begin{equation} \label{ESRI_sec}
	{L}_{\text{firm}}^k(\psi) = \sum_{i=1}^{n} \frac{s^\text{out}_i }{\sum_{j=1}^n s^\text{out}_j \delta_{p_j, k} }\big(1-h_i(T, \psi) \big) \delta_{p_i, k} \quad .
\end{equation}
It is the fraction of revenue (measured in out-strength) that firms in sector $k$ lost due to the direct and in-direct effects of the shock. The \textit{IPN-based industry-specific production loss}, $ {L}_{\text{ind}}^k(\psi)$, is defined accordingly.

In Fig. \ref{boxplot_emp_shock_may_GL_nace2} we show for each NACE2 industry the distribution of \textit{FPN-based production-losses}, ${L}_{\text{firm}}^k(\Psi)$, caused by the 1,000 synthetic shocks as boxplot. The  \textit{IPN-based production-losses}, ${L}_{\text{ind.}}^k(\phi)$ are indicated by the blue `+'es, the \textit{FPN-based production-losses} for the original COVID-19 shock, ${L}_{\text{firm}}^k(\psi)$, are given by red `x'es. It is clearly visible that for many industries losses vary strongly across the identically sized shocks, but also the variation  between industries is noteworthy. For all but two industries (M73, N82), the production loss distributions are right skewed, few industries (B06, C15, K65, M75, Q87, and R92) have  substantial outliers (grey dots) above 3 times the inter-quartile range. This means that for some particular shock realizations these sectors can suffer extremely large losses. The minimum and maximum values of production losses for different initial shocks can differ by factors of up to 9.5 (B06), 6.0 (B07), 5.7 (C12), 6.2 (J61), 41 (K65), or 25.9 (Q87). The median (mean) ratios of maximum to minimum loss is 2 (3.2). This variation in production losses across different shocks is inaccessible when using aggregated IPN data; it can not be inferred from the blue `+'es.  The large variations emerge as different shocks affect firms at different positions in the supply networks that have different systemic relevance~\citep{diem2022quantifying}. IPN-based losses (`+'es), lie frequently below the lowest FPN-based loss, while FPN-based COVID-19 losses (`x'es) lie within boxplots. The industries where IPN-based shock propagation under-estimates output losses most are  C26 (-59.5\%), C28 (-53.5\%), J58 (-51.3\%), C25 (-50.3\%), J63 (-47.8\%), and C20 (-42.1\%). Over-estimation of production losses from using IPN-based losses are highest for sectors, K66 (150\%), C19 (87.4\%), R91 (83.3\%), Q88 (80\%), S94 (65.4\%) and E39 (42\%). For other sectors, see \ref{SI_section_details_on_sector_losses}. We calculate for each industry the \textit{mean absolute deviation} and take the average across industries, which yields 30.2\%. 

Last, we consider the hypothetical case that shocks propagate 
on the same PN, but assuming that all firms have linear production functions, see \ref{SI_lin_shocks}. We find that the distribution of economy-wide production losses, $L_{\text{firm}}(\Psi)$, ranges from 9.5\% to 10.8\%. This is substantially less variation than when realistic non-linear production functions are used. As expected, the linear production function assumption makes the economy-wide production losses less dependent on which exact firms within industries are impacted by shocks. However, the variations of industry-specific production-losses, $L^k_{\text{firm}}(\Psi)$, are still very large for several sectors. This emphasizes immediately that in order to correctly estimate sector-level production losses it is crucial which firms are affected by shocks, even in the best-of-all worlds, where shocks would propagate linearly.

\section*{Discussion}

Production networks are fundamental for explaining and predicting  dynamical economic phenomena. For almost a century, these were only accessible as aggregated industry-level production networks (IPNs), usually represented as input-output tables (IOTs). Only recently, large scale firm-level production networks (FPNs), covering entire economies have become available. Based on a unique firm-level production network data set, containing almost all buyer-supplier links of the Hungarian economy, we demonstrated on the one hand that the aggregation of production networks to the industry-level can not be expected to yield anything close to correct predictions of dynamical processes, such as the propagation of short term shocks through  production networks. On the other hand we showed that using firm-level supply networks instead, a much more realistic picture can emerge. 

We first showcased that industries are not sufficiently representative of the firms they include, because firms within industries are highly heterogeneous wrt. the industries they buy from and sell to. Specifically, two firms within the same industry spend on average only 23.5\% on inputs from the same industry, and sell on average only 19.3\% of their revenues to the same industry. Even when two firms belong to the same industry their industry-level input and output vectors will differ substantially. Therefore, using industry-level production network data will likely cause substantial mis-estimations of dynamic processes actually occurring on the firm-level.

We next demonstrated that the aggregation of FPNs causes indeed large mis-estimations for economic shock propagation dynamics and the resulting production losses. The demonstration is based on a COVID-19 shock that is realistically calibrated with firm-level employment data and 1,000 synthetic COVID-19 shocks of the same size. While economy wide production losses, in response to the 1,000 shock scenarios, simulated on the FPN range from 10.5\% to 15.3\% (mean 11.9\%), the corresponding IPN-based production losses are 9.6\%. In the worst case scenario the underestimation amounts to 37.1\%. For single industries  the largest average mis-estimation of production losses range from -59.5\%  to  150\%.

\paragraph{\textbf{Implications for economic modelling and policy making}}
The presented results imply a range of immediate consequences for economic modelling, in particular for short-term economic dynamics such as shock propagation, but also more generally for the reliability of industry-level IO-models in the context of testing policy implications.

First, our findings make it crystal clear that the size of losses from shock propagation depends crucially on which exact firms are affected by the initial shock. Crises such as COVID-19, the war in Ukraine, or large natural disasters can affect firms within the same industry sectors and regions very differently and, hence, the exact materialization of the shock can lead to significantly different in-direct economic losses. Aggregated industry-level models, such as IO-models, can by design not account for this, potentially underestimating tail losses that appear when a group of systemically important firms receive shocks at the same time. Modelling impact propagation on firm-level production networks might significantly improve economic assessments of crises of this kind. 

Second, our method for creating an ensemble of synthetic shock scenarios that are identical on the industry-level, but affect firms differently can be used to estimate realistic confidence intervals for economic impacts of crises. Experts can define a shock on the industry-level (as done routinely for IO-models) and obtain distributions of the quantity of interest for each firm, specific sectors or the whole production network. This approach could reveal which combination of shocks to  individual firms causes particularly  dangerous scenarios that would go unnoticed with industry-level models. This is useful for designing scenarios in economic stability stress tests.

Third, the presented framework extends well beyond shock propagation. Other forms of network dynamics that are certainly  distorted by industry-level aggregation include economic growth, the estimation of CO2 emissions of economic activity, or the spread of price increases. Detailed future research on these topics, considering the details of firm-level production networks is necessary. These topics happen on larger time-scales and will be overlaid with other dynamics that were not covered here. These dynamics are most likely more complicated than the ones of short-term shock propagation and therefore it is reasonable to assume that the effects of aggregation are even stronger in these situations.

Fourth, specifically, for estimating CO2 emissions of industry sectors and countries, aggregating input-output tables causes substantial errors in emission estimates  \citep{su2010input, SU2010spatialaggregation,  lenzen2011aggregation}. Our results indicate, firms in the same NACE industries use very different inputs and sell to very different industries and therefore their resulting scope-3 emissions (indirect CO2 emissions along supply chains) can differ substantially. Firm-level data will be crucial for reliable and targeted CO2 emission estimates and for designing green transition enhancing economic policies that can target problematic firms \citep{stangl2023Using}. 

Fifth, in the past economic models, e.g. for assessing economic effects of natural disasters such,  as \citep{henriet2012firm}, have worked with the simplifying assumption that firms within an industry are the same wrt. their input and output vectors.
Our results suggest that for estimating and predicting effects of natural disasters in the future more reliably,  production network models should carefully feature firm-level heterogeneity within industries.

\paragraph{\textbf{Limitations and future research}}
There is a list of limitations of the presented material. For self-consistency, the industry-level production network used here is simply the aggregation of a firm-level production network. IO tables are constructed with extensive survey methodologies and the available tables can differ \citep{borsos2020unfolding}. However, also IO tables are aggregations of underlying firm- and establishment-level networks and are likely to be affected by the same problems and to a comparable extent. 

Secondary NACE categories of firms are not contained in our dataset. Larger firms producing several different types of products (in potentially several establishments) are fully aggregated to their primary NACE category. This could lead to an over-estimation of heterogeneity of input and output vectors within industries. Future research should quantify the heterogeneity of input and output vectors of establishments used for creating IO-tables. 

A potentially strong limitation is that we do not have information of firms' international import and export links. Consider two firms in one sector, one imports a specific input and the other sources it domestically we would over-estimate the heterogeneity of their input vectors. However, for the Belgium production network it has been shown that only a small fraction of firms have direct import and export linkages \citep{dhyne2021trade}.

In practice high quality economic data to calibrate industry-level economic models is widely available and some have achieved good forecasting performance \citep{pichler2022forecasting}. To calibrate firm-level models, substantially larger amounts of data are needed. For example, quantifying how a shock (e.g. a natural disaster) affects hundreds of thousands of firms is substantially harder than for a few dozens of sectors. Firms within sectors do react differently, modelling their  behavior realistically, involves many assumptions, but up to now data for calibration is scarce. 

We demonstrated that for how shocks propagate details do matter. In our simulation model important non-linearities appear in the generalized Leontief production functions (GLPF) of companies. The calibration of firms' GLPFs is currently a rough approximation combining firms' NACE4 industry affiliation with an expert based survey for 56 industry sectors conducted in \cite{pichler2022forecasting}. The calibration of the GLPF needs refinement in the future, e.g., with large scale firm-level surveys. 

Our results point out relevant open questions. 
\cite{duprez2018price} find large idiosyncrasies in price changes of producers within the same product categories. It would be interesting to see, whether these could be explained by the heterogeneity of firms' input and output vectors. In the direction of IO tables, differences of Leontief multipliers for different aggregation levels of IO-tables with potential implications for predicting economic growth were reported \citep{mcnerney2022production}. It would be of interest to see how this extends across all scales to the firm-level.  \cite{heinrich2022levels} show that correlation structures found on the sector level do not hold on the firm-level. Also for this phenomenon the intra-sector heterogeneity of firms could be part of the explanation. The effects of  of heterogeneities should also be checked for establishment level supply networks \citep{schueller2022propagation}.

General equilibrium models \citep{acemoglu2012network, carvalho2019production, magerman2016heterogeneous} were shown to depend on network measures such as Leontief multipliers or the `influence vector'. These are likely to be distorted from aggregating production networks to the industry-level. The sensitivity of results to aggregation could be investigated under a similar framework as the present one.

It has been shown that both industry \citep{acemoglu2012network} and firm-level \citep{borsos2020unfolding} production network exhibit power-law scaling patterns. It would be fascinating to find out under what conditions they preserved under aggregation and --- if not --- would that explain the differences in shock propagation and other network dynamics? Another open question is, which network modules are particularly affected by aggregation? 
And, finally, since our data shows that input and output vectors of firms remain relatively stable from one year to another. This raises the question of how fast can production networks adapt to technological change? And would an aggregate perspective of production networks under- or over-estimate the speed of adaption in the network?

Further remaining questions include input combinations. The reported large heterogeneity of inputs and outputs  of firms within the same sectors implies that the same output can be produced from different input combinations. If one input is no-longer available this might affect a certain company, while others continue production. In the longer term, if one input is becoming structurally more expensive firms could change the production to mimic competitors that use a different input mix to produce the same good. This raises the question if this large amount of heterogeneity in input and output vectors is actually a source of resilience in the production network, or just an inefficiency in knowledge transfer?

To conclude, in this work we showed the importance of modelling production networks on the firm-level. However, currently data on firm-level production networks exist only in  very few countries, and is rarely available to research. This work shows how necessary it is to make these data usable for  researchers and policy institutions. Complementing traditional industry-level models with new models that are specifically designed for firm-level data is a great opportunity forward for both  reliable policy making and progress of scientific research on resilience and transformability of the current economy.

\section*{Data and Methods}

\paragraph{\textbf{Data}}
The Hungarian FPN, $W$, is based on the 2019 VAT micro-data of the Hungarian Central Bank  \citep{borsos2020unfolding, diem2022quantifying}. 
Supply links between two firms are present if the tax content of the transactions was above 1 million Forint for 2018Q1-Q2 and 100,000 Forint for 2018Q3-2019Q4  (approx. 250 euros). The link weight, $W_{ij}$, represents the monetary value of all transactions between the two firms in the given year. We filter the data for stable supply links and 
keep a link if at least two supply transactions occurred in two different quarters, i.e. we exclude one-off transactions. The filtering reduces the number of links from approx. 2 millions to 1.1 millions, but the transaction volume drops only by approx. 10\%. The number of firms drops from 315,259 to 243,339 in 2019 and for 2018 from  296,992 to 185,322. Imports and exports are not contained in the data set. The industry affiliation of firms, $p_i$, correspond to the NACE classifications contained in the Hungarian corporate tax registry. On the NACE2 level 86 different classes are present, on the NACE 4 level 587. In 2019 the NACE affiliation is missing for 62,782 firms; in 2018 for 42,385 firms. We treat them as a residual NACE class. 

\paragraph{\textbf{Constructing firm- and industry-level COVID-19 shocks}}
The employment data (collected by the Hungarian tax authority available at the central bank) contains the number of employees, $e_i(\tau)$, firm $i$ employed in the respective month $\tau$. We assume that labor is an essential (Leontief-style) input to a firm's production (Eq. [\ref{eq_glpf}]), and that after a shock firms only keep the amount of employees needed to operate at the new reduced production level. Therefore, we treat the empirical reduction of employees as a signal for how strong the firm was affected by the consequences of the pandemic in beginning of 2020. No furlough schemes were in place in Hungary. Note that January is sufficiently distant from COVID-19 affecting Europe and May is the time when the initial shock should be fully incorporated in the employment data; there is a two months leave notice period in Hungary. The Hungarian labor data is available for approx. 160,000 firms. For the firms with no data we impute the value by drawing the fraction of employment from firms in the respective NACE4 category where the data is available. We conduct the imputation 1,000 times and receive 1,000 completed vectors. For each of them, we calculate  the value of economy-wide lost production, ${L}_{\text{firm}}$, (see Eq. [\ref{ESRI_net}]). We choose the completed shock vector that yields the median loss of production as $\psi$.
The corresponding industry-level COVID-19 shock is calculated by aggregating the vector $\psi$, to the NACE2 industry-level. As firms within a sector mostly have different ratios of in- and out-strength --- i.e., $s^\text{in}_i / s^\text{out}_i \neq s^\text{in}_j / s^\text{out}_j$ ---, we aggregate the firm-level production capacities to a vector of downstream-constrained, $\phi^\text{d}$, and a upstream-constrained remaining production capacity, $\phi^\text{u}$. For industry $k$, $\phi^\text{u}_k,\phi^\text{d}_k$ are calculated as
\begin{equation} \label{eq_aggr_psi}
	\phi^\text{u}_k = \frac{\sum_{i=1}^{n} \psi_i \; s^\text{in}_i \; \delta_{p_i, k} }{ \sum_{i=1}^{n} s^\text{in}_i \; \delta_{p_i, k} } , \qquad 
	\phi^\text{d}_k = \frac{\sum_{i=1}^{n} \psi_i \; s^\text{out}_i\; \delta_{p_i, k} }{ \sum_{i=1}^{n} s^\text{out}_i \; \delta_{p_i, k} } \quad .
\end{equation} 
We use the notation $\phi = (\phi^\text{u}, \phi^\text{d})$. We show the aggregated shock, $\phi$, for each NACE2 class in see \ref{SI_section_details_on_sector_losses}. Creating synthetic shocks,  $\psi^1, \psi^2, \dots, \psi^{1,000}$, --- that when aggregated to the industry-level are identical to $\phi^\text{d}$ and $\phi^\text{u}$ --- can be achieved by $\psi^1, \psi^2, \dots, \psi^{1,000}$ fulfilling Eq. [\ref{eq_aggr_psi}]. This implies that the aggregated firm-level shocks all fulfil $\phi^\text{u,1} = \phi^\text{u,2} = \dots = \phi^\text{u,1000}$ and $\phi^\text{d,1} = \phi^\text{d,2} = \dots = \phi^\text{d,1000}$. For details, see \ref{SI_shock_construction}. 

\paragraph{\textbf{Shock propagation model}}
The production process of each firm $i$ is represented by a generalized Leontief production function, defined as
	\begin{equation} \label{eq_glpf}
		x_i = \min\Bigg[
		\min_{k \in \mathcal{I}_i^\text{es}} \Big[ \frac{1}{\alpha_{ik}}\Pi_{ik}\Big], \: 
		\beta_{i} + \frac{1}{\alpha_i} \sum_{k \in  \mathcal{I}_i^\text{ne}} \Pi_{ik} ,  \; \frac{1}{\alpha_{l_i}}l_i, \; \frac{1}{\alpha_{c_i}}c_i \;   \Bigg] \, .
	\end{equation}
$\Pi_{ik}$ is the amount of input $k$ firm $i$ uses for production, $\mathcal{I}_i^\text{es}$ is the set of  essential inputs,   $\mathcal{I}_i^\text{ne}$ is the set of non-essential inputs of firm $i$; $l_i$ and $c_i$ are $i$'s labor and capital inputs. The essential and non-essential input types of firms are assigned according to their industry affiliation (NACE4) and an expert based survey for 56 industry sectors conducted by \citep{pichler2022forecasting}. The parameters $\alpha_{ik}$ are technologically determined coefficients, $\beta_{i}$ is the maximum production level possible without non-essential inputs $k \in \mathcal{I}_i^\text{ne}$ and  $\alpha_{i}$ is chosen to interpolate between the full production level (with all inputs) and $\beta_{i}$. All parameters are determined by $W$, $\mathcal{I}_i^\text{es}$ and $\mathcal{I}_i^\text{ne}$. The COVID-19 shock, $\psi$,  propagates through the Hungarian production network in the following way. Initially, at time $t=0$ the network, $W$, is stable and the production amount of each firm $i$ corresponds to its out-strength, $x_i(0) = s^\text{out}_i$, where $s^\text{out}_i$ corresponds to firm $i$' original revenue from its activity in the FPN, $W$. We denote firm $i$'s remaining fraction of production, at time $t$ as $h_i(t)= x_i(t)/x_i(0)$, hence at time $t=0$ before any shocks occur $h_i(0) = 1 \; \forall i$. At time $t=1$ the initial shock materializes and production levels of each firm $i$ drop to the remaining production capacity, $h_i(1) = \psi_i$. Then, we simulate how firms propagate the received shock upstream by reducing their demand to suppliers and downstream by reducing their supply to customers.  Missing non-essential inputs cause production reductions in a linear fashion, while a lack of essential inputs affects output in the non-linear Leontief way, i.e. downstream shocks can have strong negative impacts on production, depending on the supplier-buyer industry pair. The loss of a customer leads to a production reduction proportional to the customers' revenue-share, i.e. upstream shocks have only linear impacts. For each firm, $i$, we update the production output, $x^\text{d}_i(t+1)$, at $t+1$, given the downstream constrained production levels of its suppliers, $h^\text{d}_j(t)$, at time $t$ as
\begin{eqnarray} \label{eq_downstream}
	x_i^{\text{d}}(t+1) & = &   \min\Bigg[
	\min_{k \in \mathcal{I}_i^\text{es}} \left( \frac{1}{\alpha_{ik}}\sum_{j=1}^{n} W_{ji} h_j^\text{d}(t) \delta_{p_j,k}\right), \\
	  & &  \beta_{i} + \frac{1}{\alpha_i} \sum_{k \in  \mathcal{I}_i^\text{ne}}  \sum_{j=1}^{n}   W_{ji} h_j^\text{d}(t) \delta_{p_j,k}, \, \psi_i x_i^\text{}(0)
	\Bigg] \quad .  \notag
\end{eqnarray}
The production output, $x^\text{u}_i(t+1)$, of firm $i$ at $t+1$, given the upstream constrained production level of its customers, $h^\text{u}_l(t)$, at time $t$ is computed as
\begin{equation}\label{eq_upstream}
	x_i^{\text{u}}(t+1) = \min\Bigg[ \sum_{l=1}^{n} W_{il}h_l^\text{u}(t) , \, \psi_i x_i^\text{}(0) \Bigg] \quad. 
\end{equation}
The algorithm converges at time $T$, yielding final production levels, $h_i(T, \psi)$, for each firm $i$. The dependence of the final production level on the initial shock is made explicit by writing $h_i$ as a function of $\psi$. Note that the quantity,  $ s_i^\text{out} \big(1-h_i(T, \psi)\big)$, is the amount of lost revenue of firm $i$ due to the initial shock and its propagation. For a complete description of the algorithm, see \citep{diem2022quantifying}.
For simulating shocks on the industry-level network, $Z$, in Eqs. [\ref{eq_downstream}]-[\ref{eq_upstream}] we replace $W$ with $Z$, in Eq. [\ref{eq_downstream}], $\psi_i$ with $\phi^\text{d}_i$, and in Eq. [\ref{eq_upstream}] $\psi_i$ with $\phi^\text{u}_i$. 
This results in the final production levels, $h_k(T, \phi)$, for each sector, $k$, and we set ${L}_{\text{ind.}}(\phi) = 1- h_k(T, \phi )$. The overall production loss, ${L}_{\text{ind.}}^k(\phi)$, is calculated analogously as in Eq. [\ref{ESRI_net}], based on the out strengths, $s_k^\text{out}$, of sectors, $k$, as 
\begin{equation} \label{ESRI_net_ind}
	{L}_{\text{ind}}(\phi) = \sum_{k=1}^{m} \frac{s^\text{out}_k}{\sum_{l=1}^m s^\text{out}_l }\big(1-h_k(T,\phi) \big) \quad .
\end{equation}

\paragraph{Acknowledgements}
This work was supported in part by  the OeNB Anniversary fund project P18696, the Austrian Science Promotion Agency FFG project 886360, and the Austrian Federal Ministry for Climate Action, Environment, Energy, Mobility, Innovation and Technology project GZ 2021-0.664.668  and
H2020 SoBigData-PlusPlus grant agreement ID 871042.

\paragraph{Author contributions}
CD and ST conceived the work. 
AB cleaned and prepared the data.
CD wrote the code. 
CD, AB performed the analysis.
All analyzed and interpreted the results. 
CD and ST wrote the paper. 
All contributed towards the final manuscript.

\bibliography{references}

\newpage 

\onecolumn
\appendix

\renewcommand{\thesection}{SI Section \arabic{section}}  
\setcounter{section}{0}

\renewcommand{\thefigure}{S\arabic{figure}}
\setcounter{figure}{0}

\renewcommand{\thetable}{S\arabic{table}}
\setcounter{table}{0}

\renewcommand{\theequation}{S.\arabic{equation}}
\setcounter{equation}{0}

\FloatBarrier

 \section*{{\Large Supplementary Information}}

\section{Calculating input and output vectors} \label{SI_othersimmeas}
In this section we show how to calculate the intra-sector heterogeneity (or similarity) of the firms' input-output vectors. 
We start aggregating every firms' firm-level in- and output vector to the NACE2 industry-level.
The $i$th column of the FPN's adjacency matrix, $W$, represents the firm-level input vector, $W_{i.}$, of firm $i$, while the $i$th row gives the firm-level output vector, $W_{.i}$. 
We compute the corresponding industry-level input vector, $\Pi^\text{in}_{i.}$, and output vector, $\Pi^\text{out}_{i.}$, of firm $i$, by aggregating all in links (purchases) of $i$'s suppliers from the same industry and all out links (sales) to $i$'s customers in the same industry, as
\begin{equation}
	\Pi^\text{in}_{ik} = \sum_{j=1}^{m} W_{ji} \delta_{p_j, k}, \qquad \Pi^\text{out}_{ik} = \sum_{j=1}^{m} W_{ij} \delta_{p_j, k} \quad .
\end{equation}
The element $\Pi^\text{in}_{ik}$, specifies the amount of input $k$ firm $i$ is buying from suppliers, $j$, of industry, $k$, i.e., all $j$ with $p_j = k$. The element $\Pi^\text{out}_{ik}$ specifies the amount firm $i$ is selling to firms, $j$ in industry, $k$, i.e., all $j$ with $p_j = k$. 
The expression $\delta_{p_j k}$ is the Kronecker delta and is equal to one if firm $j$ produces product $k$ and zero otherwise, i.e., 
$$\delta_{p_j k} = \begin{cases}
	1 \quad \text{if} \; \; p_j = k \qquad, \\
	0 \quad \text{if} \; \;  p_j \neq k \qquad .
\end{cases}  $$

We focus on the relative importance of firms' input types (industries) and customer industries, independent of firm size. 
% I would place this  paragraph in the SI
% referenced as SI text 4
To do so, we compute the normalized input-, $\bar{\Pi}^\text{in}_{i.}$, and output vectors, $\bar{\Pi}^\text{out}_{i.}$, of every firm, $i$. The k$^\text{th}$ entry of the normalized input vector,  $\bar{\Pi}^\text{in}_{ik}$, represents the fraction of inputs firm $i$ buys from firms in industry $k$. Similarly, the k$^\text{th}$ entry of  $i$'s normalized output vector, $\bar{\Pi}^\text{out}_{ik}$, represents the fraction of firm $i$'s revenue it receives by selling to firms in industry $k$. $\bar{\Pi}^\text{in}_{ik}$ and $\bar{\Pi}^\text{out}_{ik} $ are the scaled technical and allocation coefficients in classical IO analysis.
For firm $i$ the vectors $\bar{\Pi}^\text{in}_{i.}$ and $\bar{\Pi}^\text{out}_{i.}$ are computed as
\begin{equation} \label{eq_pi_bar}
	\bar{\Pi}^\text{in}_{ik} = \frac{\Pi^\text{in}_{ik}}{\sum_{k=1}^{m} \Pi^\text{in}_{ik}}  = \frac{1}{s_i^\text{in}}\sum_{j=1}^{m} W_{ji} \delta_{p_j, k} \quad , \qquad 
	\bar{\Pi}^\text{out}_{ik} = \frac{\Pi^\text{out}_{ik}}{\sum_{k=1}^{m} \Pi^\text{out}_{ik}}  \frac{1}{s_i^\text{out}} \sum_{j=1}^{m} W_{ij} \delta_{p_j, k} \quad ,
\end{equation}
where $\delta_{p_j, k} = 1$ if firm $j$ belongs to industry $k$ and $\delta_{p_j, k} = 0$ otherwise.
We quantify the similarity between input and output vectors of two firms with the \emph{overlap coefficient} (OC) due to its clear economic interpretability. To show that our results do not depend on the specific choice of the similarity measure we also look at the jaccard index (JI).

%Note that $\sum_{k=1}^{m}\bar{\Pi}^\text{in}_{ik} = 1$ and $\sum_{k=1}^{m}\bar{\Pi}^\text{out}_{ik} = 1$ for all $i$. 
%The element, $\bar{\Pi}^\text{in}_{ik}$, of the normalized input vector of firm $i$ indicates the fraction of inputs firm $i$ sources from firms in industry $k$. Similarly, the element, $\bar{\Pi}^\text{out}_{ik}$, of the normalized output vector of firm $i$ indicates the fraction of its sales firm $i$ sells to firms in industry $k$.
%We calculate the similarity of the normalized input vector and output vectors of two firms with two similarity measures, the overlap coefficient and the Jaccard Index. 

\section{Details for calculation and interpretation of the input and output overlap coefficient}  \label{SI_details_on_IOC_OOC}

In general the overlap coefficient of two vectors $x, y$ of dimension $m$ is defined as
\begin{equation} \label{eq_OC}
	\text{OC}\left( x, \; y \right) = \frac{ \sum_{k=1}^{m} \min \left[  x_k, y_k \right]}{ \min \left[ \sum_{k=1}^{m} x_k , \; \sum_{k=1}^{m} y_k\right] } \quad .
\end{equation}
We calculate the overlap coefficient of the 1-norm $||\; ||_1$ normalized input,  and output vectors, i.e. $\bar{\Pi}^\text{in}_{i}$ and $\bar{\Pi}^\text{out}_{i}$. Therefore, in each calculation both vectors sum to one, then the denominator is always equal to one and can be dropped. The overlap coefficient is closely related to the \textit{weighted} Jaccard Index, which has the same numerator, and $ \sum_{k=1}^{m} \max \left[  x_k, y_k \right]$ as the denominator. It is also called the Szymkiewicz-Simpson distance \citep{jones1987pictures, vijaymeena2016survey}.

As introduced in the main text, for our application we calculate the input overlap coefficient (IOC) and output overlap coefficient (OOC) of two firms $i$ and $j$ as, 
\begin{eqnarray}
	\text{IOC}_{ij} %\left(\bar{\Pi}^\text{in}_{i.}, \; \bar{\Pi}^\text{in}_{j.} \right)
	& = & \sum_{k=1}^{m} \min \left[  \bar{\Pi}^\text{in}_{ik}, \; \bar{\Pi}^\text{in}_{jk} \right] \quad, \label{eq_ioc_SI} \\
	\text{OOC}_{ij} %\left(\bar{\Pi}^\text{out}_{i.}, \; \bar{\Pi}^\text{out}_{j.} \right)
	& = & \sum_{k=1}^{m} \min \left[  \bar{\Pi}^\text{out}_{ik}, \; \bar{\Pi}^\text{out}_{jk} \right] \quad  \label{eq_ooc_SI}.
\end{eqnarray}
The denominator from Eq. \ref{eq_OC} can be omitted since $\sum_{k=1}^{m}\bar{\Pi}^\text{in}_{ik} = 1$ and $\sum_{k=1}^{m}\bar{\Pi}^\text{out}_{ik} = 1$ for all $i$.
We calculate the distribution of the two measures for each industry $k$, by  computing all pairwise $\text{IOC}_{ij}$ and $\text{OOC}_{ij}$ for all firms where, $p_i = p_j = k$, and $i\neq j$, in the respective industry, $k$.
The input overlap coefficient, $\text{IOC}_{ij}$, of two firms $i$ and $j$ gives the fraction of their overall inputs they source from the same industries, i.e. the overlap of their industry input shares. The output overlap coefficient,  $\text{OOC}_{ij}$, of two firms $i$ and $j$ specifies the fraction of their overall sales they sell to the same industries, i.e. the overlap of their industry sales shares.
Note that  $\text{IOC}_{ij}$ also quantifies $i$'s and $j$'s overlap of exposures to other economic dynamics, like price increases or innovations of supplying industries. Similarly, $\text{OOC}_{ij}$ measures the common exposure to, e.g., innovation in the buyer industry that makes the input of firms obsolete.

In the example of Fig. \ref{fig1}, the relative input vector of firm 10 is $\bar{\Pi}^\text{in}_{10} = (0,0,0.5,0.5,0)$ and for firm 11  $\bar{\Pi}^\text{in}_{11} = (0,0,1,0,0)$, hence $\text{IOC}_{10, 11} = 0.5$. If a demand shock affects firm $10$, 50\% of the shock spreads upstream to sectors 3 and 4, respectively, while if the shock affects firm $11$, 100\% of the shock spreads upstream to sector 3. This means that the shock spreading dynamics overlap by only 50\% ($\text{IOC}_{10, 11} = 0.5$),  while the other 50\% spread to distinct sectors. For firms $6$ and $7$ the output vectors are $\bar{\Pi}^\text{out}_{6} = (1,0,0,0,0)$ and $\bar{\Pi}^\text{out}_{7} = (0,0,0,0,1)$, hence, $\text{OOC}_{6, 7} = 0$. This means that if either, $6$ or  $7$, receives a shock,  0\% of the shock would affect the same industry and 100\% would spread to different sectors, in one case towards sector $1$ and in the other to sector $5$. Also their exposure to demand shocks has no overlap. Firm $6$ is only exposed to industry $1$ while firm $6$ is only exposed to sector $5$.

Since we use industry-level input and output vectors, which neglect firm-level differences within industries, the real level of heterogeneity could be even larger. In our dataset cross border import and export links of firms are not available. This could lead to a potential underestimation of overlaps, but a study for Belgium \citep{dhyne2021trade} shows that firms' import and export links are few in relation to national import and export 
links.

\section{Further results on input and output overlaps} \label{SI_further_details_on_IOC_OOC}

\subsection*{Further results for NACE C26}
%% PUT THIS PARAGRAPH IN THE SI - IT DOES NOT BELONG TO RESULTS. IT WOULD ALSO FIT TO THE DOISCUSSION BUT WILL MAKE IT TOO LONG 
Even though, in all four in-degree (out-degree) groups there are firms with very similar input (output) vectors, the results clearly show that in general firms have surprisingly small overlaps with respect to to their suppliers' industries (inputs) and customers' (output) industries. This implies that if two random firms in in-degree (out-degree) bin $>$35  receive the same absolute size shock, on average only 34\% (14\%) of the shock's volume is propagated to firms of the same industry while 66\% (86\%) of the shock is propagated to firms in other industries. At the same time it means that two firms in this industry have on average 66\% (86\%) of their upstream (downstream) exposures to different supplier (buyer) industries. The low level of similarity of input and output vectors clearly shows that aggregating these firms into a single industry is not representative of the single firms' input-output vectors and will lead to large biases and mis-estimations of economic dynamics.

\subsection*{Further results on output overlaps across industries}
The highest median OOC$_{ij}$ are found in Veterinary activities (M75), Manufacture of beverages (C11), Manufacture of other transport equipment (C30), Forestry and logging (A2), Manufacture of leather and related products (C15), Manufacture of basic pharmaceutical products (C21), Telecommunications (J61), whereas the lowest  median OOC$_{ij}$ are found in service sectors such as Public administration and defence; compulsory social security (O84), Travel agency and related activities (N79), or Scientific research and development (M72), but also non-service sectors such as Remediation activities and other waste management services (E39), Other manufacturing (C32), or Manufacture of textiles (C13) are among the lowest output overlap sectors.  The average standard deviation is 0.17, the standard deviation of standard deviations is 0.047, and the error bar length appears to be relatively homogeneous across sectors. This indicates that the variation of pairwise output overlaps,  OOC$_{ij}$,  within sectors is relatively similar across sectors. For the other degree bins see SI Fig. \ref{SI_fig11a_nace4_IOC_other_buckets}.

The same results are shown for the other three out-degree bins 1-5, 6-15, and 16-35 in  \ref{SI_other_degree_buckets}  Fig. \ref{SI_fig3b_nace2_other_deg_buckets}. As for industry C26, the output overlaps are smaller for lower degree bins; the averages over the mean (median) output overlaps, are 0.110, (0.021) 0.157 (0.135), 0.223 (0.215), for the bins 1-5, 6-15, and 16-35, respectively. The averages over the standard deviations of output overlaps, are 0.266, 0.129, 0.109, respectively and therefore the variation of output overlaps within is on average decreasing with the number of out-links. Figure \ref{fig4nace2_medians_for_buckets}b illustrates this relationship more clearly by showing for each in-degree size bin (1-5, 6-15, 16-35, $>$35) the boxplot of the industries' median OOC values. It is clearly visible that output vectors of firms within industries become more homogeneous with the number of suppliers. 
\ref{SI_nace4_simmeas} shows that OOC$_{ij}$ are even lower when computed for at NACE 4 level. SI Fig. \ref{SI_Figure10a_nace4_OC_across_secs}b shows the average of the mean (median) output overlaps, across NACE4 industries is 0.231 (0.207). The standard deviation of mean (median) output overlaps is 0.179 (0.19), i.e. higher than for the NACE2 level.  This indicates that the variation of average output vector overlaps is higher at the NACE 4 level. The average standard deviation is 0.126 and the standard deviation of standard deviations is 0.056.  Note that the average IOC and OOC levels seem to be more similar on the NACE 4 level than at the NACE 2 level where the average IOC is higher than OOC. For the other degree bins see SI Fig. \ref{SI_fig11b_nace4_IOC_other_buckets}.
%~/WorkII/SupplyRank/HungaryProject/Heterogeneity_pnas/figures/SI_Figure10b.png 
%1 
%mean_of_means sd_of_means mean_of_medians sd_of_medians mean_of_sds sd_of_sds
%> 35   0.231       0.179           0.207          0.19       0.135     0.068
%
SI Fig. \ref{SI_fig2_nace2_JI_overlaps} in \ref{SI_jaccard_index} shows qualitatively similar results for the Jaccard Index for the degree bin $>$35. The pairwise input Jaccard Index (IJI) distributions are slightly shifted towards higher similarity values with a average mean (median), 0.398 (0.394) and  slightly less variation with a standard deviation of means of 0.07 (0.067). The pairwise output Jaccard Index (OJI) distributions are also shifted towards slightly higher similarity values with a average over means (medians), of 0.301 (0.291) and  slightly less variation with a standard deviation of means of 0.076 (0.077).

\FloatBarrier

\section{Overlap coefficients across industries for other degree bins} \label{SI_other_degree_buckets}

This section shows the results of the pairwise input overlap coefficient, IOC,  and output overlap coefficient, OOC,  distributions across all  NACE2 industries for the three degree bins 1-5, 6-15, and 16-35 that are not shown in Fig. \ref{fig3_nace2main}.
As for Fig. \ref{fig3_nace2main} 
we calculate the summary statistics --- mean, 5\%, 25\%, 50\% (median) 75\% and 95\% percentiles --- for the pairwise IOC (SI Fig. \ref{SI_fig3a_nace2_other_deg_buckets}) and OOC (SI Fig. \ref{SI_fig3b_nace2_other_deg_buckets}) distributions for all  NACE2 industries. Again these statistics are visualized as boxplots. The x-axis shows the 86 NACE2 codes present in the data set; the y-axis denotes the overlap coefficients, each boxplot corresponds to a NACE2 class. The dark thick horizontal bars correspond to the median, ($p_{5\%}$), the interquartile range ($p_{25\%}$ -- $p_{75\%}$) is shown as thick dark vertical lines, and the error bars ($p_{5\%}$ -- $p_{95\%}$) are indicated by thin light vertical lines. The thin vertical black lines separate NACE2 classes by their NACE1 affiliation.

The results for the IOC distributions are shown in SI Fig. \ref{SI_fig3a_nace2_other_deg_buckets}.
SI Fig. \ref{SI_fig3a_nace2_other_deg_buckets}a shows the pairwise IOC$_{ij}$ for firms with in-degree between one and five, $1 \leq k_i^{in} \leq 5$. The mean over the industries' mean (median) IOC is 0.132 (0.009), the standard deviation of mean (median) IOCs is 0.081 (0.062). The mean standard deviation is 0.262.
SI Fig. \ref{SI_fig3a_nace2_other_deg_buckets}b shows the pairwise IOC$_{ij}$ for firms with in-degree between one and five, $6 \leq k_i^{in} \leq 15$. The mean over the industries' mean (median) IOC is 0.202 (0.148), the standard deviation of mean (median) IOCs is 0.081 (0.088). The mean standard deviation is 0.192.
SI Fig. \ref{SI_fig3a_nace2_other_deg_buckets}c shows the pairwise IOC$_{ij}$ for firms with in-degree between one and five, $16 \leq k_i^{in} \leq 35$. The mean over the industries' mean (median) IOC is 0.269 (0.241), the standard deviation of mean (median) IOCs is 0.083 (0.091). The mean standard deviation is 0.168. 
As for NACE C26 in the main text we see that on average input vector overlaps increase with the number of suppliers. 
% SI_fig3a_nace2
% IOC     mean_of_means sd_of_means mean_of_medians sd_of_medians mean_of_sds sd_of_sds
%1-5           0.132       0.081           0.009         0.062       0.262     0.064
%6-15          0.202       0.081           0.148         0.088       0.192     0.041
%16-35         0.269       0.083           0.241         0.091       0.168     0.050

The results for the OOC distributions are shown in SI Fig. \ref{SI_fig3b_nace2_other_deg_buckets}.
SI Fig. \ref{SI_fig3b_nace2_other_deg_buckets}a shows the pairwise OOC$_{ij}$ for firms with out-degree between one and five, $1 \leq k_i^{out} \leq 5$. The mean over the industries' mean (median) OOC is 0.110 (0.021), the standard deviation of mean (median) OOCs is 0.094 (0.118). The mean standard deviation is 0.226.
SI Fig. \ref{SI_fig3b_nace2_other_deg_buckets}b shows the pairwise OOC$_{ij}$ for firms with out-degree between one and five, $6 \leq k_i^{out} \leq 15$. The mean over the industries' mean (median) OOC is 0.157 (0.135), the standard deviation of mean (median) OOCs is 0.078 (0.074). The mean standard deviation is 0.129.
SI Fig. \ref{SI_fig3b_nace2_other_deg_buckets}c shows the pairwise OOC$_{ij}$ for firms with out-degree between one and five, $16 \leq k_i^{out} \leq 35$. The mean over the industries' mean (median) OOC is 0.223 (0.215), the standard deviation of mean (median) OOCs is 0.078 (0.078). The mean standard deviation is 0.109.
Again average overlaps seem to increase with degree (number of customers). 
Further, output overlaps are on average slightly lower than input overlaps. 
% SI_fig3b_nace2
%OOC     mean_of_means sd_of_means mean_of_medians sd_of_medians mean_of_sds sd_of_sds
%1-5           0.110       0.094           0.021         0.118       0.226     0.081
%6-15          0.157       0.078           0.135         0.074       0.129     0.041
%16-35         0.223       0.078           0.215         0.078       0.109     0.024

\begin{figure}[h]
	\centering
	\includegraphics[width=0.33\columnwidth]{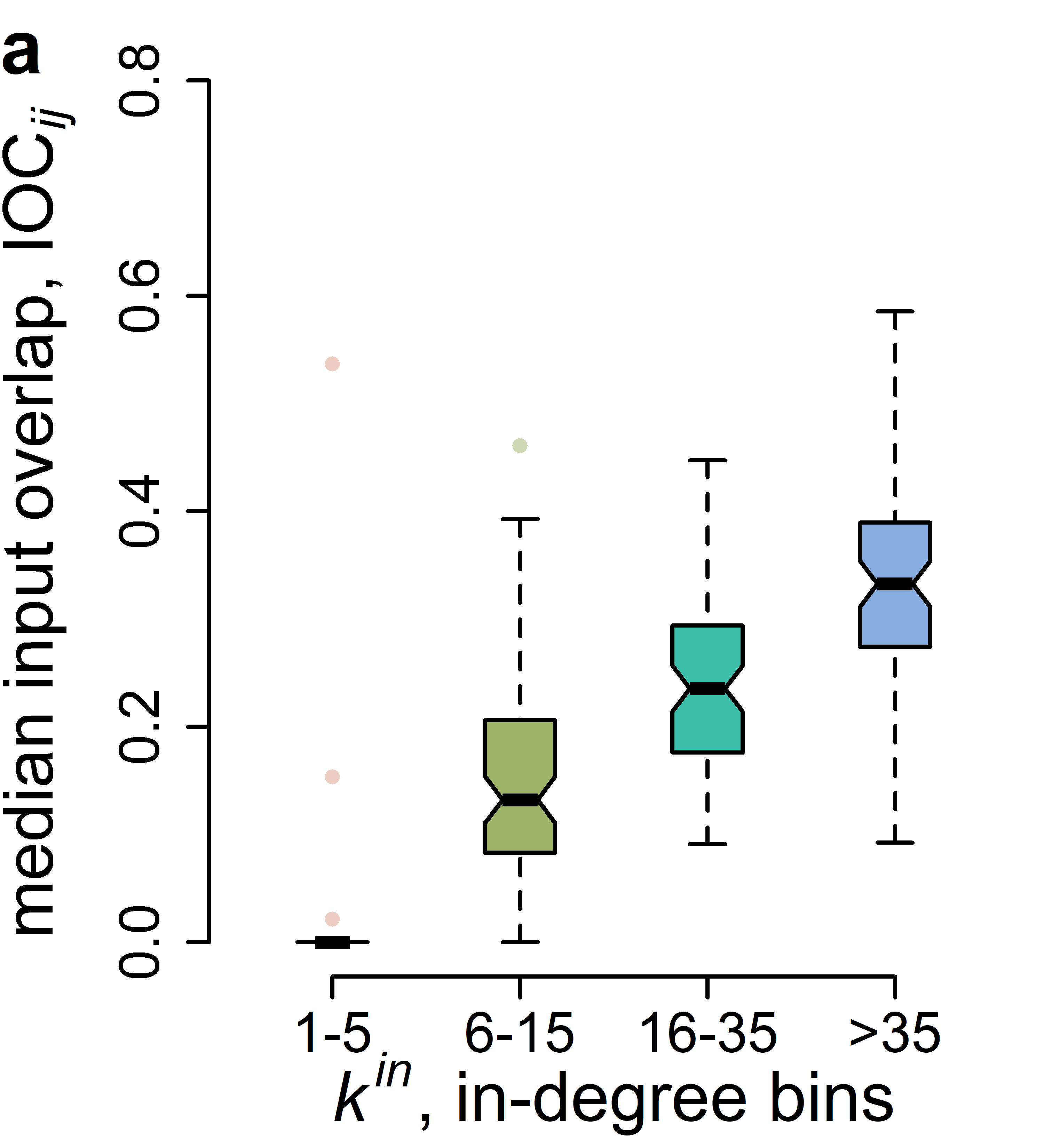} 
	\includegraphics[width=0.33\columnwidth]{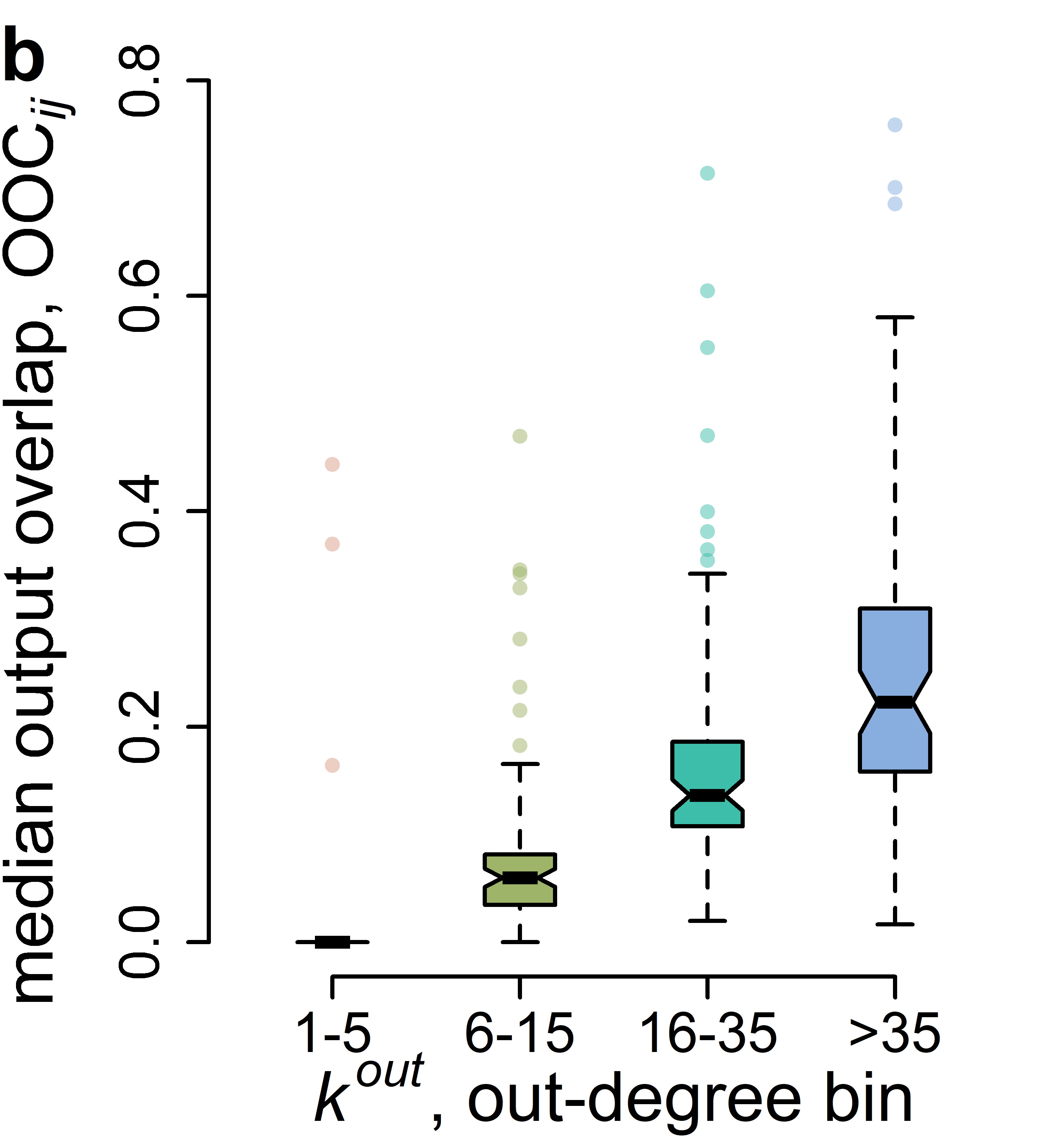} 
	\caption{Increase of input- and output-vector similarity with increasing in-degree, $k^{in}$, and out-degree, $k^{out}$, bins (1-5, 6-15, 16-35, $>$35).
		a) boxplots of the median input overlap coefficients for all NACE2 industries for each in-degree bins, respectively. 
		b) boxplots of the median out overlap coefficients for all NACE2 industries for each out-degree bins, respectively. 
		It is clearly visible that input- and output-vectors of firms within industries become on average more similar (higher median IOC and OOC values) with the number of suppliers and buyers.}
	\label{fig4nace2_medians_for_buckets}
\end{figure}

\begin{figure*}[t]
	\centering
	\includegraphics[width=1\textwidth]{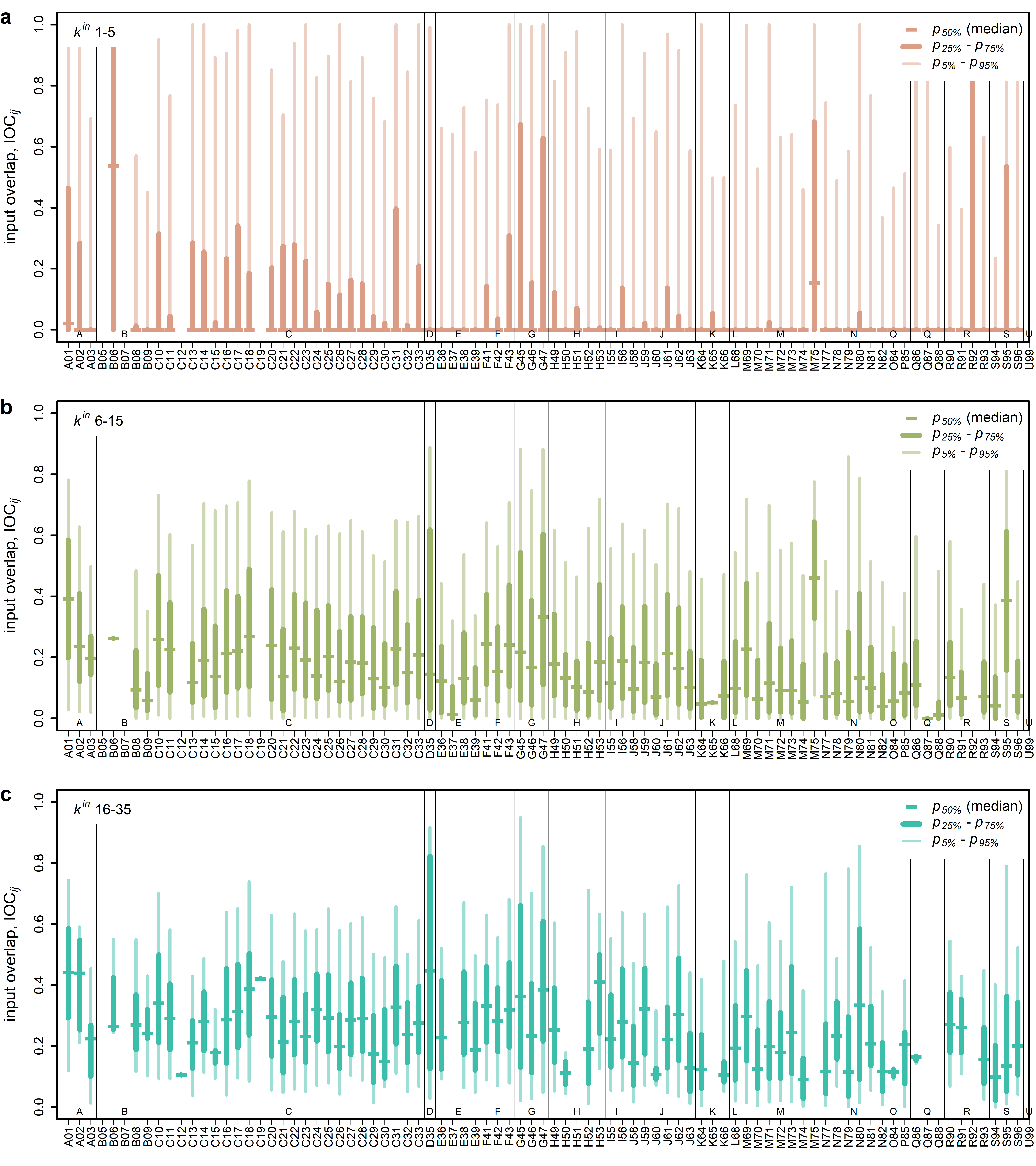} 
	\caption{Distributions of pairwise input vector overlaps, IOC$_{ij}$, of firms across NACE 2 industries for three in-degree size bins. NACE2 classes are on the x-axis; overlap coefficients on the y-axis. 
		a) pairwise IOC$_{ij}$ for firms with in-degree between one and five, $1 \leq k_i^{in} \leq 5$. The mean over the industries' mean (median) IOC is 0.132 (0.009), the standard deviation of mean (median) IOCs is 0.081 (0.062). The mean standard deviation is 0.262.
		b) pairwise IOC$_{ij}$ for firms with in-degree between 6 and 15, $6 \leq k_i^{in} \leq 15$. The mean over the industries' mean (median) IOC is 0.202 (0.148), the standard deviation of mean (median) IOCs is 0.081 (0.088). The mean standard deviation is 0.192.
		c) pairwise IOC$_{ij}$ for firms with in-degree between 16 and 35, $16 \leq k_i^{in} \leq 35$. The mean over the industries' mean (median) IOC is 0.269 (0.241), the standard deviation of mean (median) IOCs is 0.083 (0.091). The mean standard deviation is 0.168. 
	}
	\label{SI_fig3a_nace2_other_deg_buckets}
\end{figure*}

\begin{figure*}[t]
	\centering
	\includegraphics[width=1\textwidth]{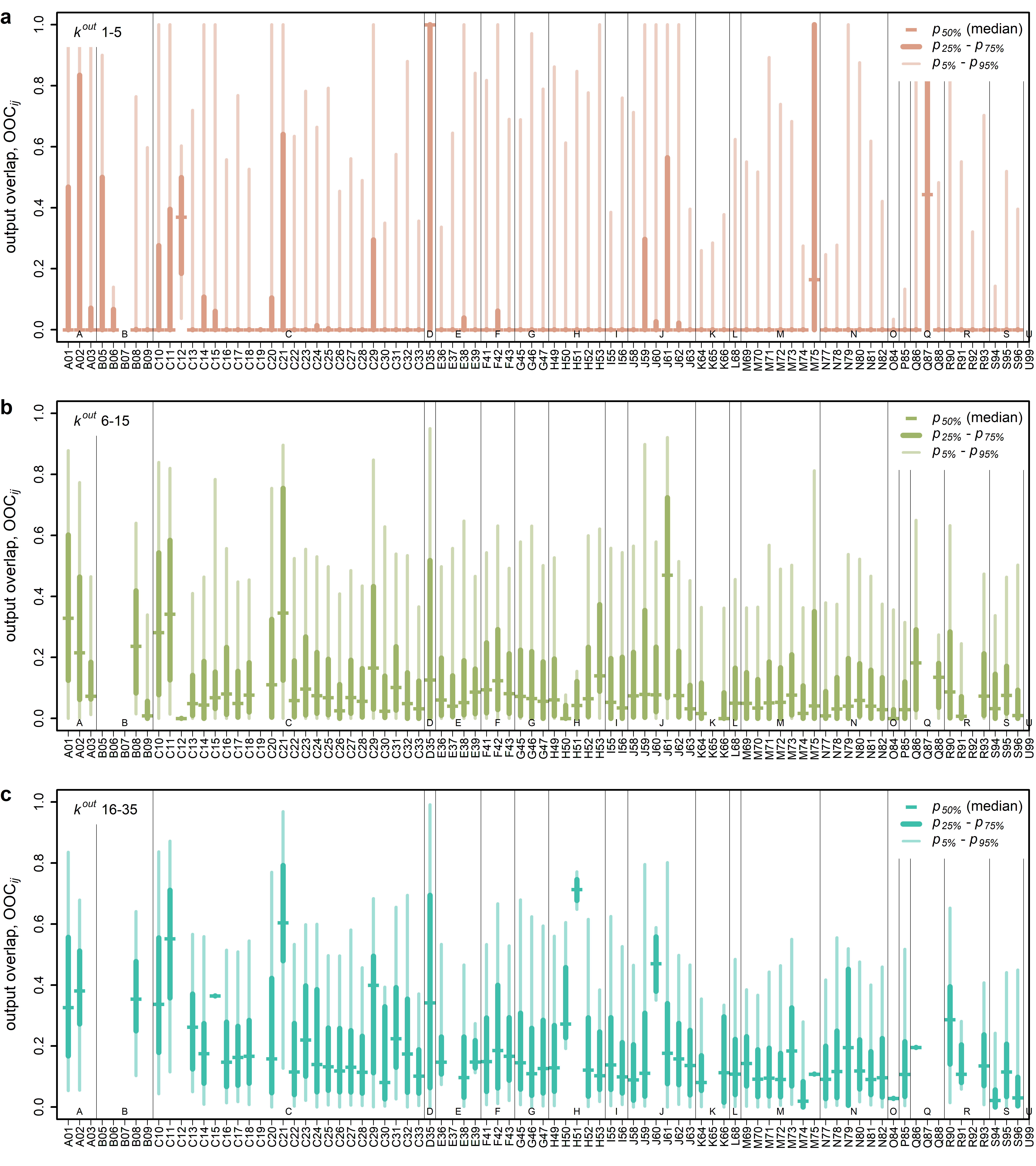} 
	\caption{Distributions of pairwise output vector overlaps, OOC$_{ij}$, of firms across NACE 2 industries for three in-degree size bins. NACE2 classes are on the x-axis; overlap coefficients on the y-axis. 
		a) pairwise OOC$_{ij}$ for firms with out-degree between one and five, $1 \leq k_i^{out} \leq 5$. The mean over the industries' mean (median) OOC is 0.110 (0.021), the standard deviation of mean (median) OOCs is 0.094 (0.118). The mean standard deviation is 0.226.
		b) pairwise OOC$_{ij}$ for firms with out-degree between 6 and 15, $6 \leq k_i^{out} \leq 15$. The mean over the industries' mean (median) OOC is 0.157 (0.135), the standard deviation of mean (median) OOCs is 0.078 (0.074). The mean standard deviation is 0.129.
		c) pairwise OOC$_{ij}$ for firms with out-degree between 16 and 35, $16 \leq k_i^{out} \leq 35$. The mean over the industries' mean (median) OOC is 0.223 (0.215), the standard deviation of mean (median) OOCs is 0.078 (0.078). The mean standard deviation is 0.109. 
	}
	\label{SI_fig3b_nace2_other_deg_buckets}
\end{figure*}

\begin{figure*}[ht]
	\centering
	\includegraphics[width=0.99\textwidth]{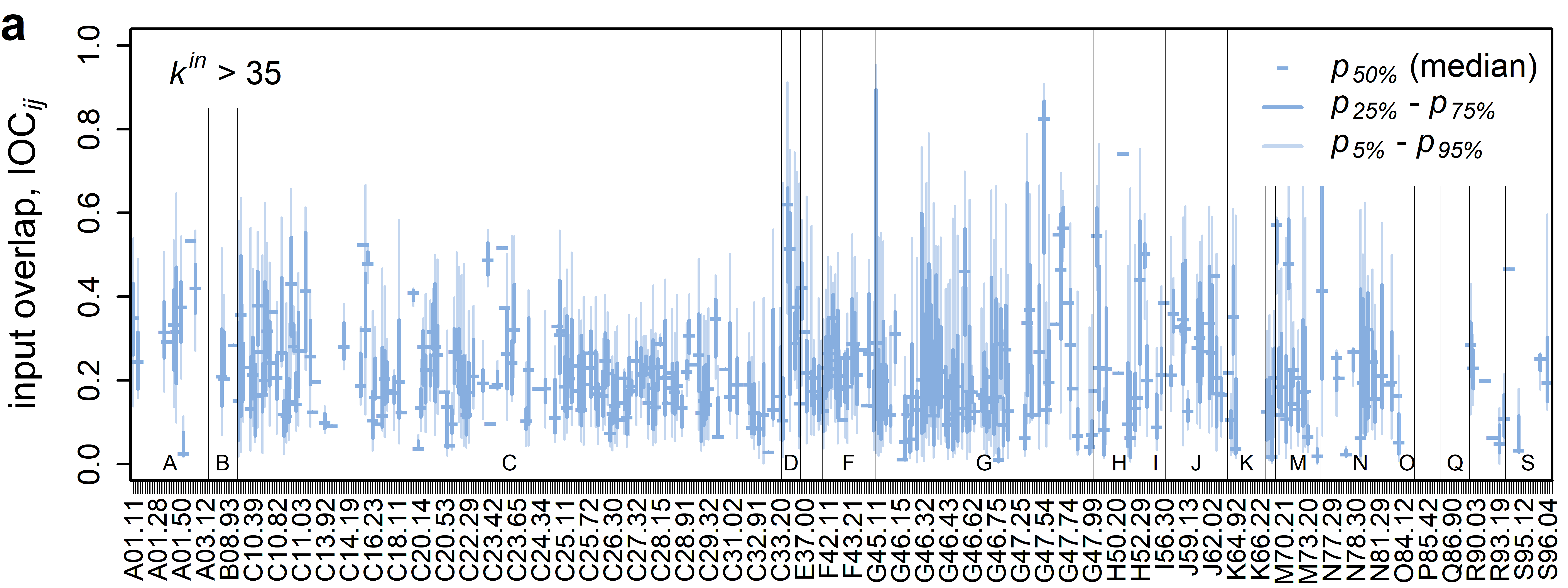} 
	\includegraphics[width=0.99\textwidth]{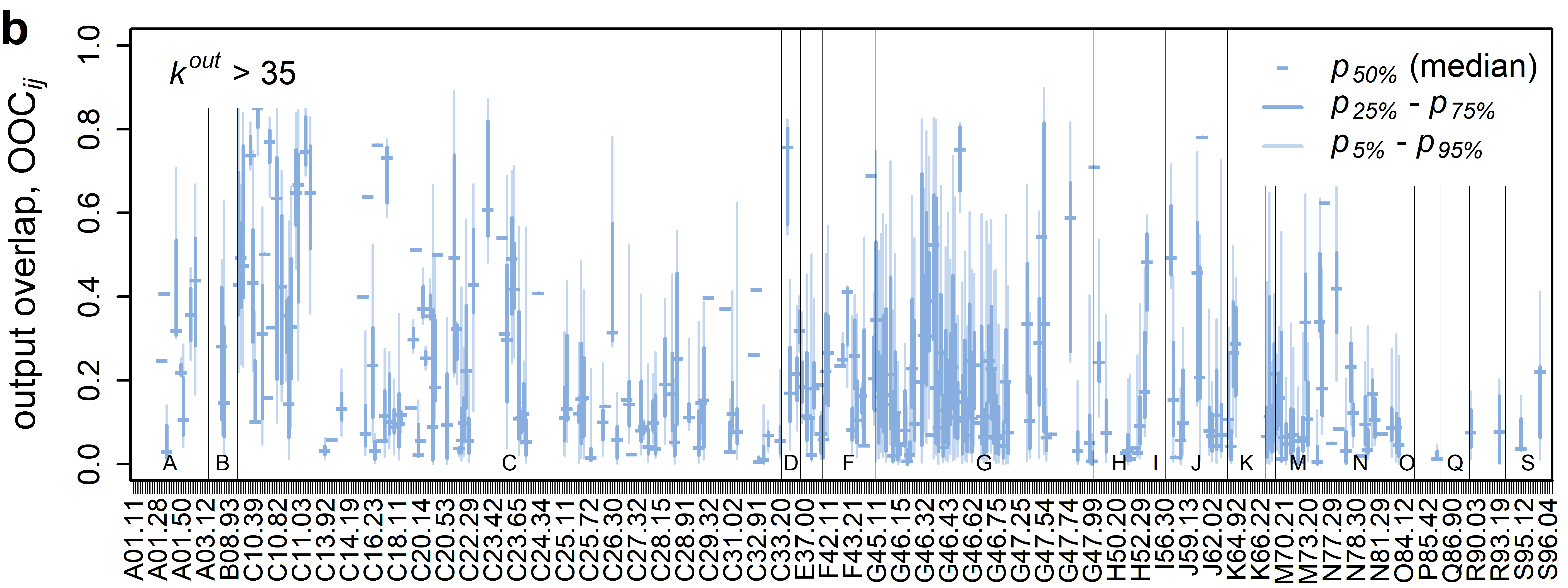} 
	\caption{Pairwise similarity distributions of input- and output-vectors of firms within each NACE4 industry. Similarity is measured with the overlap coefficient for firms with more than 35 suppliers (a) and buyers (b), respectively. 
	The y-axis denotes the overlap coefficients, the x-axis shows the NACE4 code for the respective boxplots. The dark blue horizontal bars correspond to the median, ($p_{50\%}$), dark blue vertical lines to the interquartile range ($p_{25\%}$ -- $p_{75\%}$), and thin light blue vertical lines to error bars ($p_{5\%}$ -- $p_{95\%}$). Thin black vertical lines separate NACE1 classes. Empty columns indicate  sectors with less than two firms in this degree bin. 
	a) distributions of pairwise intra-industry input overlap coefficients, IOC$_{ij}$. The average of the mean (median) input overlaps, across NACE2 industries is 0.237 (0.216) and the standard deviation of mean (median) input overlaps is 0.11 (0.12). The average standard deviation is 0.126. This indicates that relatively low input overlaps are the norm, but there are several outliers with higher similarities.	
	b) distributions of pairwise intra-industry output overlap coefficients, OOC$_{ij}$. The average of the mean (median) output overlaps, across NACE2 industries is 0.231 (0.207) and the standard deviation of mean (median) output overlaps is 0.179 (0.19), indicating  that relatively low output overlaps are the norm, but there are relatively many outliers with higher similarities. The average standard deviation is 0.135. Output overlaps are on average only slightly lower than the input overlaps, but there is more variation across industries. 
    If industry-level aggregation were fully representative for the IO-vectors of firms in both panels all distributions would correspond to a single bar at the value 1.}
	\label{SI_Figure10a_nace4_OC_across_secs}
\end{figure*}

Next we illustrate how average similarity increases with the degree bins. 
Fig. \ref{fig4nace2_medians_for_buckets} illustrates this relationship more clearly by showing for each degree size bin (1-5, 6-15, 16-35, $>$35) the boxplot of the industries' median IOC and OOC values.
Fig. \ref{fig4nace2_medians_for_buckets}a shows boxplots of the median input overlap coefficients, IOC, for all NACE2 industries for each in-degree bin, respectively. We see that for the bin with 1 to 5 suppliers almost all medians are zero. Then the distribution of medians is substantially shifted upwards for the bin of 6-15 suppliers and it continues to increase for the other two in-degree bins with 16-35 and more than 35 suppliers, respectively. 
Fig. \ref{fig4nace2_medians_for_buckets}b shows boxplots of the median output overlap coefficients, OOC, for all NACE2 industries for each out-degree bin, respectively. We see that for the bin with 1 to 5 buyers almost all medians are zero. Then the distribution of medians is slightly shifted upwards for the bin of 6-15 buyers, but there are several outlier industries with higher output overlaps. The median OOC continue to increase for the other two out-degree bins with 16-35 and more than 35 buyers, respectively. 
It is visible that the upper tails of the median OOC distributions are longer than for the median IOC distributions. Overall median OOCs are lower than median IOCs.

% fig4nace2_medians_for_buckets
% median IOC   median_of_medians mean_of_medians sd_of_medians
%1-5               0.000           0.009         0.062
%6-15              0.132           0.148         0.088
%16-35             0.236           0.241         0.091
%>35               0.333           0.330         0.102
%
% median OOC   median_of_medians mean_of_medians sd_of_medians
%1-5               0.000           0.024         0.126
%6-15              0.060           0.082         0.086
%16-35             0.137           0.179         0.131
%>35               0.223           0.257         0.161

\FloatBarrier

\clearpage

%~/WorkII/SupplyRank/HungaryProject/Heterogeneity_pnas/figures/SI_Figure10a.png 
%
%mean_of_means sd_of_means mean_of_medians sd_of_medians mean_of_sds sd_of_sds
%> 35   0.237        0.11           0.216          0.12       0.126     0.056

%~/WorkII/SupplyRank/HungaryProject/Heterogeneity_pnas/figures/SI_Figure10b.png 
%1 
%mean_of_means sd_of_means mean_of_medians sd_of_medians mean_of_sds sd_of_sds
%> 35   0.231       0.179           0.207          0.19       0.135     0.068

\section{Overlap coefficients for NACE 4 level input output vectors} \label{SI_nace4_simmeas}

In this section we show that the pairwise input overlaps, IOC$_{ij}$, and  output overlaps, OOC$_{ij}$,   are lower for all pairs of firms within NACE 4 industries for the NACE 4 level input and output vectors. Remember, in the previous analysis we have computed the overlaps for all pairs of firms within a NACE2 industry and on the NACE2 level input and output vectors.

In the following figures we show the pairwise overlap coefficient distributions of input- and output-vectors of firms within each NACE4 industry for the respective degree-bins 1-5, 6-15, 6-35, and $>$35. 
The y-axis denotes the overlap coefficients, the x-axis shows the NACE4 code for the respective boxplots. The dark  horizontal bars correspond to the median, ($p_{50\%}$), dark  vertical lines to the interquartile range ($p_{25\%}$ -- $p_{75\%}$), and thin light  vertical lines to error bars ($p_{5\%}$ -- $p_{95\%}$). Thin black vertical lines separate NACE1 classes. Empty columns indicate  sectors with less than two firms in this degree bin. 

First, we show the distributions of the input overlap coefficients, IOC$_{ij}$. 
SI Fig. \ref{SI_fig11a_nace4_IOC_other_buckets}a shows the distributions of pairwise intra-industry input overlap coefficients, IOC$_{ij}$, for firms with more than 35 suppliers, $k^\text{in} > 35$. The average of the mean (median) input overlaps, across NACE2 industries is 0.237 (0.216) and the standard deviation of mean (median) input overlaps is 0.11 (0.12). The average standard deviation is 0.126. This indicates that relatively low input overlaps are the norm, but there are several outliers with higher similarities.		
SI Fig. \ref{SI_fig11a_nace4_IOC_other_buckets}a shows the distributions of pairwise  input overlap coefficients,  IOC$_{ij}$, for firms with in-degree between one and five, $1 \leq k_i^{in} \leq 5$. The mean over the industries' mean (median) IOC$_{ij}$ is 0.063 (0.005), the standard deviation of mean (median) IOCs is 0.074 (0.057). The mean standard deviation is 0.168.
SI Fig. \ref{SI_fig11a_nace4_IOC_other_buckets}b shows the distributions of  pairwise input overlap coefficients,  IOC$_{ij}$, for firms with in-degree between 6 and 15, $6 \leq k_i^{in} \leq 15$. The mean over the industries' mean (median) IOC$_{ij}$ is 0.112 (0.063), the standard deviation of mean (median) IOCs is 0.084 (0.083). The mean standard deviation is 0.139.
SI Fig. \ref{SI_fig11a_nace4_IOC_other_buckets}c shows the distributions of  pairwise  input overlap coefficients,  IOC$_{ij}$, for firms with in-degree between 16 and 35, $16 \leq k_i^{in} \leq 35$. The mean over the industries' mean (median) IOC$_{ij}$ is 0.165 (0.140), the standard deviation of mean (median) IOCs is 0.107 (0.112). The mean standard deviation is 0.130. 

Second, we show the distributions of the output overlap coefficients, OOC$_{ij}$. 
SI Fig. \ref{SI_fig11a_nace4_IOC_other_buckets}b shows the distributions of pairwise intra-industry output overlap coefficients, OOC$_{ij}$, for more than 35 buyers, $k^\text{out} > 35$. The average of the mean (median) output overlaps, across NACE2 industries is 0.231 (0.207) and the standard deviation of mean (median) output overlaps is 0.179 (0.19), indicating  that relatively low output overlaps are the norm,  but there are relatively many outliers with higher similarities. The average standard deviation is 0.135. Output overlaps are on average only slightly lower than the input overlaps, but there is more variation across industries. 
SI Fig. \ref{SI_fig11b_nace4_IOC_other_buckets}a shows the distributions of  pairwise output overlap coefficients, OOC$_{ij}$, for firms with out-degree between one and five, $1 \leq k_i^{out} \leq 5$. The mean over the industries' mean (median) OOC$_{ij}$ is 0.056 (0.005), the standard deviation of mean (median) OOCs is 0.075 (0.054). The mean standard deviation is 0.148.
SI Fig. \ref{SI_fig11b_nace4_IOC_other_buckets}b shows the distributions of  pairwise output overlap coefficients, OOC$_{ij}$, for firms with out-degree between 6 and 15, $6 \leq k_i^{out} \leq 15$. The mean over the industries' mean (median) OOC$_{ij}$ is 0.081 (0.057), the standard deviation of mean (median) OOCs is 0.063 (0.068). The mean standard deviation is 0.087.
SI Fig. \ref{SI_fig11b_nace4_IOC_other_buckets}c shows the distributions of  pairwise output overlap coefficients, OOC$_{ij}$, for firms with out-degree between 16 and 35, $16 \leq k_i^{out} \leq 35$. The mean over the industries' mean (median) OOC$_{ij}$ is 0.127 (0.116), the standard deviation of mean (median) OOCs is 0.080 (0.082). The mean standard deviation is 0.078. 
Note that if industry-level aggregation was fully representative for the IO-vectors of firms in all figures all distributions would correspond to a single bar at the value 1.

\begin{figure}[ht]
	\centering
	\includegraphics[width=0.3\columnwidth]{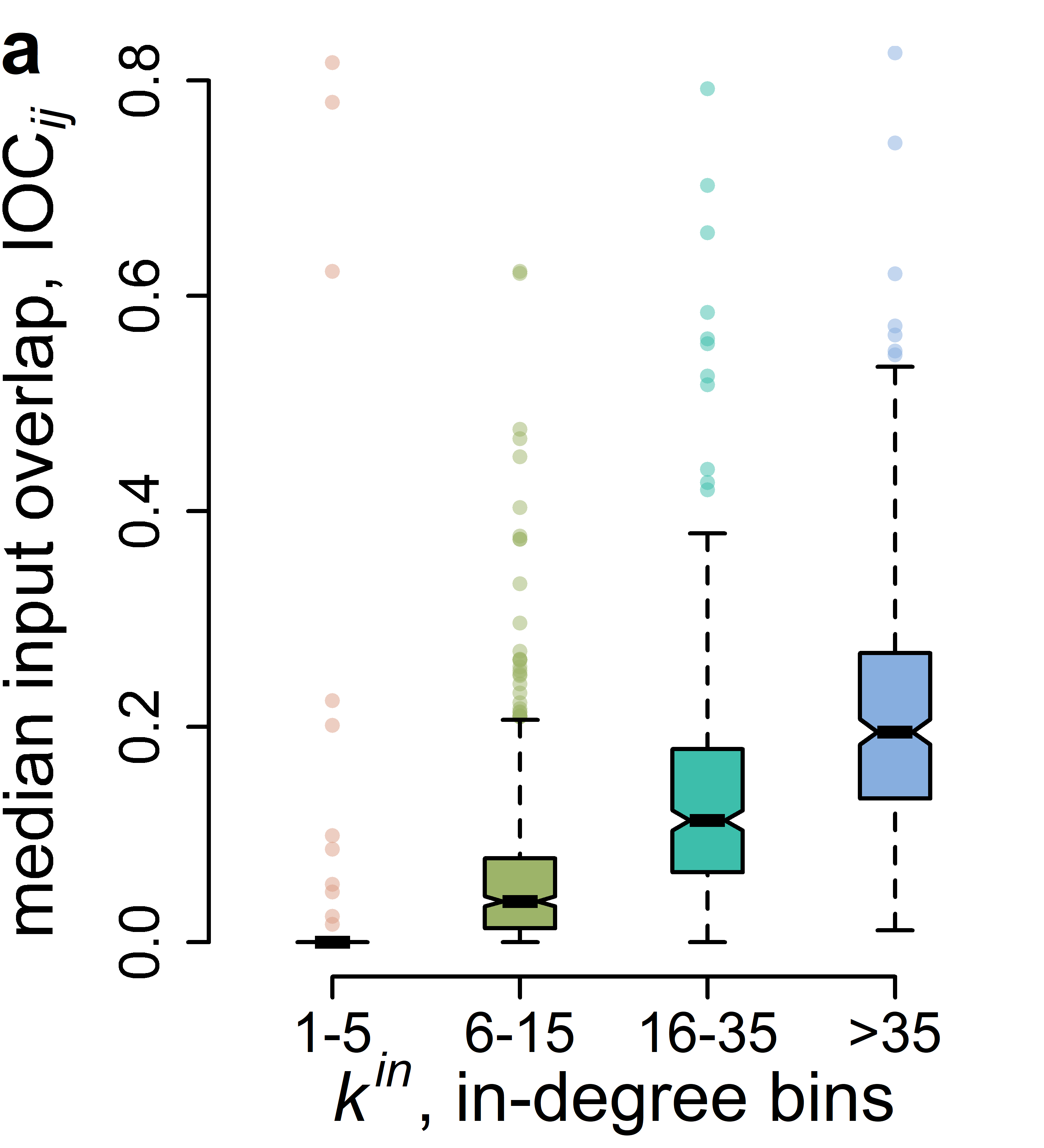} 
	\includegraphics[width=0.3\columnwidth]{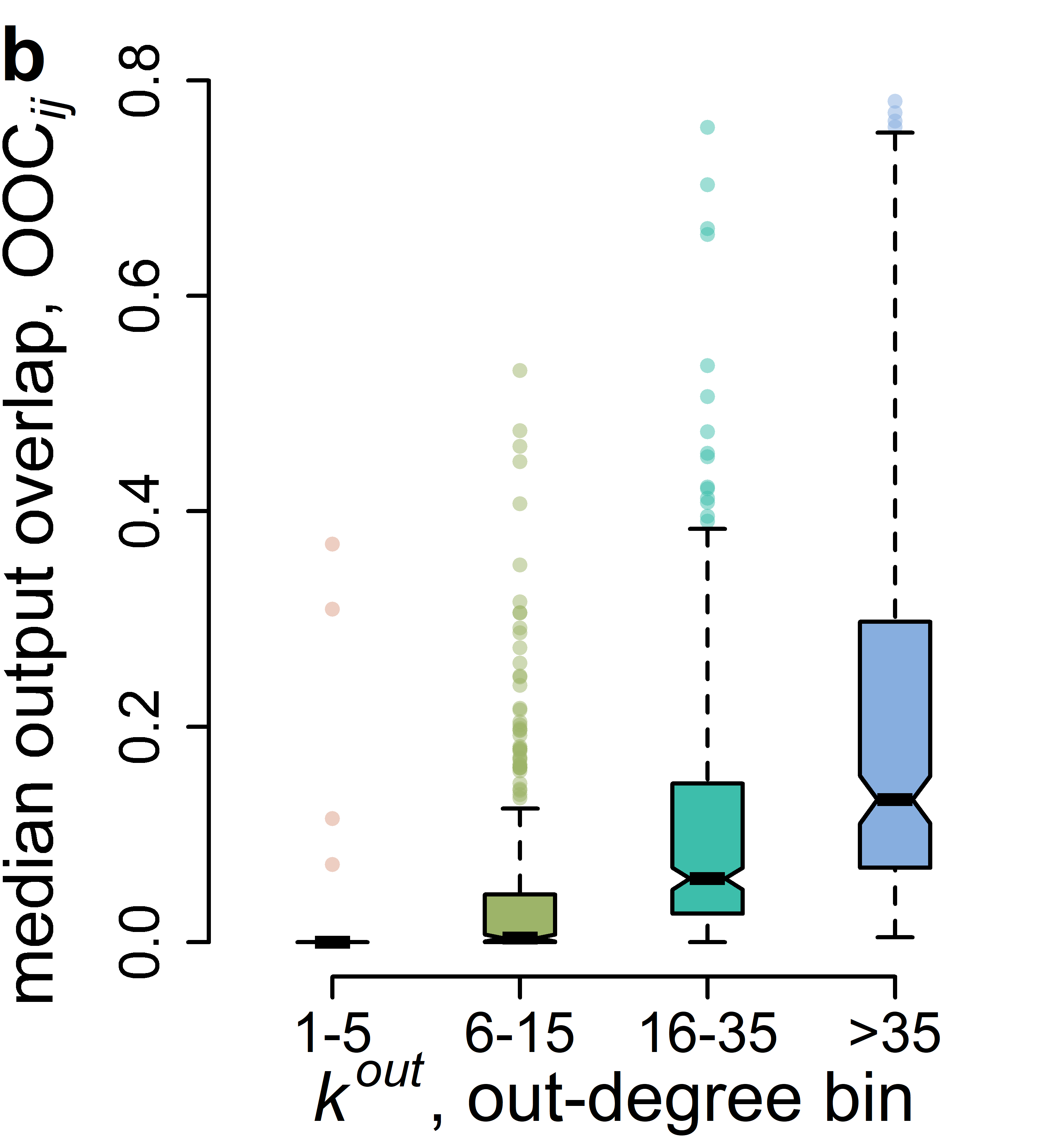} 
	\caption{Increase of input- and output-vector similarity with increasing in-degree, $k^{in}$, and out-degree, $k^{out}$, bins (1-5, 6-15, 16-35, $>$35).
		a) boxplots of the median input overlap coefficients for all NACE 4 industries for each in-degree bin, respectively. 
		b) boxplots of the median out overlap coefficients for all NACE 4 industries for each out-degree bin, respectively. 
		It is clearly visible that input- and output-vectors of firms within industries become on average more similar (higher median IOC and OOC values) with the number of suppliers and buyers.}
	\label{SI_fig}
\end{figure}
%
%~/WorkII/SupplyRank/HungaryProject/Heterogeneity_pnas/figures/SI_Figure9a.png 
% IOC nace4 median_of_medians mean_of_medians sd_of_medians
%1-5               0.000           0.005         0.057
%6-15              0.038           0.063         0.083
%16-35             0.113           0.140         0.112
%>35               0.195           0.216         0.120
%
%~/WorkII/SupplyRank/HungaryProject/Heterogeneity_pnas/figures/SI_Figure9b.png 
% OOC nace4 median_of_medians mean_of_medians sd_of_medians
%1-5               0.000           0.007         0.071
%6-15              0.004           0.039         0.076
%16-35             0.059           0.112         0.137
%>35               0.132           0.207         0.190

Next we show specifically how the average overlap coefficients increase with the degree of firms. 
Fig. \ref{SI_fig} illustrates this relationship by showing for each degree size bin (1-5, 6-15, 16-35, $>$35) on the x-axis, the boxplot of the NACE 4 industries' median overlap coefficients on the y-axis.
Fig. \ref{SI_fig}a shows boxplots of the median input overlap coefficients, IOC$_{ij}$, for all NACE4 industries for each in-degree bin, respectively. We see that for the bin with 1 to 5 suppliers most medians are zero. Then the distribution of medians is  shifted upwards for the bin of 6-15 suppliers and it continues to increase for the other two in-dgree bins with 16-35 and more than 35 suppliers, respectively. Note that even for two highest degree bins medians can range from almost zero to above 0.8. 
Fig. \ref{SI_fig}b shows boxplots of the median output overlap coefficients, OOC$_{ij}$, for all NACE2 industries for each out-degree bin, respectively. We see that for the bin with 1 to 5 buyers almost all medians are zero. Then the distribution of medians is slightly shifted upwards for the bin of 6-15 buyers, but there are several outlier industries with higher output overlaps. The median OOC continue to increase for the other two out-dgree bins with 16-35 and more than 35 buyers, respectively. 
Note that even for two highest degree bins medians can range from  zero to around 0.8. 
It is visible that the tails of the median OOC distributions are longer than for the median IOC distributions. Overall median OOCs appear lower than median IOCs.

\begin{figure*}[t]
	\centering
	\includegraphics[width=1\textwidth]{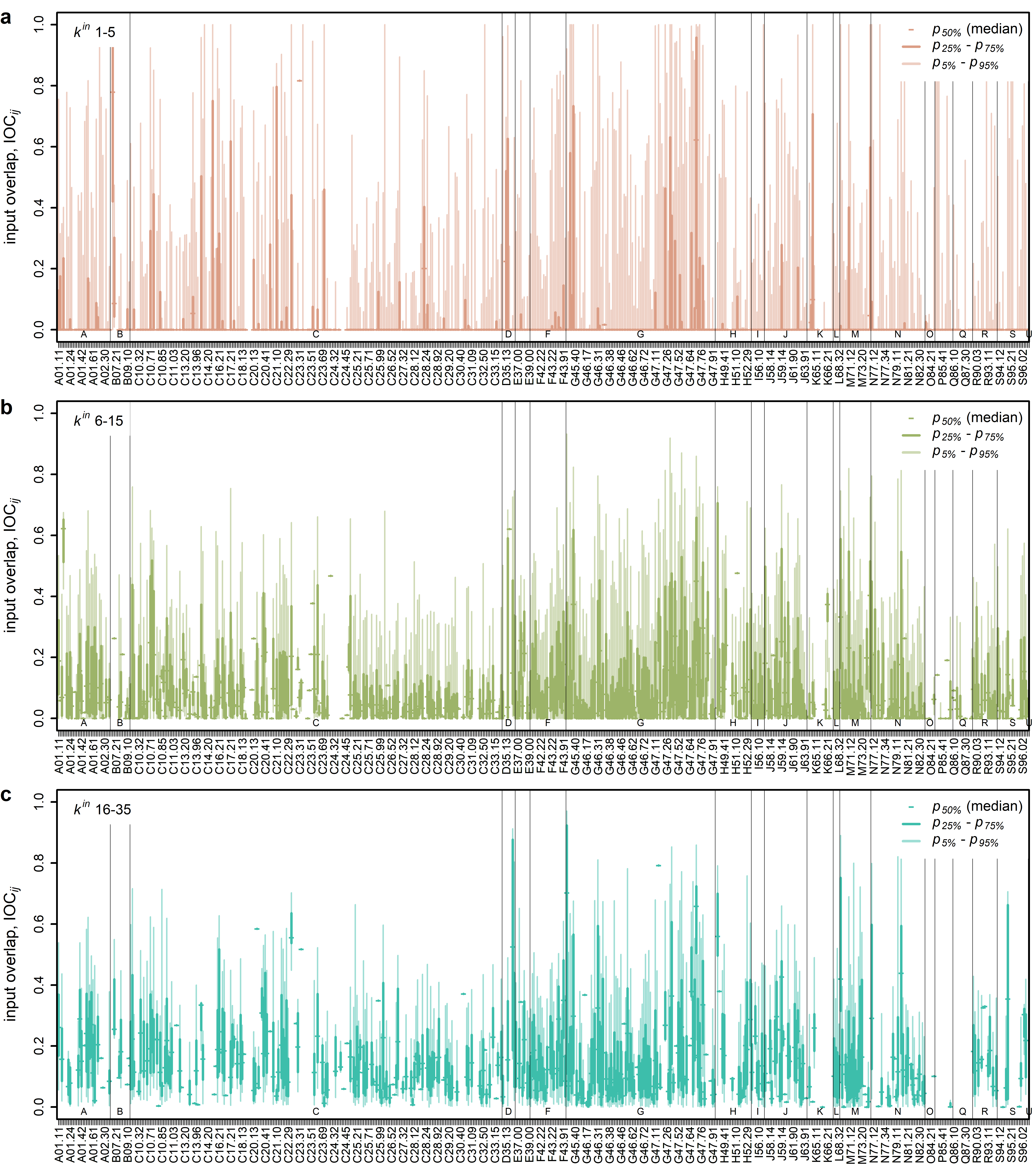} 
	\caption{	Distributions of pairwise input vector overlaps, IOC$_{ij}$, of firms across NACE 4 industries for three in-degree size bins. NACE 4 classes are on the x-axis; overlap coefficients on the y-axis. 
		a) pairwise IOC$_{ij}$ for firms with in-degree between one and five, $1 \leq k_i^{in} \leq 5$. The mean over the industries' mean (median) IOC is 0.063 (0.005), the standard deviation of mean (median) IOCs is 0.074 (0.057). The mean standard deviation is 0.168.
		b) pairwise IOC$_{ij}$ for firms with in-degree between 6 and 15, $6 \leq k_i^{in} \leq 15$. The mean over the industries' mean (median) IOC is 0.112 (0.063), the standard deviation of mean (median) IOCs is 0.084 (0.083). The mean standard deviation is 0.139.
		c) pairwise IOC$_{ij}$ for firms with in-degree between 16 and 35, $16 \leq k_i^{in} \leq 35$. The mean over the industries' mean (median) IOC is 0.165 (0.140), the standard deviation of mean (median) IOCs is 0.107 (0.112). The mean standard deviation is 0.130. 
	}
	\label{SI_fig11a_nace4_IOC_other_buckets}
\end{figure*}
%~/WorkII/SupplyRank/HungaryProject/Heterogeneity_pnas/figures/SI_Figure11a.png 
% 
%mean_of_means sd_of_means mean_of_medians sd_of_medians mean_of_sds sd_of_sds
%1-5           0.063       0.074           0.005         0.057       0.168     0.089
%6-15          0.112       0.084           0.063         0.083       0.139     0.064
%16-35         0.165       0.107           0.140         0.112       0.130     0.060

\begin{figure*}[t]
	\centering
	\includegraphics[width=1\textwidth]{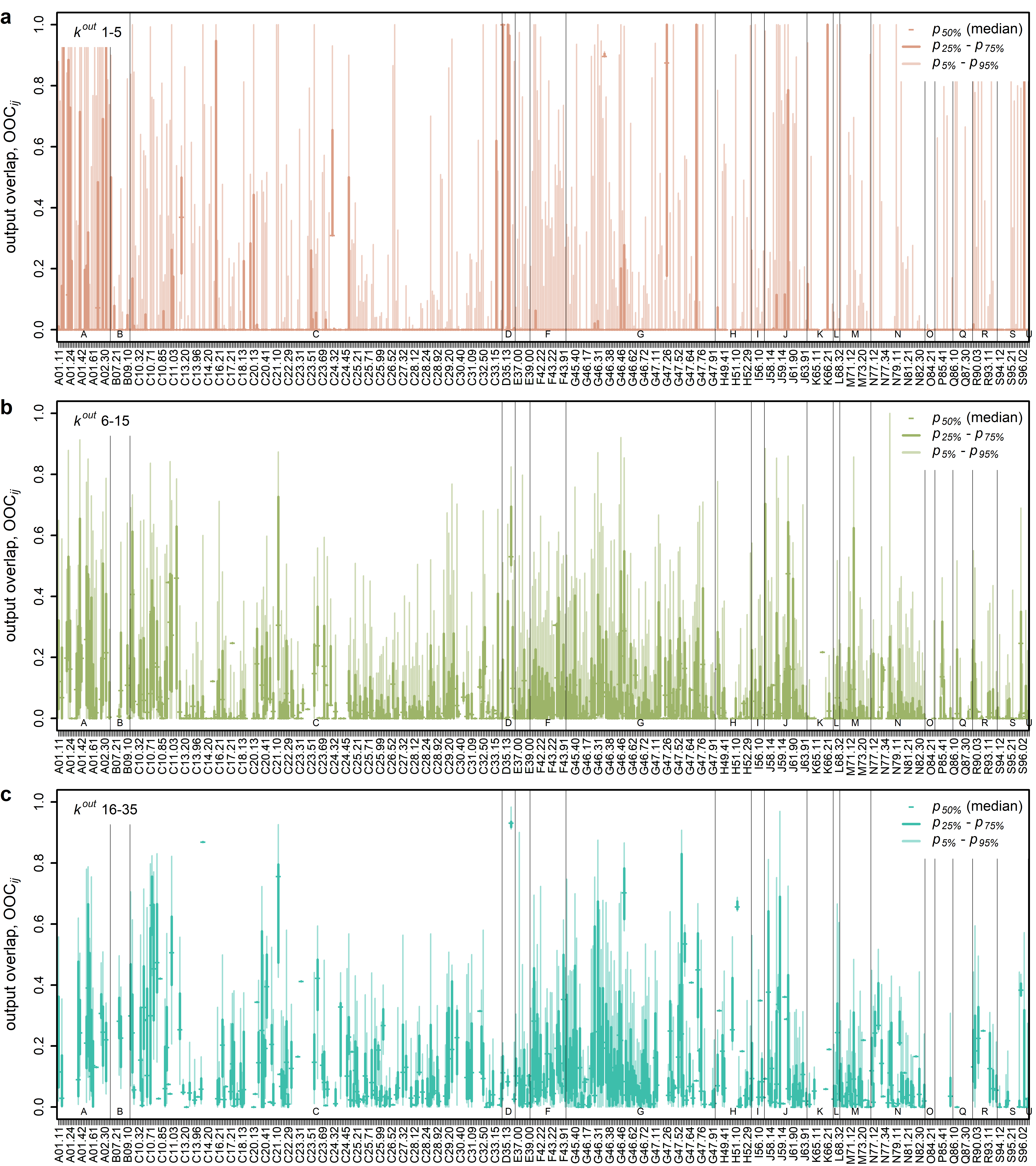} 
	\caption{	Distributions of pairwise output vector overlaps, OOC$_{ij}$, of firms across NACE 4 industries for three in-degree size bins. NACE 4 classes are on the x-axis; overlap coefficients on the y-axis. 
		a) pairwise OOC$_{ij}$ for firms with out-degree between one and five, $1 \leq k_i^{out} \leq 5$. The mean over the industries' mean (median) OOC is 0.056 (0.005), the standard deviation of mean (median) OOCs is 0.075 (0.054). The mean standard deviation is 0.148.
		b) pairwise OOC$_{ij}$ for firms with out-degree between 6 and 15, $6 \leq k_i^{out} \leq 15$. The mean over the industries' mean (median) OOC is 0.081 (0.057), the standard deviation of mean (median) OOCs is 0.063 (0.068). The mean standard deviation is 0.087.
		c) pairwise OOC$_{ij}$ for firms with out-degree between 16 and 35, $16 \leq k_i^{out} \leq 35$. The mean over the industries' mean (median) OOC is 0.127 (0.116), the standard deviation of mean (median) OOCs is 0.080 (0.082). The mean standard deviation is 0.078. 
	}
	\label{SI_fig11b_nace4_IOC_other_buckets}
\end{figure*}
%~/WorkII/SupplyRank/HungaryProject/Heterogeneity_pnas/figures/SI_Figure11b.png 
%
%mean_of_means sd_of_means mean_of_medians sd_of_medians mean_of_sds sd_of_sds
%1-5           0.056       0.075           0.005         0.054       0.148     0.091
%6-15          0.081       0.063           0.057         0.068       0.087     0.042
%16-35         0.127       0.080           0.116         0.082       0.078     0.035

\clearpage

\FloatBarrier

\begin{figure*}[ht]
	\centering
	\includegraphics[width= 0.85\textwidth]{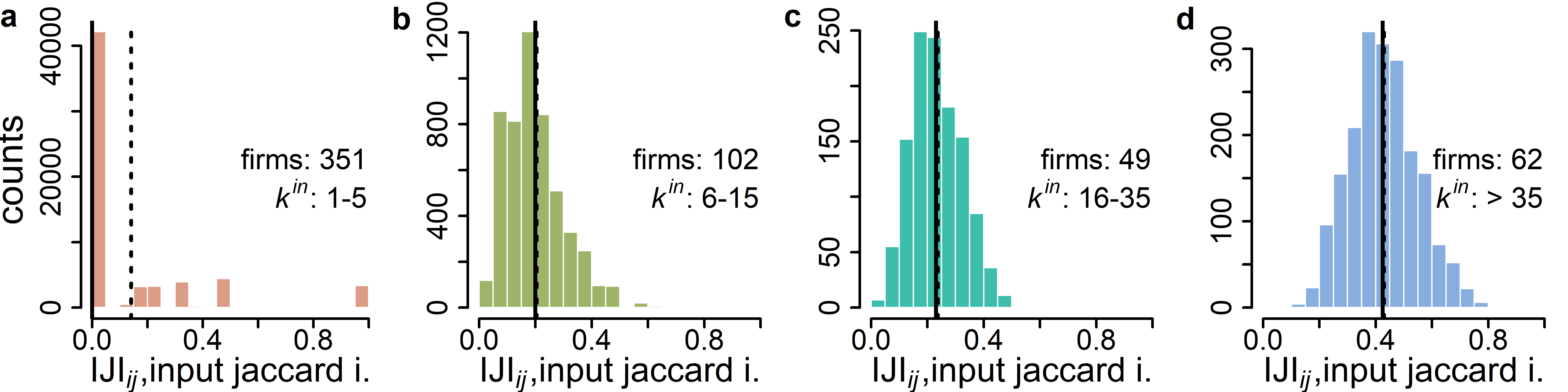} 
	\includegraphics[width= 0.85\textwidth]{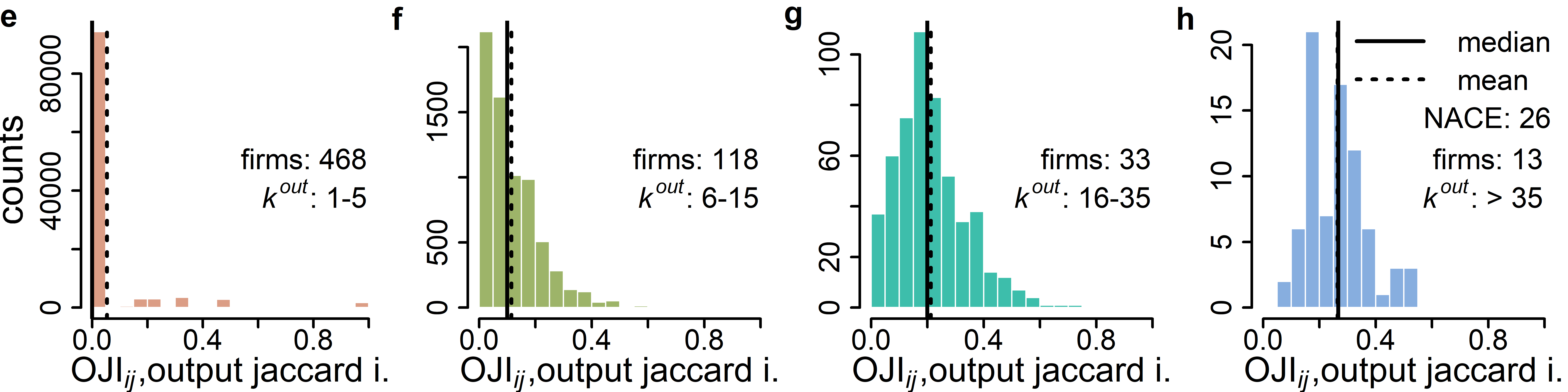} 
	\caption{		Pairwise similarity distributions of input and output vectors for firms of NACE class 26, Manufacture of computer, electronic and optical products measured with the Jaccard Index.  
	a-d) show input Jaccard Indices, IJI$_{ij}$, and e-h) output Jaccard Indices, OJI$_{ij}$, visualized as histograms, for four in-degree, $k_i^{in}$, (number of suppliers) and out-degree, $k_i^{out}$ (number of buyers), bins, respectively. Jaccard Index values are on the x-axis in bins of width 0.05; the y-axis shows the frequency to fall in the respective bin. Vertical solid lines correspond to median and dashed lines to mean overlap coefficients.
	a) pairwise IJI$_{ij}$ for 351 firms with %in-degree between one and five, 
	$1 \leq k_i^{in} \leq 5$. The median and mean input Jaccard Index is 0 and 0.141, respectively; the standard deviation is 0.261.  
	b) pairwise IJI$_{ij}$ for 102 firms with %in-degree between one and five, 
	$6 \leq k_i^{in} \leq 15$. The median and mean input Jaccard Index is 0.2 and 0.204, respectively; the standard deviation is 0.109.
	c) pairwise IJI$_{ij}$ for 49 firms with %in-degree between one and five, 
	$16 \leq k_i^{in} \leq 35$. The median and mean input Jaccard Index is 0.231 and 0.237, respectively; the standard deviation is 0.091.  
	d) pairwise IJI$_{ij}$ for 62 firms with %in-degree between one and five, 
	$35 < k_i^{in}$. The median and mean input Jaccard Index is 0.425 and 0.43, respectively; the standard deviation is 0.119.   
	It is clearly visible that the similarity of input vectors is low for all size bins, but increases on average with the number of suppliers.
	e) pairwise OJI$_{ij}$ for 468 firms with %out-degree between one and five, 
	$1 \leq k_i^{out} \leq 5$. The median and mean output  Jaccard Index is 0 and 0.054, respectively; the standard deviation is  0.163.  
	f) pairwise OJI$_{ij}$ for 118 firms with %out-degree between one and five,
	$6 \leq k_i^{out} \leq 15$. The median and mean output  Jaccard Index  is 0.1 and 0.115, respectively; the standard deviation is 0.109. 
	g) pairwise OJI$_{ij}$ for 33 firms with %out-degree between one and five, 
	$16 \leq k_i^{out} \leq 35$. The median and mean output  Jaccard Index is 0.2 and 0.212, respectively; the standard deviation is  0.127 .  
	h) pairwise OJI$_{ij}$ for 13 firms with %in-degree between one and five, 
	$ 35 < k_i^{out} $. The median and mean output  Jaccard Index  is 0.267 and 0.265, respectively; the standard deviation is  0.106. 
	The similarity of output vectors is even lower than for input vectors, and also increases on average with the number of buyers. 
	If industry-level aggregation were fully representative for the IO-vectors of firms in NACE C26 in all panels the distributions would correspond to a single bar at the value 1.}
	\label{SI_fig1nace2_jaccard_C26}
\end{figure*}

\begin{figure*}[t]
	\centering
	\includegraphics[width=0.9\textwidth]{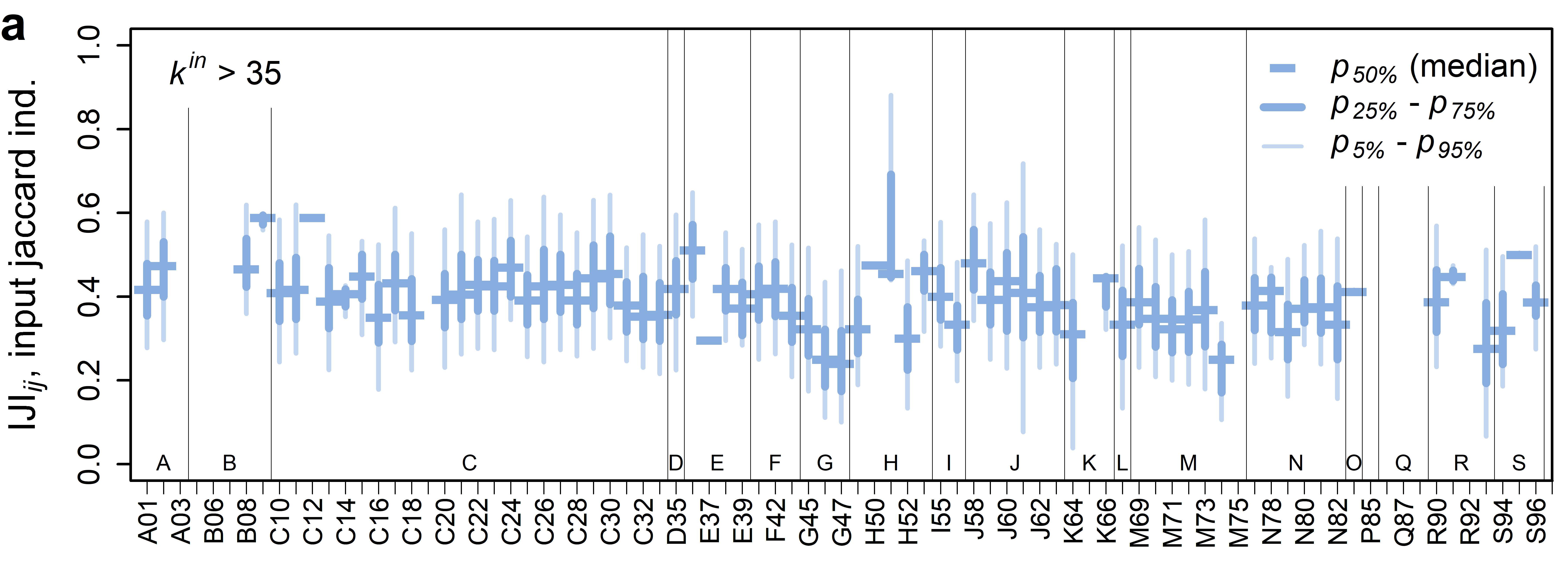} 
	\includegraphics[width=0.9\textwidth]{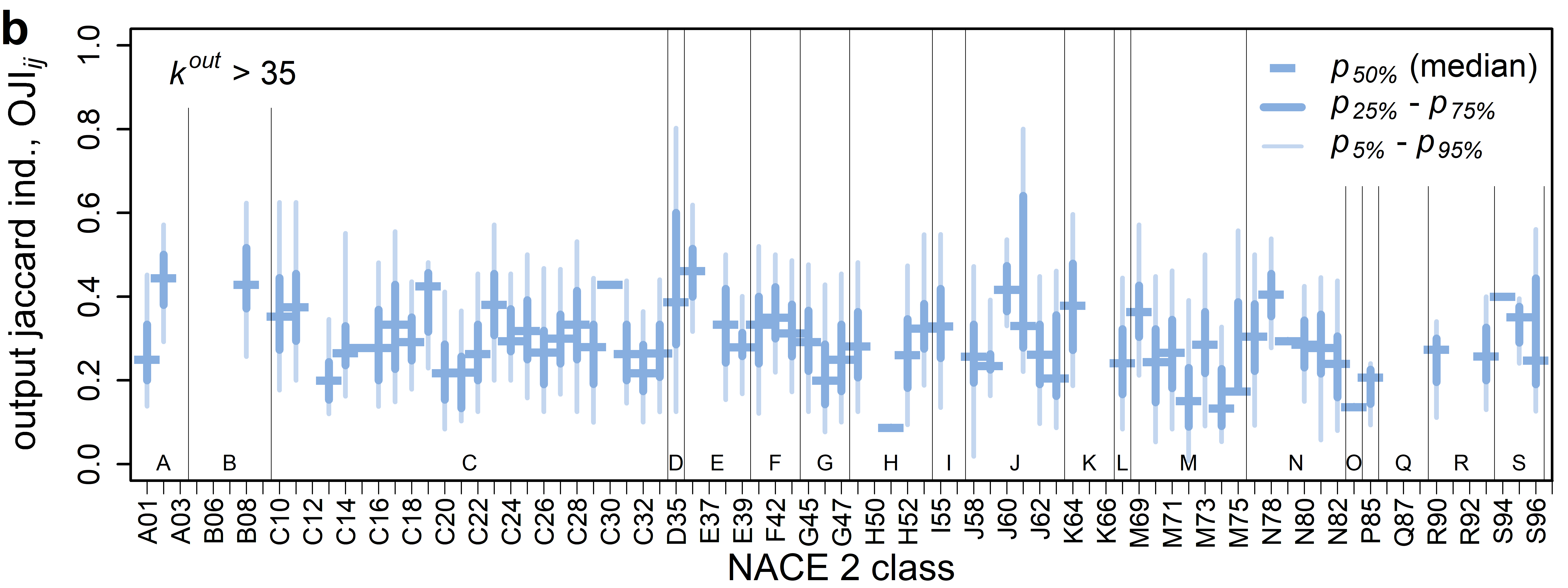} 
	\caption{Pairwise similarity distributions of input- and output-vectors of firms within each NACE2 industry. Similarity is measured with the Jaccard Index for firms with more than 35 suppliers (a) and buyers (b), respectively. 
	The y-axis denotes the Jaccard Index, the x-axis shows the NACE2 code for the respective boxplots. The dark blue horizontal bars correspond to the median, ($p_{50\%}$), dark blue vertical lines to the interquartile range ($p_{25\%}$ -- $p_{75\%}$), and thin light blue vertical lines to error bars ($p_{5\%}$ -- $p_{95\%}$). Thin black vertical lines separate NACE1 classes. Empty columns indicate sectors with less than two firms in this degree bin. 
	a) distributions of pairwise intra-industry input Jaccard Index, IJI$_{ij}$. The average of the mean (median) input Jaccard Index, across NACE2 industries is 0.398 (0.394) and the standard deviation of mean (median) input Jaccard Index is 0.07 (0.067). The average standard deviation is 0.031. This indicates that relatively low input similarity is the norm with few outliers.	
	b) distributions of pairwise intra-industry output vector Jaccard Index, OOC$_{ij}$. The average of the mean (median) output overlaps, across NACE2 industries is 0.301 (0.291) and the standard deviation of mean (median) output Jaccard Index values is 0.076 (0.077), indicating  that relatively low output similarities are the norm with few outliers. The average standard deviation is 0.031. Output overlaps are on average lower than the input overlaps, but there is only slightly more variation across industries. 
    If industry-level aggregation were fully representative for the IO-vectors of firms in both panels all distributions would correspond to a single bar at the value 1.
%~/WorkII/SupplyRank/HungaryProject/Heterogeneity_pnas/figures/SI_Figure2a.png 
% IOC    mean_of_means sd_of_means mean_of_medians sd_of_medians mean_of_sds sd_of_sds
%> 35         0.398        0.07           0.394         0.067         0.1     0.031
% OOC   mean_of_means sd_of_means mean_of_medians sd_of_medians mean_of_sds sd_of_sds
%> 35         0.301       0.076           0.291         0.077       0.114     0.031
	}
	\label{SI_fig2_nace2_JI_overlaps}
\end{figure*}

\section{Jaccard Index confirms low similarities} \label{SI_jaccard_index}
The Jaccard Index for two binary vectors $x, y$ of dimension $m$ can be defined as
\begin{equation} \label{eq_JI}
	\text{JI}\left( x, \; y \right) = \frac{ \sum_{k=1}^{m} \min \left[  x_k, y_k \right]}{ \sum_{k=1}^{m} \max \left[  x_k, y_k \right] } \quad .
\end{equation}

For firm $i$ we define the binary input vector, $\pi^\text{in}_{i}$,  as $\pi^\text{in}_{ik} =1$ if $\bar{\Pi}^\text{in}_{ik} > 0 $ and the binary output vector, $\pi^\text{out}_{i}$, as $\pi^\text{out}_{ik} =1$ if $\bar{\Pi}^\text{out}_{ik} > 0 $. 
Analogously to the IOC and OOC we define the pairwise input vector Jaccard Index, IJI, and the pairwise output vector Jaccard Index, OJI, of two firms $i$ and $j$ as
\begin{equation}
	\text{IJI}_{ij} %\left(\bar{\Pi}^\text{in}_{i.}, \; \bar{\Pi}^\text{in}_{j.} \right)
	 =  \frac{\sum_{k=1}^{m} \min \left[ \pi^\text{in}_{ik}, \; \pi^\text{in}_{jk} \right]}{\sum_{k=1}^{m} \max \left[  \pi^\text{in}_{ik}, \; \pi^\text{in}_{jk} \right]} \quad, \label{eq_iji_SI}  \qquad
	\text{OJI}_{ij} %\left(\bar{\Pi}^\text{out}_{i.}, \; \bar{\Pi}^\text{out}_{j.} \right)
	 =  \frac{\sum_{k=1}^{m} \min \left[  \pi^\text{out}_{ik}, \; \pi^\text{out}_{jk} \right]}{\sum_{k=1}^{m} \max \left[  \pi^\text{out}_{ik}, \; \pi^\text{out}_{jk} \right]} \quad . %\label{eq_oji_SI}.
\end{equation}

\subsection*{Results for the Jaccard Index}

We show that the results from the main text do not depend on the specific similarity measure. We show the results of Fig. \ref{fig2nace2} and  Fig. \ref{fig3_nace2main} are qualitatively similar when using IJI and OJI instead of IOC and OOC.

First, we show the pairwise similarity distributions of input and output vectors for firms of NACE class 26, Manufacture of computer, electronic and optical products measured with the Jaccard index.  
SI Fig. \ref{SI_fig1nace2_jaccard_C26}a-d show input Jaccard Indices, IJI$_{ij}$, and SI Fig. \ref{SI_fig1nace2_jaccard_C26}e-h output Jaccard Indices, OJI$_{ij}$, visualized as histograms, for four in-degree, $k_i^{in}$, (number of suppliers) and out-degree, $k_i^{out}$ (number of buyers), bins, respectively. Jaccard Index values are on the x-axis in bins of width 0.05; the y-axis shows the frequency to fall in the respective bin. Vertical solid lines correspond to median and dashed lines to mean overlap coefficients.
SI Fig. \ref{SI_fig1nace2_jaccard_C26}a-d shows that the median and mean similarities of input vectors measured by the IJI are slightly higher than for the IOC. The medians are 0, 0.2, 0.231 and 0.425 for the IJI and 0, 0.121, 0.199, and 0.343 for the IOC. The differences in means is slightly smaller. Further, the standard deviation is smaller for the IJI values than for the IOC values (0.261, 0.109, 0.091, 0.119, vs. 0.282, 0.192, 0.161, 0.148). For smaller size bins the IJI distribution is also right skewed, but less so and the distribution becomes symmetric faster than for the IOC. As indicated by the lower standard deviations the distributions are narrower. In general the similarities are relatively low and far away from the value of one, which would indicate that industry-level aggregation is representative for firm-level input vectors. 
%
%in JI  median  mean sd  itqr
%1-5    0.000 0.141 0.261 0.250
%6-15   0.200 0.204 0.109 0.148
%16-35  0.231 0.237 0.091 0.133
%> 35   0.425 0.430 0.119 0.166
%
% IOC      median  mean    sd  itqr
%1-5    0.000 0.141 0.282 0.113
%6-15   0.121 0.196 0.192 0.227
%16-35  0.199 0.239 0.161 0.176
%> 35   0.343 0.343 0.148 0.230

Fig. \ref{SI_fig1nace2_jaccard_C26}e-h shows that the median and mean similarities of output vectors measured by the OJI are slightly higher than for the OOC. The medians are 0, 0.1, 0.2 and 0.267 for the OJI and 0, 0.025, 0.119, and 0.119 for the OOC. The differences in the means are smaller. Further, the standard deviation is smaller for the OJI values than for the OOC values (0.163, 0.109, 0.127, 0.106, vs. 0.190, 0.141, 0.156, 0.123). For smaller size bins the OJI distribution is also right skewed, but less so and the distribution becomes symmetric faster. As indicated by the lower standard deviations the distributions are narrower. In general the similarities are relatively low and far away from the value of one, which would indicate that industry-level aggregation is representative for firm-level input vectors. 
%
%out JI  median  mean    sd  itqr
%1-5    0.000 0.054 0.163 0.000
%6-15   0.100 0.115 0.109 0.167
%16-35  0.200 0.212 0.127 0.168
%> 35   0.267 0.265 0.106 0.128
%
%OOC   median  mean sd  itqr
%1-5    0.000 0.054 0.190 0.000
%6-15   0.025 0.087 0.141 0.111
%16-35  0.119 0.169 0.156 0.208
%> 35   0.119 0.143 0.123 0.091

The  patterns of increasing similarity with degree also holds true for IJI and OJI. 
SI Fig. \ref{SI_fig2_nace2_JI_overlaps} shows the IJI and OJI for the degree bins of firms with more than 35 suppliers ($k^\text{in} > 35$) and more than 35 customers ($k^\text{out} > 35$), respectively. 
SI Fig. \ref{SI_fig2_nace2_JI_overlaps}a shows that as for NACE class C26 the average similarity is slightly higher for the IJI than for the IOC. 
For the IJI$_{ij}$ the average of the mean (median) input overlaps, across NACE2 industries is 0.398 (0.394) and the standard deviation of mean (median) input overlaps is 0.07 (0.067). 
For the IOC$_{ij}$ the average of the mean (median) input overlaps, across NACE2 industries is 0.35 (0.33) and the standard deviation of mean (median) input overlaps is 0.084 (0.102). 
The average standard deviation for IJI is 0.031, which is substantially lower than the average standard deviation for the IOC of 0.156. This implies that the distributions are on average more concentrated for the jaccard index based input vector similarity. 
This is not surprising as both measures have a similar numerator, but the binary counting of input vectors in the Jaccard Index probably reduces the range of possible lower range outliers. This is because the binary counting of the JI tends to give overlaps that are small when measured with the OC a higher weight  (the JI denominator divides in the best case by the number of joint inputs and in the worst case by the number of different inputs of both firms added up). The same reasoning could explain the slightly higher average similarity values of JI over OC.
%~/WorkII/SupplyRank/HungaryProject/Heterogeneity_pnas/figures/SI_Figure2a.png 
% IOC    mean_of_means sd_of_means mean_of_medians sd_of_medians mean_of_sds sd_of_sds
%> 35         0.398        0.07           0.394         0.067         0.1     0.031
SI Fig. \ref{SI_fig2_nace2_JI_overlaps}b shows the results for the pairwise output vector similarity based on the Jaccard Index. 
For the OJI$_{ij}$ the average of the mean (median) output vector Jaccard Index, across NACE2 industries is 0.301 (0.291) and the standard deviation of mean (median) input Jaccard Index is 0.076 (0.077). 
OOC$_{ij}$. The average of the mean (median) output overlaps, across NACE2 industries is 0.282 (0.257) and the standard deviation of mean (median) output overlaps is 0.147 (0.161). 
Again the average Jaccard Index based similarity, OJI, is slightly higher than the average output overlap coefficient OOC. 
The average standard deviation for OJI is 0.031, which is substantially lower than  the average standard deviation for the OOC of 0.17. This implies that the distributions are on average more concentrated for the Jaccard Index based output vector similarity. 
% OOC   mean_of_means sd_of_means mean_of_medians sd_of_medians mean_of_sds sd_of_sds
%> 35         0.301       0.076           0.291         0.077       0.114     0.031

Overall we observe a qualitatively similar degree of similarity when using the Jaccard Index instead of the overlap coefficient. 

\FloatBarrier

\clearpage

\begin{figure*}[t]
	\centering
	\includegraphics[width= 1\textwidth]{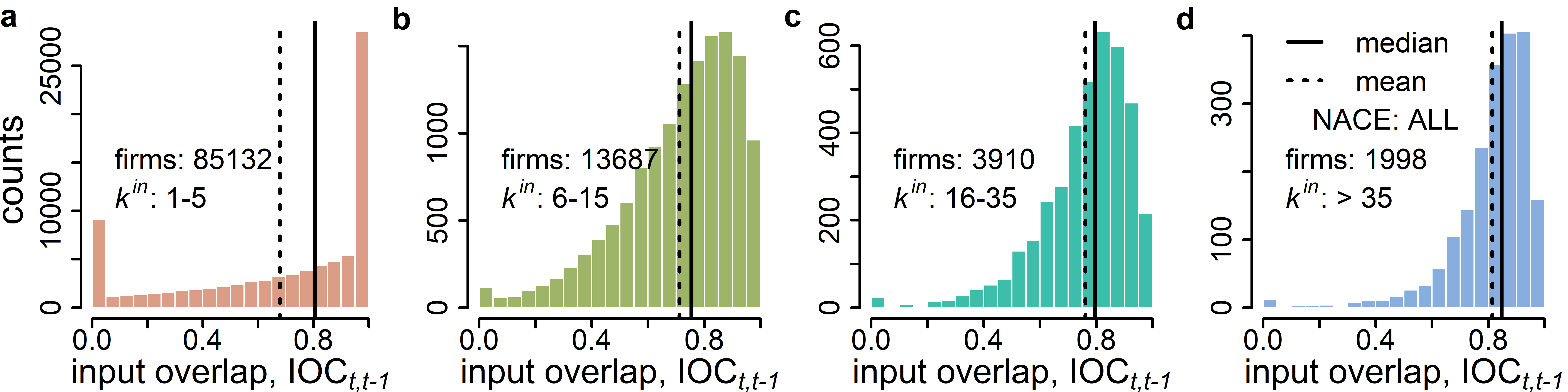} 
	\includegraphics[width= 1\textwidth]{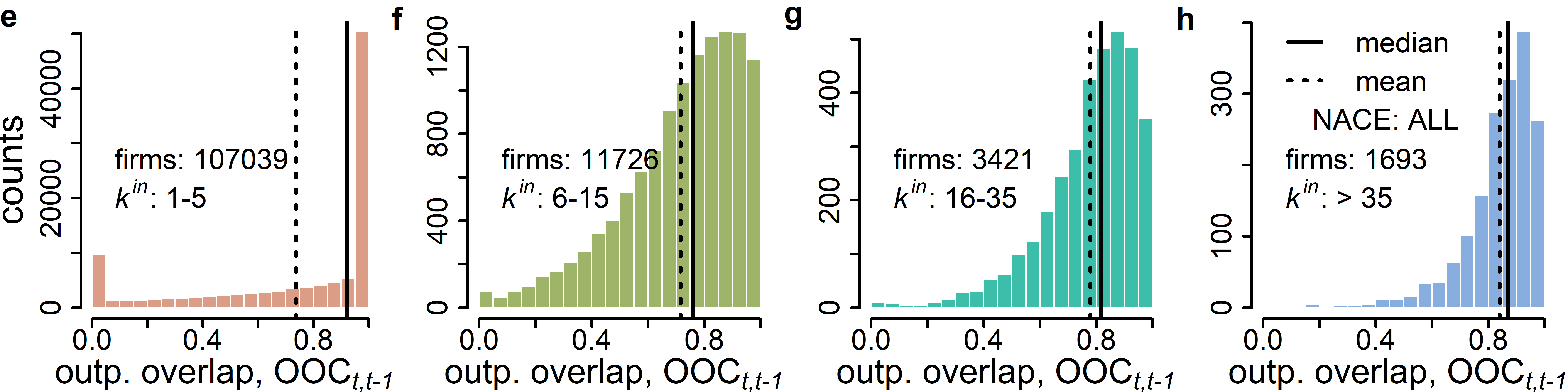} 
	\caption{Distribution of input and output overlap coefficients of firms' input- and output-vectors across the years 2019 and 2018 over all NACE2 industries. 
        The overlap coefficients, OC,  are on the x-axis and counts for the respective OC-value bin on the y-axis. 
		a-d) illustrate the distributions of, $\text{IOC}_{t, t-1}$, across all NACE 2 industries for the four in-degree bins (1-5, 6-15, 16-35, $>$35) as histograms. The median and mean IOCs over time, $\text{IOC}_{t, t-1}$, are 0.805 (0.678) , 0.755 (0.712), 0.797 (0.761) and 0.847 (0.814), respectively, indicated by the vertical solid (dashed) lines. The standard deviations for the in-degree bins are 0.345, 0.203, 0.161 and 0.142, respectively, and decreasing with the number of in-links. 
		e-h) illustrate the distributions of, $\text{OOC}_{t, t-1}$, for the respective out-degree bins. The median and (mean) OOCs over time, $\text{OOC}_{t, t-1}$, are 0.922 (0.737) , 0.816 (0.778), 0.869 (0.841) and 0.847 (0.814), The standard deviations for the out-degree bins are 0.34, 0.209, 0.163 and 0.128; again decreasing with the in-link number.
		The similarity of firms input- and output-vectors over time is substantially higher than for the pairwise intra-industry similarities. }
	\label{SI_fig4nace2_OC_hist_over_time}
\end{figure*}

\begin{figure*}[t]
	\centering
	\includegraphics[width= 0.95\textwidth]{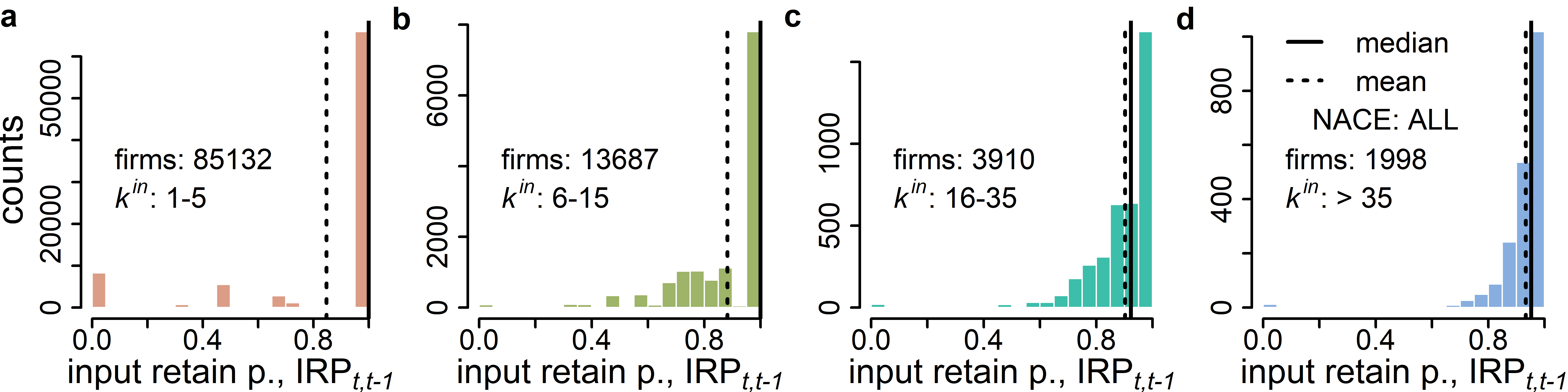} 
	\includegraphics[width= 0.95\textwidth]{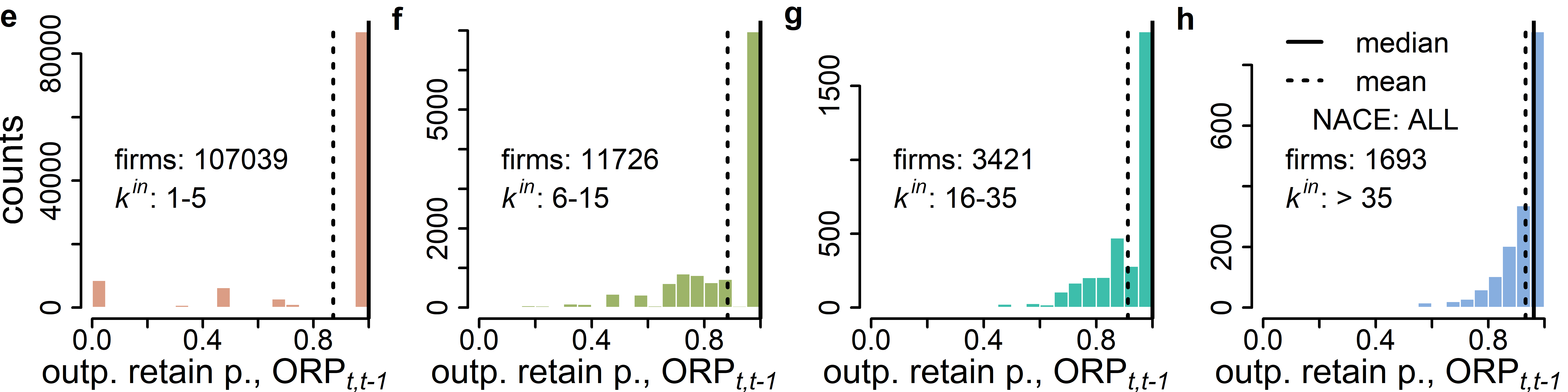} 
	\caption{Distribution of input and output retention probabilities (IRPs and ORPs) of firms for 2019 and 2018 across all industries. 
        The retention probabilities, RPs,  are on the x-axis and counts for the respective RP-value bins on the y-axis. 
		a-d) illustrate the distributions of, $\text{IRP}_{t, t-1}$, across all NACE 2 industries for the four in-degree bins (1-5, 6-15, 16-35, $>$35) as histograms. The median and mean IRPs over time, $\text{IRP}_{t, t-1}$, are 1 (0.847) , 1 (0.882), 0.923 (0.902) and 0.952 (0.933), respectively, indicated by the vertical solid (dashed) lines. The means increase with the in-degree. The standard deviations for the in-degree bins are 0.316, 0.173, 0.133 and 0.108, respectively, and decreasing with the number of in-links. With increasing in-degree the distributions become more concentrated on the value 1, i.e. most firms retain almost all NACE2 input types. 
		e-h) illustrate the distributions of, $\text{ORP}_{t, t-1}$, for the respective out-degree bins. The median and (mean) ORPs over time, $\text{ORP}_{t, t-1}$, are 1 (0.872) , 1 (0.883), 1 (0.913) and 0.962 (0.933), i.e. means increase with in-degree. The standard deviations for the out-degree bins are 0.295, 0.177, 0.130 and 0.100; again decreasing with the in-link number. 
        With increasing out-degree the distributions become more concentrated on the value 1, i.e. most firms retain almost all NACE2 customer industries. 
		The similarity of firms input- and output-vectors over time is slightly higher than the intra-industry similarities.}
	\label{SI_fig4nace2_RetProb}
\end{figure*}

%~/WorkII/SupplyRank/HungaryProject/Heterogeneity_pnas/figures/SI_Figure7a.png 
% IRP    median  mean    sd  itqr
%1-5    1.000 0.847 0.316 0.000
%6-15   1.000 0.882 0.173 0.200
%16-35  0.923 0.902 0.133 0.143
%> 35   0.952 0.933 0.108 0.091

%~/WorkII/SupplyRank/HungaryProject/Heterogeneity_pnas/figures/SI_Figure7b.png 
%ORP    median  mean    sd  itqr
%1-5    1.000 0.872 0.295 0.000
%6-15   1.000 0.883 0.177 0.200
%16-35  1.000 0.913 0.130 0.143
%> 35   0.962 0.933 0.100 0.100

\section{Input and output vectors are similar over time} \label{SI_section_auto_sim}

In this section we show that the low pairwise IOC and OOC values for firms within the same industries are not a generic feature of the micro-level data. 
The similarity of firms input and output vectors over time is substantially higher than the intra-industry similarities. 
To show this we calculate for each firm the overlap coefficient of its relative input vector in the year $t$ with its input vector in the previous year $t-1$ as
\begin{equation}
    \text{IOC}_{t, t-1} =  \sum_{k=1}^{m} \min \left[  \bar{\Pi}^\text{in}_{ik}(t), \; \bar{\Pi}^\text{in}_{ik}(t-1) \right]
\end{equation}
Analgously, we compute the output overlap coefficient between two years $t$ and $t-1$ as
\begin{equation}
    \text{OOC}_{t, t-1} = \sum_{k=1}^{m} \min \left[  \bar{\Pi}^\text{out}_{ik}(t), \; \bar{\Pi}^\text{out}_{ik}(t-1) \right]
\end{equation}
The two measures indicate the fraction of total inputs (outputs) that is spent  on (sold to) the same industry in the two year. We calculate the overlap coefficients over time for the years 2019 and 2018. Firms are allocated into the respective in- and out-degree bins based on their number of suppliers or customers in the year 2018. 

SI Fig. \ref{SI_fig4nace2_OC_hist_over_time} we show the distribution of input and output overlap coefficients of firms' input- and output-vectors across the years 2019 and 2018 over all NACE2 industries. 
The overlap coefficients, OC,  are on the x-axis and counts for the respective OC-value bin on the y-axis. 
SI Fig. \ref{SI_fig4nace2_OC_hist_over_time}a-d illustrates the distributions of $\text{IOC}_{t, t-1}$ across all NACE 2 industries for the four in-degree bins (1-5, 6-15, 16-35, $>$35) as histograms. The median and mean IOCs over time, $\text{IOC}_{t, t-1}$, are 0.805 (0.678) , 0.755 (0.712), 0.797 (0.761) and 0.847 (0.814), respectively, and thus substantially higher than for the intra-industry IOCs. The standard deviations for the in-degree bins are 0.345, 0.203, 0.161 and 0.142, respectively and decreasing with the number of in-links. The distributions are left skewed, i.e. very low overlap coefficients are outliers and for the smallest in-degree bin bi-modal. In all four bins there are firms having almost zero input overlap in the two years. While this number is relatively high for the smallest in-degree bin it decreases strongly for higher in-degree bins. For firms with few suppliers this is most likely due to the change of a single or the primary supplier. For the few cases where firms with many suppliers have almost no overlap the likely explanation is that they went out of business between the two years and did not source inputs anymore in the second year. As the network is growing --- due to a reduction of the link reporting threshold in mid-2018 --- the overlaps over time shown here might be smaller than in practice. Therefore, we check also the probability of retaining an input type from the year 2018 in the year 2019 and find that these are even higher than the overlap coefficients, for details see SI Fig. \ref{SI_fig4nace2_RetProb}a-d.  
%
%all IOC  median  mean    sd  itqr
%1-5    0.805 0.678 0.345 0.556
%6-15   0.755 0.712 0.203 0.269
%16-35  0.797 0.761 0.161 0.187
%> 35   0.847 0.814 0.142 0.141
%

Analogously Fig. \ref{SI_fig4nace2_OC_hist_over_time}e-h illustrates the distributions of $\text{OOC}_{t, t-1}$ for the respective out-degree bins. The median and (mean) OOCs over time, $\text{OOC}_{t, t-1}$, are 0.922 (0.737) , 0.816 (0.778), 0.869 (0.841) and 0.847 (0.814), respectively, and thus substantially higher than for the intra-industry OOCs and slightly higher than the IOCs over time. The standard deviations for the out-degree bins are 0.34, 0.209, 0.163 and 0.128; again decreasing with the number of out-links. The distributions are left skewed and for the smallest out-degree bin bi-modal. In all four bins there are firms having almost zero output overlap in the two years, but substantially less so than for the $\text{IOC}_{t, t-1}$. The probability of retaining an output type (buyer industry) from the year 2018 in the year 2019 is again higher than the overlap coefficients, for details see SI Fig. \ref{SI_fig4nace2_RetProb}a-d.
%~/WorkII/SupplyRank/HungaryProject/Heterogeneity_pnas/figures/SI_Figure5b.png 
% OOC   median  mean    sd  itqr
%1-5    0.922 0.737 0.340 0.460
%6-15   0.761 0.716 0.209 0.280
%16-35  0.816 0.778 0.163 0.201
%> 35   0.869 0.841 0.128 0.137

To show that firms overwhelmingly keep existing inputs and buyer industries we calculate for each firm the input retention probability and the output retention probability from the binary input and output vectors of a year $t$ with the previous year $t-1$. 
Recall the binary input vector, $\pi^\text{in}_{i}$, is defined as $\pi^\text{in}_{ik} =1$ if $\bar{\Pi}^\text{in}_{ik} > 0 $ and the binary output vector, $\pi^\text{out}_{i}$, as $\pi^\text{out}_{ik} =1$ if $\bar{\Pi}^\text{out}_{ik} > 0 $. 

We define the input retention probability, IRP$_{t, t-1}$, for a firm $i$ between two years $t$ and $t-1$ as
\begin{equation}
    \text{IRP}_{t, t-1} =  \frac{\sum_{k=1}^{m} \min \left[  \pi^\text{in}_{ik}(t), \; \pi^\text{in}_{ik}(t-1) \right]}{\sum_{k=1}^{m} \pi^\text{in}_{ik}(t-1) } \quad .
\end{equation}
$ \text{IRP}_{t, t-1} $ is the probability that a random input contained in the input vector of firm $i$ in year $t-1$ is still present in the input vector of firm $i$ at time $t$. 
Analogously, we compute the output retention probability, ORP$_{t, t-1}$, for a firm $i$ between two years $t$ and $t-1$ as
\begin{equation}
    \text{ORP}_{t, t-1} = \frac{\sum_{k=1}^{m} \min \left[  \pi^\text{out}_{ik}(t), \; \pi^\text{out}_{ik}(t-1) \right]}{\sum_{k=1}^{m} \min \pi^\text{out}_{ik}(t-1) } \quad . 
\end{equation}
$ \text{ORP}_{t, t-1} $ is the probability that a random buyer industry contained in the output vector vector of firm $i$ in year $t-1$ is still present in the output vector of firm $i$ at time $t$. 

We calculate IRP and ORP over time for the years 2019 and 2018 for each firm. Firms are allocated into the respective in- and out-degree bins based on their number of suppliers or customers in the year 2018. 
The results are shown as histograms  SI Fig. \ref{SI_fig4nace2_RetProb}, where the retention probabilities are on the x-axis and counts for the respective RP-value bins on the y-axis. 
SI Fig. \ref{SI_fig4nace2_RetProb}a-d illustrate the distributions of, $\text{IRP}_{t, t-1}$, across all NACE 2 industries for the four in-degree bins (1-5, 6-15, 16-35, $>$35) as histograms. The median and mean IRPs over time, $\text{IRP}_{t, t-1}$, are 1 (0.847) , 1 (0.882), 0.923 (0.902) and 0.952 (0.933), respectively, indicated by the vertical solid (dashed) lines. The means increase with the in-degree. The standard deviations for the in-degree bins are 0.316, 0.173, 0.133 and 0.108, respectively, and decreasing with the number of in-links. With increasing in-degree the distributions become more concentrated on the value 1, i.e. most firms retain almost all NACE2 input types. 
SI Fig. \ref{SI_fig4nace2_RetProb}e-h illustrate the distributions of, $\text{ORP}_{t, t-1}$, for the respective out-degree bins. The median and (mean) ORPs over time, $\text{ORP}_{t, t-1}$, are 1 (0.872) , 1 (0.883), 1 (0.913) and 0.962 (0.933). The means increase with the out-degree. The standard deviations for the out-degree bins are 0.295, 0.177, 0.130 and 0.100; again decreasing with the out-link number.   With increasing out-degree the distributions become more concentrated on the value 1, i.e. most firms retain almost all NACE2 customer industries. 
The similarity of firms input- and output-vectors over time is substantially higher than the intra-industry similarities.

\subsection*{Overlaps over time for industries}

\begin{figure*}[t]
	\centering
	\includegraphics[width= 0.95\textwidth]{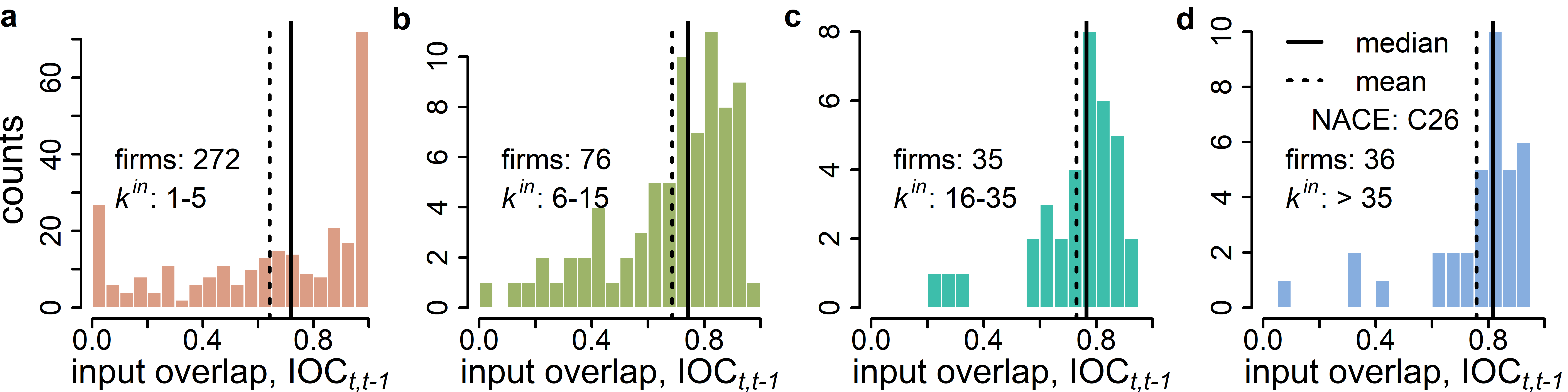} 
	\includegraphics[width= 0.95\textwidth]{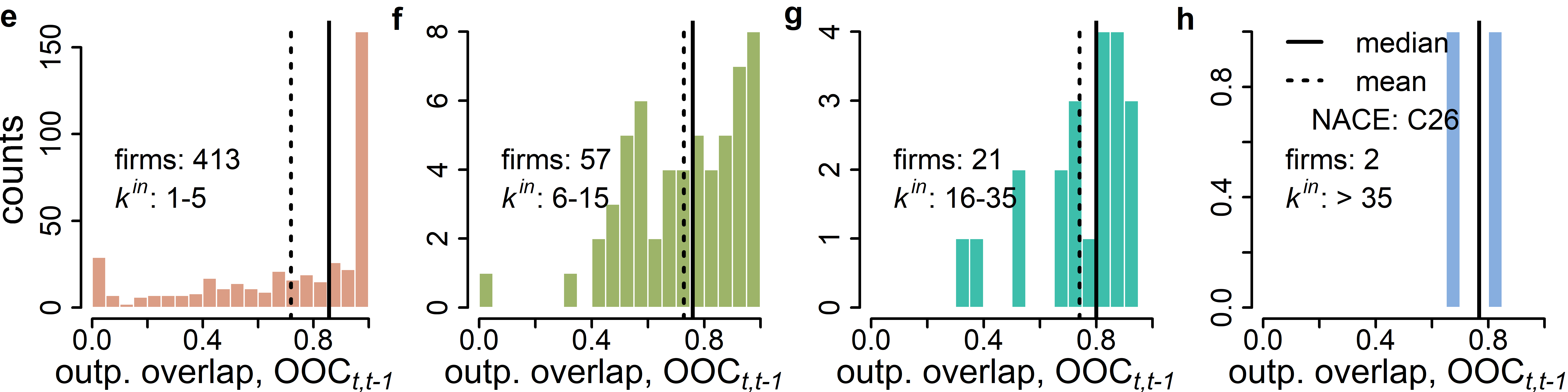} 
	\caption{Similarity of firms' input- and output-vectors for 2019 and 2018 for NACE2 class C26, measures with the overlap coefficient (OC). The OC is on the x-axis and the counts for the respective OC-value bin on the y-axis. 
		a-d) illustrate the distributions of, $\text{IOC}_{t, t-1}$, across all NACE 2 industries for the four in-degree bins (1-5, 6-15, 16-35, $>$35) as histograms. The median and mean IOCs over time, $\text{IOC}_{t, t-1}$, are 0.718 (0.642) , 0.743 (0.686), 0.765 (0.730) and 0.818 (0.759), respectively, indicated by the vertical solid (dashed) lines, and increasing with in-degree. The standard deviations for the in-degree bins are 0.34, 0.217, 0.167 and 0.188, respectively, and decreasing with the number of in-links. 
		e-h) illustrate the distributions of, $\text{OOC}_{t, t-1}$, for the respective out-degree bins. The median and (mean) OOCs over time, $\text{OOC}_{t, t-1}$, are 0.857 (0.719) , 0.759 (0.727), 0.801 (0.740) and 0.768 (0.768). Only the means are increasing, but not the medians. The standard deviations for the out-degree bins are 0.321, 0.206, 0.176 and 0.115; again decreasing with the out-link number.
		The similarity of firms input- and output-vectors over time is substantially higher than the intra-industry similarities. }
	\label{SI_fig4nace2_OC_C26}
\end{figure*}

% OC over time
%C26 IOC median  mean  sd  itqr
%1-5    0.718 0.642 0.340 0.570
%6-15   0.743 0.686 0.217 0.255
%16-35  0.765 0.730 0.167 0.155
%> 35   0.818 0.759 0.188 0.132
%
%C26 OOC      median  mean    sd  itqr
%1-5    0.857 0.719 0.321 0.490
%6-15   0.759 0.727 0.206 0.339
%16-35  0.801 0.740 0.176 0.183
%> 35   0.768 0.768 0.115 0.081
In this section we show the distribution of input and output overlap coefficients over time for specific NACE2 industries. 
For completeness we illustrate the similarity over time for NACE2 industry C26 in SI Fig. \ref{SI_fig4nace2_OC_C26}. 
The overlap coefficient, OC,  is on the x-axis and counts for the respective OC-value bin on the y-axis. 
SI Fig. \ref{SI_fig4nace2_OC_C26}a-d illustrate the distributions of, $\text{IOC}_{t, t-1}$, across all NACE 2 industries for the four in-degree bins (1-5, 6-15, 16-35, $>$35) as histograms. The median and mean IOCs over time, $\text{IOC}_{t, t-1}$, are 0.718 (0.642) , 0.743 (0.686), 0.765 (0.730) and 0.818 (0.759), respectively, indicated by the vertical solid (dashed) lines, and increasing with in-degree The standard deviations for the in-degree bins are 0.34, 0.217, 0.167 and 0.188, respectively, and decreasing with the number of in-links. 
SI Fig. \ref{SI_fig4nace2_OC_C26}e-h illustrate the distributions of, $\text{OOC}_{t, t-1}$, for the respective out-degree bins. The median and (mean) OOCs over time, $\text{OOC}_{t, t-1}$, are 0.857 (0.719) , 0.759 (0.727), 0.801 (0.740) and 0.768 (0.768). Only the means are increasing, but not the medians. The standard deviations for the out-degree bins are 0.321, 0.206, 0.176 and 0.115; again decreasing with the in-link number.
The similarity of firms input- and output-vectors over time is substantially higher, than the intra-industry similarities. For NACE C26 neither input or output overlaps are consistently larger across degree bins.

\begin{figure*}[t]
	\centering
	\includegraphics[width=0.9\textwidth]{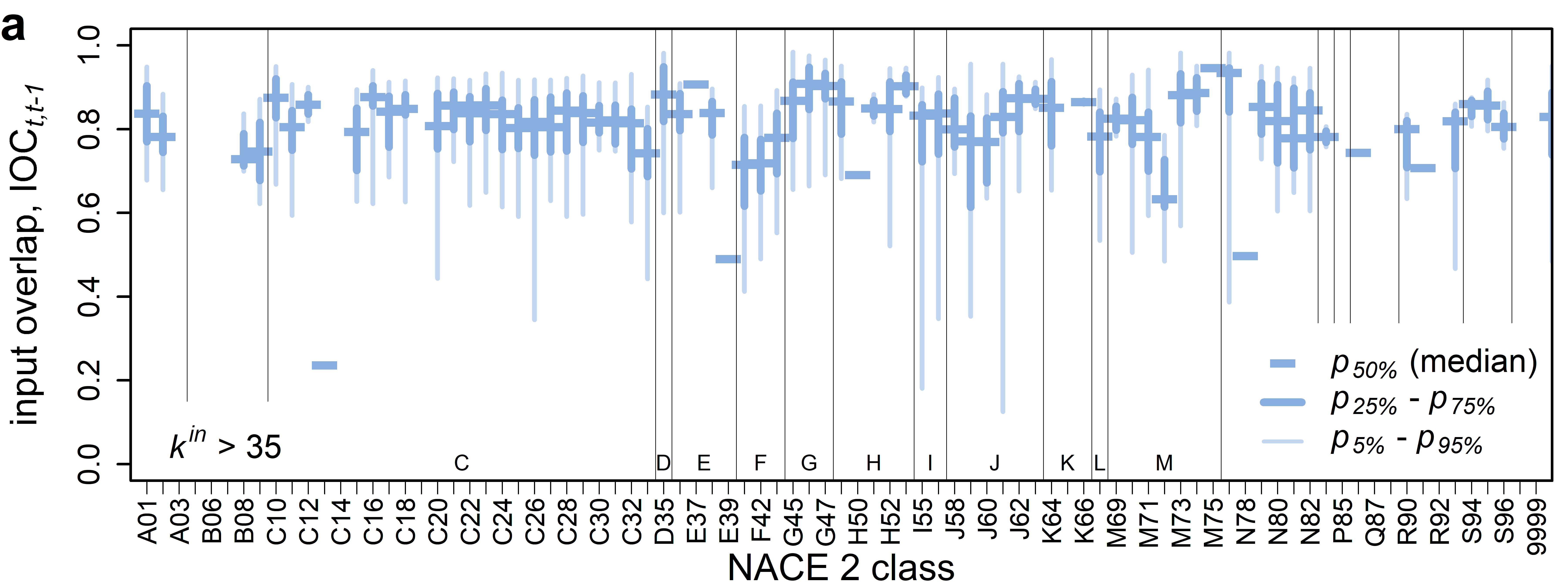} 
	\includegraphics[width=0.9\textwidth]{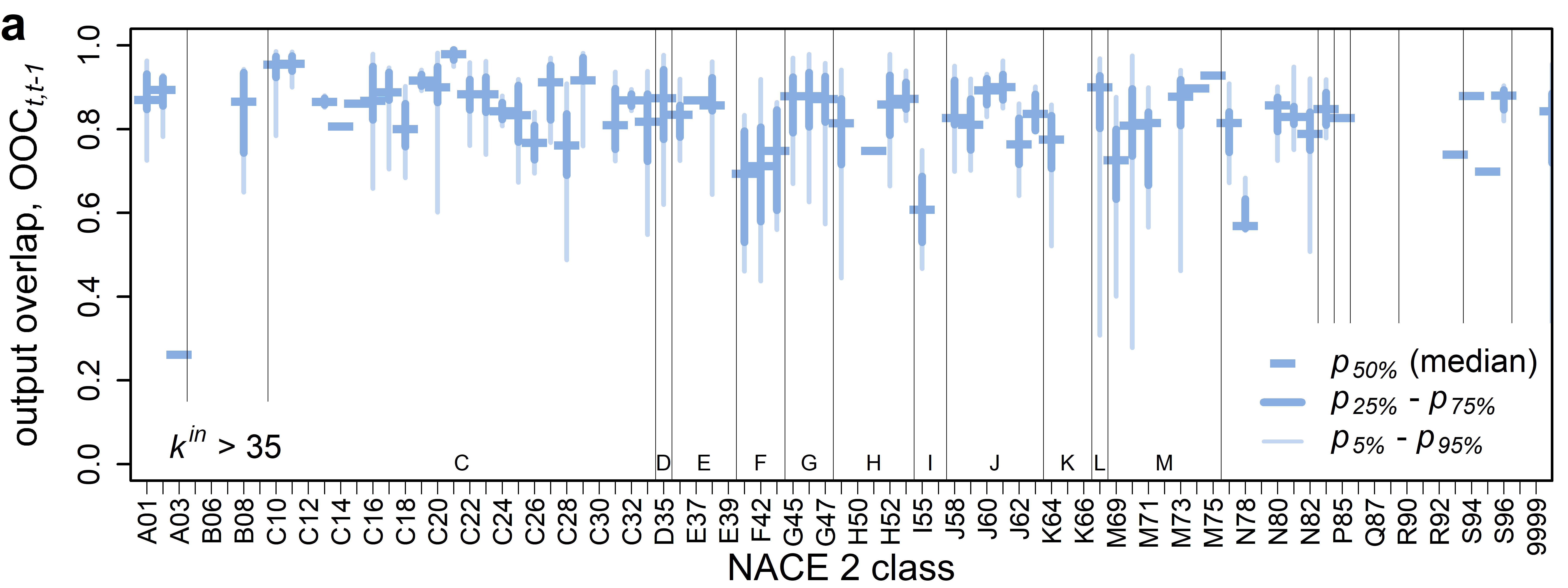} 
	\caption{Similarity distributions of input- and output-vectors of firms between 2019 and 2018 for each NACE2 industry. Similarity is measured with the overlap coefficient for firms with more than 35 suppliers (a) and buyers (b), respectively. 
	The y-axis denotes the overlap coefficients between the two years, the x-axis shows the NACE2 code for the respective boxplots. The dark blue horizontal bars correspond to the median, ($p_{50\%}$), dark blue vertical lines to the interquartile range ($p_{25\%}$ -- $p_{75\%}$), and thin light blue vertical lines to error bars ($p_{5\%}$ -- $p_{95\%}$). Thin black vertical lines separate NACE1 classes. Empty columns indicate  no firms in this degree bin. 
	a) distributions of firms input overlap coefficients, IOC$_{t, t-1}$. The average of the mean (median) input overlaps, across NACE2 industries is 0.784 (807) and the standard deviation of mean (median) input overlaps is 0.099 (0.102). The average standard deviation is 0.124. This indicates that high input overlaps are the norm with few outliers.	
	b) distributions of pairwise intra-industry output overlap coefficients, OOC$_{t, t-1}$. The average of the mean (median) output overlaps, across NACE2 industries is 0.813 (0.828) and the standard deviation of mean (median) output overlaps is 0.101 (0.103), indicating  that relatively low output overlaps are the norm with few outliers. The average standard deviation is 0.108. Output overlaps are on average slightly higher than input overlaps.
	}
	\label{SI_Figure10a_nace4}
\end{figure*}

%~/WorkII/SupplyRank/HungaryProject/Heterogeneity_pnas/figures/SI_Figure8_2a.png 
%1 
%mean_of_means sd_of_means mean_of_medians sd_of_medians mean_of_sds sd_of_sds
%> 35         0.784       0.099           0.807         0.102       0.124     0.058
%~/WorkII/SupplyRank/HungaryProject/Heterogeneity_pnas/figures/SI_Figure8_2b.png 
%1 
%mean_of_means sd_of_means mean_of_medians sd_of_medians mean_of_sds sd_of_sds
%> 35         0.813       0.101           0.828         0.103       0.108     0.053

Next we look at the distributions of $\text{IOC}_{t, t-1}$ and $\text{OOC}_{t, t-1}$ across NACE2 industries. 
For the following figures, the y-axis denotes the overlap coefficients between the two years, the x-axis shows the NACE2 code for the respective boxplots. The dark  horizontal bars correspond to the median, ($p_{50\%}$), dark vertical lines to the interquartile range ($p_{25\%}$ -- $p_{75\%}$), and thin light  vertical lines to error bars ($p_{5\%}$ -- $p_{95\%}$). Thin black vertical lines separate NACE1 classes. Empty columns indicate  sectors with less than two firms in this degree bin. 

\begin{figure*}[t]
	\centering
	\includegraphics[width=1\textwidth]{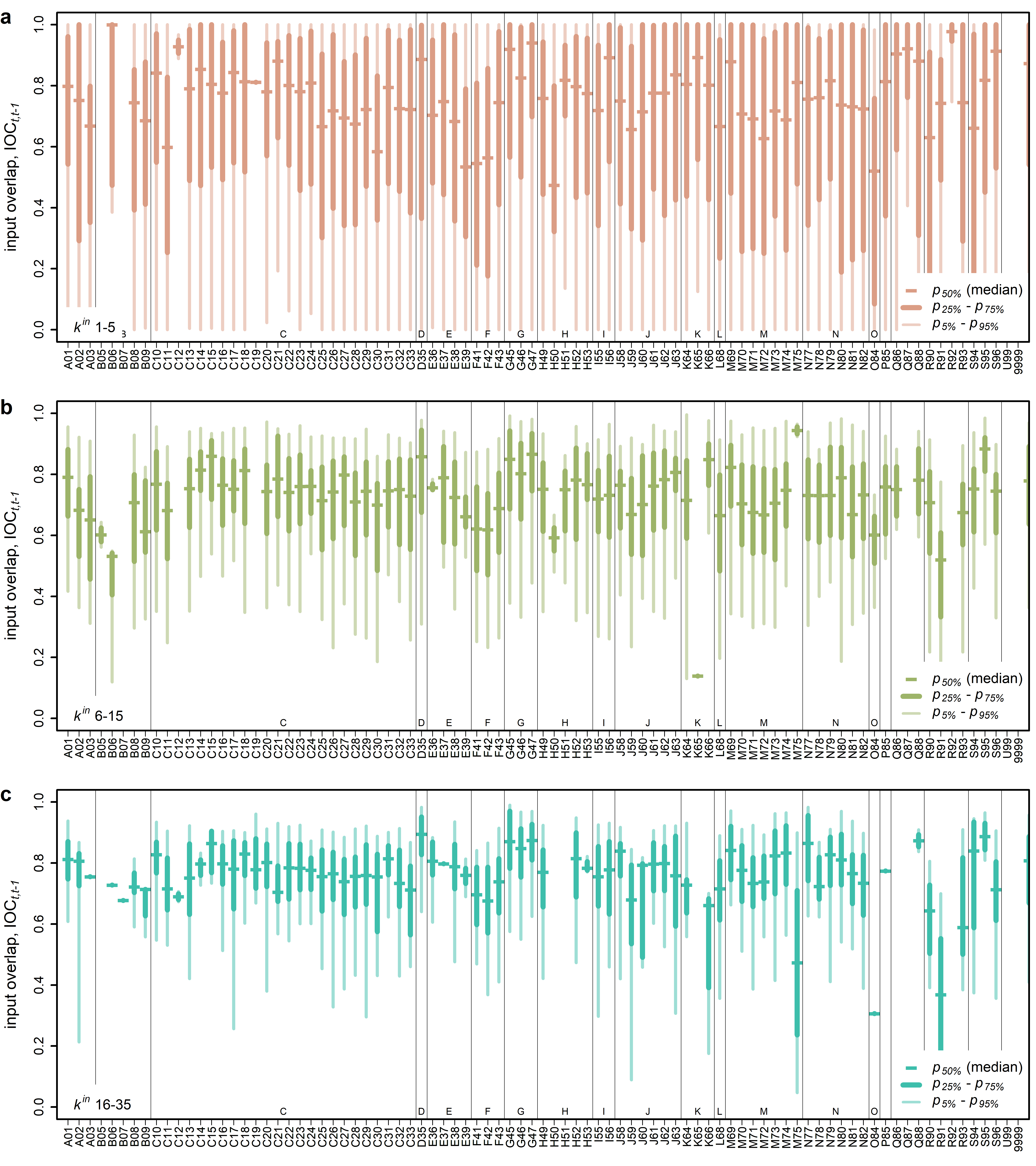} 
	\caption{	Distributions of input vector overlaps, IOC$_{t, t-1}$, of firms across NACE2 industries for the years 2019 and 2018. NACE2 classes are on the x-axis; overlap coefficients on the y-axis. 
		a)  distributions of input overlap coefficients, IOC$_{t, t-1}$, for firms with in-degree between one and five, $1 \leq k_i^{in} \leq 5$. The mean over the industries' mean (median) IOC$_{t, t-1}$ is 0.660 (0.763), the standard deviation of mean (median) IOCs is 0.079 (0.104). The mean standard deviation is 0.334.
		b) distributions of input overlap coefficients, IOC$_{t, t-1}$, for firms with in-degree between 6 and fifteen, $6 \leq k_i^{in} \leq 15$. The mean over the industries' mean (median) IOC$_{t, t-1}$ is 0.692 (0.729), the standard deviation of mean (median) IOCs is 0.097 (0.100). The mean standard deviation is 0.184.
		c) distributions of input overlap coefficients, IOC$_{t, t-1}$, for firms with in-degree between 16 and 35, $16 \leq k_i^{in} \leq 35$. The mean over the industries' mean (median) IOC$_{t, t-1}$ is 0.730 (0.094), the standard deviation of mean (median) IOCs is 0.094 (0.096). The mean standard deviation is 0.162. 
	}
	\label{SI_fig11a_nace2}
\end{figure*}

First, we focus on the distributions of input overlaps for the years 2019 and 2018, $\text{IOC}_{t, t-1}$, in 
SI Fig. \ref{SI_Figure10a_nace4}a and SI Fig. \ref{SI_fig11a_nace2}.
SI Fig. \ref{SI_Figure10a_nace4}a shows the distributions of firms input overlap coefficients, IOC$_{t, t-1}$, for firms with more than 35 suppliers,  $k_i^{in} > 35$. The average of the mean (median) input overlaps, across NACE2 industries is 0.784 (807) and the standard deviation of mean (median) input overlaps is 0.099 (0.102). The average standard deviation is 0.124. This indicates that high input overlaps are the norm with few outliers.	
SI Fig. \ref{SI_fig11a_nace2}a shows the  distributions of input overlap coefficients, IOC$_{t, t-1}$, for firms with in-degree between one and five, $1 \leq k_i^{in} \leq 5$. The mean over the industries' mean (median) $\text{IOC}_{t, t-1}$ is 0.660 (0.763), the standard deviation of mean (median) IOCs is 0.079 (0.104). The mean standard deviation is 0.334.
SI Fig. \ref{SI_fig11a_nace2}b shows the distributions of input overlap coefficients, IOC$_{t, t-1}$, for firms with in-degree between 6 and fifteen, $6 \leq k_i^{in} \leq 15$. The mean over the industries' mean (median) $\text{IOC}_{t, t-1}$ is 0.692 (0.729), the standard deviation of mean (median) IOCs is 0.097 (0.100). The mean standard deviation is 0.184.
SI Fig. \ref{SI_fig11a_nace2}c shows the distributions of input overlap coefficients, IOC$_{t, t-1}$, for firms with in-degree between 16 and 35, $16 \leq k_i^{in} \leq 35$. The mean over the industries' mean (median) $\text{IOC}_{t, t-1}$ is 0.730 (0.094), the standard deviation of mean (median) IOCs is 0.094 (0.096). The mean standard deviation is 0.162.

\begin{figure*}[t]
	\centering
	\includegraphics[width=1\textwidth]{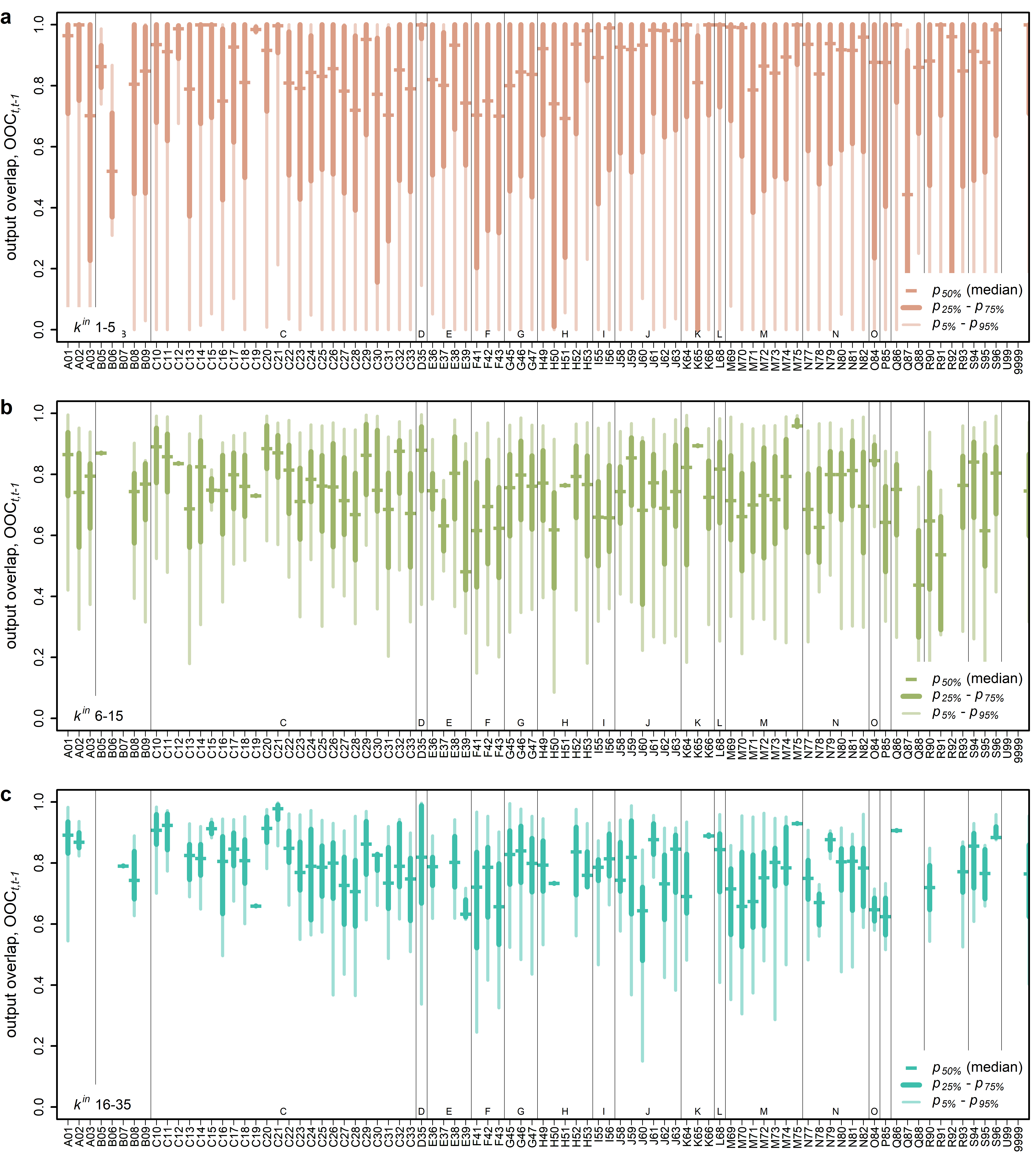} 
	\caption{Distributions of output vector overlaps, OOC$_{t, t-1}$, of firms across NACE 2 industries for the years 2019 and 2018. NACE2 classes are on the x-axis; overlap coefficients on the y-axis. 
		a) distributions of output overlap coefficients, OOC$_{t, t-1}$, for firms with out-degree between one and five, $1 \leq k_i^{out} \leq 5$. The mean over the industries' mean (median) OOC$_{t, t-1}$ is 0.727 (0.876), the standard deviation of mean (median) OOCs is 0.085 (0.110). The mean standard deviation is 0.331.
		b) distributions of output overlap coefficients, OOC$_{t, t-1}$, for firms with out-degree between 6 and 15, $6 \leq k_i^{out} \leq 15$. The mean over the industries' mean (median) OOC$_{t, t-1}$ is 0.714 (0.749), the standard deviation of mean (median) OOCs is 0.084 (0.093). The mean standard deviation is 0.208.
		c) distributions of output overlap coefficients, OOC$_{t, t-1}$, for firms with out-degree between 16 and 35, $16 \leq k_i^{out} \leq 35$. The mean over the industries' mean (median) OOC$_{t, t-1}$ is 0.760 (0.780), the standard deviation of mean (median) OOCs is 0.119 (0.120). The mean standard deviation is 0.139.
	}
	\label{SI_fig11b_nace2}
\end{figure*}

Second, we focus on the distributions of output overlaps for the years 2019 and 2018, $\text{OOC}_{t, t-1}$, in 
SI Fig. \ref{SI_Figure10a_nace4}b and SI Fig. \ref{SI_fig11b_nace2}.
SI Fig. \ref{SI_Figure10a_nace4}b shows the distributions of output overlap coefficients, OOC$_{t, t-1}$, for firms with more than 35 customers, $ k_i^{out} > 35$. The average of the mean (median) output overlaps, across NACE2 industries is 0.813 (0.828) and the standard deviation of mean (median) output overlaps is 0.101 (0.103), indicating  that relatively low output overlaps are the norm with few outliers. The average standard deviation is 0.108. Output overlaps are on average slightly higher, than input overlaps for the degree bin $>$35.
SI Fig. \ref{SI_fig11b_nace2}a shows the  distributions of output overlap coefficients, OOC$_{t, t-1}$, for firms with out-degree between one and five, $1 \leq k_i^{out} \leq 5$. The mean over the industries' mean (median) OOC$_{t, t-1}$ is 0.727 (0.876), the standard deviation of mean (median) OOCs is 0.085 (0.110). The mean standard deviation is 0.331.
SI Fig. \ref{SI_fig11b_nace2}b shows the  distributions of output overlap coefficients, OOC$_{t, t-1}$, for firms with out-degree between 6 and 15, $6 \leq k_i^{out} \leq 15$. The mean over the industries' mean (median) OOC$_{t, t-1}$ is 0.714 (0.749), the standard deviation of mean (median) OOCs is 0.084 (0.093). The mean standard deviation is 0.208.
SI Fig. \ref{SI_fig11b_nace2}c shows the  distributions of output overlap coefficients, OOC$_{t, t-1}$, for firms with out-degree between 16 and 35, $16 \leq k_i^{out} \leq 35$. The mean over the industries' mean (median) OOC$_{t, t-1}$ is 0.760 (0.780), the standard deviation of mean (median) OOCs is 0.119 (0.120). The mean standard deviation is 0.139.

Overall overlap coefficients of firms input- and output vectors for the years 2018 and 2019 are substantially higher than the pairwise overlap coefficients within industries.

%~/WorkII/SupplyRank/HungaryProject/Heterogeneity_pnas/figures/SI_Figure8_3a.png 
%mean_of_means sd_of_means mean_of_medians sd_of_medians mean_of_sds sd_of_sds
%1-5           0.660       0.079           0.763         0.104       0.334     0.048
%6-15          0.692       0.097           0.729         0.100       0.184     0.043
%16-35         0.730       0.094           0.758         0.096       0.162     0.093

%~/WorkII/SupplyRank/HungaryProject/Heterogeneity_pnas/figures/SI_Figure8_3b.png 
%mean_of_means sd_of_means mean_of_medians sd_of_medians mean_of_sds sd_of_sds
%1-5           0.727       0.085           0.876         0.110       0.331     0.062
%6-15          0.714       0.084           0.749         0.093       0.208     0.050
%16-35         0.760       0.119           0.780         0.120       0.139     0.057

%correlate output similarity change with input similarity changes / scatter plot.

\FloatBarrier

\section{Constructing synthetic firm-level shocks with same sector level impacts}\label{SI_shock_construction}

Recall that $\zeta$ is based on the actual employment reductions in the course of the early phase of the COVID-19 pandemic. % 
$\zeta_i$ is the reduction of the labor input for firm $i$ between January and May 2020, $\zeta_i=\max[1-e_i(may)/e_i(jan), \:0]$, where $e_i$ is the number of employees in the respective month.
In this section we describe the algorithm for constructing new synthetic firm-level shock vectors, $\zeta^1, \zeta^2, \dots , \zeta^{1,000}$,  that differ in how firms within industries are affected, but are of the exactly same size when aggregated to the industry-level. From these shocks we derive the remaining production level vectors, $\Psi=\{\psi^1,\psi^2, \dots, \psi^{1,000} \}$ that enter the shock propagation algorithm described in the Data and Methods section. 
We show how to construct a new shock vector, $\zeta^l$.
This problem can be solved sequentially for all industries $k \in \{1,2 \dots, m \}$, in our case the 593 NACE 4 classes contained in the data and the additional industry we introduce for all firms without NACE information.

We start by specifying additional industry-level notation.
For a given industry $k \in \{1,2 \dots, m \}$ we denote the number of firms within the industry as $n^k$. The indices of the $n^k$ firms in sector $k$ are denoted as $I^k = \{i \; |  p_i = k\}$. The in- and out-strength of industry $k$ is defined as 
$s^{\text{in},k} = \sum_{i=1}^{n} s^\text{in}_i \; \delta_{p_i, k}$ and $ s^{\text{out},k} = \sum_{i=1}^{n} s^\text{out}_i \; \delta_{p_i, k} $. 

The initial shock to  sector $k$, can be defined either through aggregating the shock vector $\zeta$, or through aggregating the remaining production levels vector $\psi$, since $\psi = 1- \zeta$. 
In  Eq. [\ref{eq_aggr_psi}] in the Data and Methods Section, we derived the shock to sector $k$ by aggregating the vector $\psi$, according to firms in- and out-strengths as
\begin{equation} \label{eq_aggr_psi_si}
	\phi^\text{u}_k = \frac{\sum_{i=1}^{n} \psi_i \; s^\text{in}_i \; \delta_{p_i, k} }{ \sum_{i=1}^{n} s^\text{in}_i \; \delta_{p_i, k} } , \qquad 
	\phi^\text{d}_k = \frac{\sum_{i=1}^{n} \psi_i \; s^\text{out}_i\; \delta_{p_i, k} }{ \sum_{i=1}^{n} s^\text{out}_i \; \delta_{p_i, k} } \quad .
\end{equation} 
$ \phi^\text{u}_k$ indicates the fraction of goods sector $k$ is still buying from its supplier industries, i.e. the  fraction of $k$'s in-strength, $s^{\text{in},k}$,  remaining after the shock. $ \xi^\text{u}_k = 1-  \phi^\text{u}_k$ is the size of the corresponding demand shock that propagates upstream. $\phi^\text{d}_k$ indicates the fraction of goods sector $k$ is still selling to its buyer industries, i.e. the  fraction of $k$'s out-strength, $s^{\text{out},k}, $,  remaining after the shock. $ \xi^\text{d}_k = 1- \phi^\text{d}_k$ is the size of the corresponding supply shock that propagates downstream. 
For the shock propagation algorithm it is more convenient to work with $\psi, \phi^\text{u}$ and $\phi^\text{d}$, but for sampling new synthetic shocks we continue to work with $\zeta, \xi^\text{u}$ and $\xi^\text{d}$. 

Our goal is to find for each firm $i$ in industry $k$ (i.e., where $p_i = k$)  an initial shock, $\zeta^l_i$, such that after aggregation, the sector level shock has the same size as the empirically defined original shock, $\zeta$. 
\begin{eqnarray} \label{shock_sampling_constraint_si} \notag
	\text{find:}  & \zeta^l_i &  \forall \; i \; \text{where} \; p_i = k    \\
	\text{s.t.:} &  &    \sum_{i=1}^{n} \zeta^l_i \;s_i^\text{in}  \; \delta_{p_i, k} = \xi^\text{u}_k \; s^{\text{in},k} \quad  , \\ \notag
	&   & \sum_{i=1}^{n} \zeta^l_i \; s_i^\text{out}  \; \delta_{p_i, k} = \xi^\text{d}_k  \; s^{\text{out},k} \quad  , \\ \notag
	& &  \zeta^l_i \in [0,1]  \quad  .
\end{eqnarray}
The two right hand side terms $\xi^\text{u}_k \: s^{\text{in},k}$ and $\xi^\text{d}_k \: s^{\text{out},k}$ are the target shock sizes the new firm-level shock, $\zeta^l$, needs to fulfil for industry $k$.
%
%$\omega_i^{\text{in},k} = s_i^\text{in} / s^{\text{in},k}$ and  $\omega_i^{\text{out},k} = s_i^\text{out} / s^{\text{out},k}$ denote the share of in- and out-strength of firm $i$ in sector $k$, respectively. 
We know that at least one solution exists always exists, the original initial shock $\zeta$. 
%We denote the fraction of lost production after the shock of firm $i$ as $\zeta^l_i = 1- \psi_i$ and of industry $k$, $\bar{ \phi^\text{u}_k} = 1- \phi^\text{u}_k$ and $\bar{\phi^\text{d}_k} = 1-\phi^\text{d}_k$.
Here, our sampled shocks fulfil Eq. [\ref{eq_aggr_psi_si}] at the NACE4 level, as the NACE2 level constraint would lead to even higher variability in the resulting production losses.
Not that if no firm-level shock is available and we want firm-level shocks that correspond to a specific industry-level shock, then the targeted shock size can be specified directly with the sector level shock vectors, $(\xi^\text{u}_k, \xi^\text{d}_k )$. In this way we can construct many random firm-level shocks and receive a distribution of production losses for the given industry-level shock.
We solve sampling problem [\ref{shock_sampling_constraint_si}] in two steps. 

\paragraph{Sampling new shocks}
The first step is shown in detail in Algorithm \ref{algo_drawing}. 
First, we define the auxiliary index set $\tilde{I}^k = \{i \; |  p_i = k\}$ that contains all indices of firms belonging to sector $k$. The firm index, $i$, refers to the row and column index firm $i$ belongs to in the adjacency matrix $W$, and the position of $i$ in the industry affiliation vector $p$. Note we use the terms ``firm $i$'' and ``index, $i$'' interchangeably.
We initialise the algorithm by setting the shock size for each firm $i$ in industry $k$ to zero, i.e. $\zeta^l_i \gets 0$ for all  $i \in I^k$. 
Then, we add shocks to the values, $\zeta^l_i$ ($ \forall i \:|\: p_i = k$),  until the new shock is larger  than the original shock target, i.e.,
\begin{equation}
    \left( \sum_{i=1}^{n} \zeta^l_i \;s_i^{\text{in}}  \; \delta_{p_i, k} \right) \geq \left( \xi^\text{u}_k \: s^{\text{in},k} \right) \quad , \qquad  \left(\sum_{i=1}^{n}\zeta^l_i \; s_i^{\text{out}}   \; \delta_{p_i, k} \right) \geq \left( \xi^\text{d}_k \: s^{\text{out},k} \right) \quad .
\end{equation}

The shocks are added in the following way. 
First we draw a firm index $i$ from the index set $\tilde{I}^k$, and delete the index $i$ from the index set, $\tilde{I}^k$. 
Then we draw a shock value, $\eta \in[0,1]$, from a specified distribution that takes values between zero and 1. Here we draw the shock values from the empirical distribution of employment shocks of sector $k$, i.e., $\eta \sim \{ \zeta_j \; |\;  j \in I^k \} $. Note that the empirical shock distribution, $\{ \zeta_j \; |\;  j \in I^k \} $, contains only values that lie between zero and 1. Note that we could also sample more general shocks, by drawing values from, e.g., the Beta distribution that is flexible enough to sample very concentrated or very evenly distributed shocks. Note that here we do not draw negative shocks that would be interpreted as production gains, or increases in production capacity, even though this would be possible for more general shocks.
We add the additional shock value, $\eta$, to the previous shock level of firm $i$, i.e., $ \zeta^l_i  \gets \min[1 , \zeta^l_i  + e ]$. The $\min[1, .]$ function is necessary, because a firm $i$ can be drawn a second time for receiving a shock, but shocks can not be larger than one --- a firm can not lose more than 100\% of its production. 
Since, this procedure is continued until the necessary aggregate shock level ($\xi^\text{u}_k \: s^{\text{in},k} $, $\xi^\text{d}_k \: s^{\text{in},k}  $) is reached, it can happen that each firm has been drawn  already and the index set $\tilde{I}^k$ is empty, i.e., $\tilde{I}^k = \emptyset$. In this case we fill up the index set  again with all firms in industry $k$, i.e. we set $\tilde{I}^k \gets \{i \; |  p_i = k\}$. This can happen when in the original shock, $\zeta$, relatively large firms received relatively large shocks, and these large firms only received small shocks in the first round of Monte Carlo draws.

\begin{algorithm}
	\caption{Drawing shocks for firms in industry $k$} \label{algo_drawing}	
	\begin{algorithmic}[1]
		\vspace*{2mm}
		\State Set $\tilde{I}^k = \{i \; |  p_i = k\}$ \Comment{Create the set, containing all firm indices of firms belonging to sector $k$.}
		\State Set $\zeta^l_i \gets 0$ for all  $i \in I^k$ \Comment{Initialize the algorithm by setting all shocks to zero.}
        \vspace*{2mm}
		\While{$ \quad 
		    \sum_{i=1}^{n} \zeta^l_i \;s_i^{\text{in}}  \; \delta_{p_i, k}  \leq  \xi^\text{u}_k \: s^{\text{in},k} \quad \text{and} \quad 
            \sum_{i=1}^{n}\zeta^l_i \; s_i^{\text{out}}   \; \delta_{p_i, k} \leq  \xi^\text{d}_k \: s^{\text{out},k}  \quad 
		$ \\}
		\vspace*{2mm}
		\State  $i \sim \tilde{I}^k$ \Comment{Draw a firm index $i$ from sector $k$.}
		\State Delete $i$ from $\tilde{I}^k$ \Comment{Remove firm index $i$ from the index set.}
		\State Draw a shock $\eta \sim \{ \zeta_j \; |\;  j \in I^k \} $ \Comment{ Draw a shock $\zeta_j$ from the empirical distribution of shocks of sector $k$.}
		\State Update shock $ \zeta^l_i  \gets \min[1 , \zeta^l_i  + \eta ]$ \Comment{Update the shock of firm $i$ with the additional drawn shock. } 
		\If{$\tilde{I}^k = \emptyset$}  
		\Comment{If each firm has received a shock and the aggregate shock is still too small.}
			\State Set $\tilde{I}^k \gets \{i \; |  p_i = k\}$ \Comment{Fill up the index set again and continue to draw shocks.}
		\EndIf
		\vspace*{2mm}
		\EndWhile
		\vspace*{2mm}
		\State \Return $\zeta^l_i $ for $i \in I^k$ . \Comment{Return the shock vector and use it as input for Algorithm \ref{algo_rescaling}.}
	\end{algorithmic}
\end{algorithm}

\paragraph{Rescaling of shocks}
In a second step we find weights to rescale the shocks, $\zeta^l_i$, such that the constraints in Eq. [\ref{shock_sampling_constraint_si}] hold exactly. 
The basic idea is to divide the firms in sector $k$ into two groups. 
The first group contains firms that have a higher ratio of in-strength to out-strength than the empirical shock, i.e., 
$$\frac{s_i^\text{in}}{s_i^\text{out}} > \frac{\xi^\text{u}_k \;  s^{\text{in},k}}{\xi^\text{d}_k \; s^{\text{out},k}} \qquad $$
We assign all firms $i$ of sector $k$ that fulfil this condition to the set $I^{\text{in},k}$. 
The second group contains firms that have a higher ratio of out-strength to in-strength than the target shock,  i.e., 
$$\frac{s_i^\text{out}}{s_i^\text{in}} > \frac{\xi^\text{d}_k \; s^{\text{out},k}}{\xi^\text{u}_k \;  s^{\text{in},k}} \qquad $$
We assign all firms $i$ of sector $k$ that fulfil this condition to the set $I^{\text{out},k}$. Edge cases having exactly the same ratio can be added to the group with fewer firms.
Then, we define a rescaling factor for the in-strength `heavy' firms, $v^\text{in}$,  that rescales all $\zeta^l_i$ where $i \in I^{\text{in},k} $,
and a rescaling factor for the out-strength `heavy' firms,  $v^\text{out}$,  that rescales all $\zeta^l_i$ where $i \in I^{\text{out},k} $. 
If we increase $v^\text{in}$ while leaving  $v^\text{out}$ untouched, the shock scenario, $\zeta^l$, will result in a higher loss of in-strength relative to the loss out-strength of sector $k$ and therefore a larger upstream shock relative to the size of the downstream shock. 
If we increase $v^\text{out}$ while leaving  $v^\text{in}$ untouched, the shock scenario, $\zeta^l$, will result in a higher loss of out-strength relative to the loss of in-strength  to of sector $k$ and therefore a larger downstream shock relative to the size of the upstream shock. 
Now we only need to determine the weights $v^\text{in}$ and  $v^\text{out}$, such that the first two constraints in problem statement [\ref{shock_sampling_constraint_si}] exactly hold.

In principle the weights $v^\text{in}$ and  $v^\text{out}$ can be found by solving the following linear system of equations, 
\begin{eqnarray} \label{lin_system_rescale_weights1_si}
	 v^\text{in}  \sum_{i \in I^{\text{in},k}} \zeta^l_i \; s_i^{\text{in}} \; &  + & \; v^\text{out} \sum_{i \in I^{\text{out},k}} \zeta^l_i \; s_i^{\text{in}}    =  \xi^\text{u}_k \: s^{\text{in},k} \quad, \\
	 \label{lin_system_rescale_weights2_si}
	  v^\text{in}  \sum_{i \in I^{\text{in},k}} \zeta^l_i \; s_i^{\text{out}} \; & + & \; v^\text{out} \sum_{i \in I^{\text{out},k}} \zeta^l_i \; s_i^{\text{out}}    =  \xi^\text{d}_k \: s^{\text{out},k} \quad .
\end{eqnarray}
The linear system [\ref{lin_system_rescale_weights1_si}-\ref{lin_system_rescale_weights2_si}] can be written in standard matrix form as 
\begin{equation}
	A v = \xi_k \: s^k  \quad , 
\end{equation}
where $v = (v^\text{in}, v^\text{out})^\top$, 
$\xi_k \: s^k = (\xi^\text{u}_k \: s^{\text{in},k} ,   \xi^\text{d}_k \: s^{\text{out},k}) $,
$$A_{11} =  \sum_{i \in I^{\text{in},k}} \zeta^l_i \; s_i^{\text{in}} \quad ,$$
$$A_{12} = \sum_{i \in I^{\text{out},k}} \zeta^l_i \; s_i^{\text{in}}  \quad ,$$
$$A_{21} =  \sum_{i \in I^{\text{in},k}} \zeta^l_i \;s_i^{\text{out}} \quad ,$$ and 
$$A_{22} = \sum_{i \in I^{\text{out},k}} \zeta^l_i \; s_i^{\text{out}} \quad .$$ 
The system is not always directly solvable for a given vector, $\zeta^l$, that results from Algorithm \ref{algo_drawing}.

In Algorithm \ref{algo_rescaling} we show how to find the rescaling weights, $v = (v^\text{in}, v^\text{out})^\top$, for a given $\zeta^l$.
For each firm in industry $k$, i.e., the set $I^k = \{i \; |  p_i = k\}$, we initialize the algorithm with the elements from the shock vector, $\zeta^l_i$, that results from Algorithm \ref{algo_drawing}.
Then, we calculate the size of the violation of the first two constraints in problem statement [\ref{shock_sampling_constraint_si}], i.e. the distance to the targeted upstream shock size,
$ o^\text{in} \gets \big\lvert \sum_{i=1}^{n} \zeta^l_i \; s_i^{\text{in}}  \; \delta_{p_i, k} - \xi^\text{u}_k \: s^{\text{in},k} \big\rvert $ and the distance to the targeted downstream shock size, $ o^\text{out} \gets \big\lvert  \sum_{i=1}^{n}\zeta^l_i \; s_i^{\text{out}}  \; \delta_{p_i, k} - \xi^\text{d}_k \: s^{\text{out},k} \big\rvert $,  
where |.| denotes the absolute value.
We define the ``available for rescaling'' indicator vector, $d$, where $d_i=0$ indicates that the shock, $\zeta^l_i$, can be  rescaled, and $d_i=1$ indicates that it can not be rescaled, because, $\zeta^l_i$, was  scaled above 1 in a previous iteration. 
Initially we set  $ d_i \gets 0 \; \forall \; i \in I^k$, i.e. all firm shocks can initially be rescaled.

We continue the following calculations until the distance to the targeted upstream and downstream shock becomes smaller than a threshold $\epsilon$, i.e. the algorithm stops when $ \left( o^\text{in}   \leq \epsilon \right) \; \text{and} \; \left(  o^\text{out} \leq \epsilon \right)$. We set the parameter epsilon to 0.01,  such that in absolute monetary terms the difference in shocks becomes smaller than 10 Forint (approx 0.025 Euros). 

First, we calculate the remaining target shock size, $b = (b^\text{in}, b^\text{out})$. $b$ is the respective upstream or downstream shock target, ($\xi^\text{u}_k, \xi^\text{d}_k$),  reduced  by the respective in-strength or out-strength of firms that are not available for rescaling anymore.  $b^\text{in} \gets \left( \xi^\text{u}_k \: s^{\text{in},k} \;  - \; \sum_{i=1}^{n} d_i \; s_i^{\text{in}} \right)$  specifies the size of the targeted in-strength shock that remains after deducting the in-strength of firms that received already a 100\% shock, i.e., where $\zeta^l_i=1$ and therefore where $d_i = 1$. 
$b^\text{out} \gets \left( \xi^\text{d}_k \: s^{\text{out},k} \; - \; \sum_{i=1}^{n} d_i \; s_i^{\text{out}}   \right)$ specifies the size of the targeted out-strength shock that remains after deducting the out-strength of firms that received already a 100\% shock, i.e., where $\zeta^l_i=1$ and therefore $d_i = 1$. The variables $b^\text{in}$ and $b^\text{out}$ need to be calculated in every iteration, because the change in the remaining target shock size, $b$, affects which firms belong to the set of ``in-strength-heavy'' firms and the set of ``out-strength-heavy'' firms.
Hence, we update these two sets by setting $I^{\text{in},k} $ to include all firms $i$ where  $\frac{s_i^\text{in}}{s_i^\text{out}} > \frac{b^\text{in} }{b^\text{out}}$ and $I^{\text{out},k}$ to include all $i$ where  $\frac{s_i^\text{out}}{s_i^\text{in}} > \frac{b^\text{out} }{b^\text{in} } $. Edge cases can again be added to the group with fewer firms.
 
Next, we need to update the values of the coefficient matrix $A$. The values are updated, because firms that have received already a full shock ($d_i = 1$) are not considered anymore for rescaling, i.e. we sum only over firms where $d_i =0$. We calculate $A_{11} =  \sum_{i \in I^{\text{in},k}} \zeta^l_i \; s_i^{\text{in}} \; \mathbb{I}_{(d_i = 0)}$, 
$A_{12} = \sum_{i \in I^{\text{out},k}} \zeta^l_i \; s_i^{\text{in}} \mathbb{I}_{(d_i = 0)} $, 
 $A_{21} =  \sum_{i \in I^{\text{in},k}} \zeta^l_i \; s_i^{\text{out}}\mathbb{I}_{(d_i = 0)}$, and 
 $A_{22} = \sum_{i \in I^{\text{out},k}} \zeta^l_i \; s_i^{\text{out}} \mathbb{I}_{(d_i = 0)}$.
$\mathbb{I}_{(d_i = 0)}$ is the indicator variable that is one if firm $i$ can be rescaled and zero if firm $i$ can not be rescaled anymore.
The system has a solution when the rank of $A$ has the same rank as the matrix $(A|b)$. 

We list the four cases when the shocks, $\zeta^l_i$, lead to a violation of the rank condition in matrix, $A$. 
First, if no firm $i$ that has  positive in-strength and is available for rescaling, (i.e., where $d_i=0$), receives a shock, then the first row would be zero. Further, if additionally $b^\text{in} > 0 $ the system has no solution. 
We can remedy this case by drawing a new shock for a firm that has previously not received a shock and has positive in-strength.
Second, if no firm that has positive out-strength and  is available for rescaling, (i.e., where $d_i=0$), receives a shock, then the second row would be zero. Further, if additionally $b^\text{out}_k > 0 $ the system has no solution. 
We can remedy this case by drawing a new shock for a firm that has previously not received a shock and has positive in-strength.
These two cases do not happen with the initially drawn shocks, $\zeta^l_i$, because of the condition in the while statement of Algorithm \ref{algo_drawing}, but they can occur during the adjustment procedure in Algorithm \ref{algo_rescaling}, because the summations depend on the indicator variable $d_i$. 
Third, if no firm from the group of high in- to out-strength ratio, $ I^{\text{in},k}$, receives a shock, then the first column of $A$ is zero, which usually leads to an unsolvable system.
We can remedy this case by drawing a new shock for a firm, $i$,  that has previously not received a shock, $\zeta^l_i=0$, and belongs to the set $I^{\text{in},k}$.
Fourth,  if no firm from the group of high out- to in-strength ratio, $ I^{\text{out},k} $, receives a shock, then the second column of $A$ is zero, which usually leads to an unsolvable system. We can remedy this case by drawing a new shock for a firm, $i$,  that has previously not received a shock, $\zeta^l_i=0$, and belongs to the set $I^{\text{out},k}$.
The last two cases can occur since we do not specifically avoid them in Algorithm \ref{algo_drawing}. 
If an additional shock was drawn, the matrix $A$ needs to be updated again. 

Next, we can solve the linear system of equations $Av=b$, by computing the generalized inverse, $A^\dagger$, of $A$ and set $v \gets A^\dagger b$.
Then, we rescale the the elements of the shock vector $\zeta^l$, in the following way. 
For the firms, $i$, that belong to the ``in-strength-heavy'' group, $i \in  I^{\text{in},k}$, and are still available for rescaling, (where $d_i = 0$), we set $ \zeta^l_i \gets v^\text{in} \: \zeta^l_i$. 
For the firms, $i$, that belong to the ``out-strength-heavy'' group, $i \in  I^{\text{out},k}$, and are still available for rescaling, $d_i = 0$, we set $ \zeta^l_i \gets v^\text{out} \: \zeta^l_i$. 
Then, we update the indicator variable by setting $d_i \gets 1$ for all $i$ with $\zeta^l_i > 0 $. 
To ensure that shocks are not larger than one we  take the maximum with 1, i.e. we set $ \zeta^l_i  \gets  \min[\zeta^l_i , \; 1]$. 
Finally, we update the distance to the target shock, $o^\text{in}$ and $o^\text{out}$. 

We have implemented algorithm \ref{algo_rescaling} sufficiently fast to sample the 1,000 shocks for each of the approx. 245,000 firms within a few hours. Note that the common rescaling of many firm shocks at the same time with the same factors $v$ might not lead to a full traversing of the space of all possible firm-level shocks that are consistent with our sampling problem [\ref{shock_sampling_constraint_si}]. This means that in practice for one specific industry-level shock the heterogeneity of production losses computed on the firm-level could be even larger. We have checked that the resulting shocks are uncorrelated on the firm-level and perfectly correlated (identical) when aggregated to the industry-level.

\begin{algorithm}
	\caption{Rescaling weights for shocks of firms in industry $k$} \label{algo_rescaling}	
	\begin{algorithmic}[1]
		\State set $I^k = \{i \; |  p_i = k\}$
		\State initialize with $\zeta^l_i$ for $i \in I^k$ 
   \vspace*{1mm}
		\State set $o^\text{in} \gets \big\lvert \sum_{i\in I^k} \zeta^l_i \; s_i^{\text{in}}  \;  - \; \xi^\text{u}_k \: s^{\text{in},k} \big\rvert \quad \text{and} \quad o^\text{out} \gets \big\lvert  \sum_{i\in I^k}\zeta^l_i \; s_i^{\text{out}}  \;  - \; \xi^\text{d}_k \: s^{\text{out},k} \big\rvert $    \Comment{ Calculate the distance from the targeted shock.}
		\State set $ d_i \gets 0 \; \forall \; i \in I^k$ \Comment{All shocks, $\zeta^l_i \; \forall i \in I^k$, are available for rescaling.  }
		\vspace*{2mm}
 		\While{$ \left( o^\text{in}   > \epsilon \right) \; \text{and} \; \left(  o^\text{out} > \epsilon \right)$ } \Comment{Iterate until the distance to target up- and downstream shock size is small.}
		\vspace*{2mm}
		\State set $b \gets \left( \xi^\text{u}_k \: s^{\text{in},k} \;  - \; \sum_{i=1}^{n} d_i \; s_i^{\text{in}}   \; , \; \;  \xi^\text{d}_k \: s^{\text{out},k} \; - \; \sum_{i=1}^{n} d_i \; s_i^{\text{out}}   \right) $ \Comment{Calculate the remaining absolute shock that is left after deducting strength of fully scaled up firms $i$ where $d_i =1$.}
   \vspace*{1mm}
		\State Set $I^{\text{in},k} $ to include all $i$ where  $\frac{s_i^\text{in}}{s_i^\text{out}} > \frac{b^\text{in} }{b^\text{out}}$  \Comment{Update ``in-strength-heavy'' group. }
		\State Set $I^{\text{out},k}$ to include all $i$ where  $\frac{s_i^\text{out}}{s_i^\text{in}} > \frac{b^\text{out} }{b^\text{in} } $  \Comment{Update ``out-strength-heavy'' group. }
   \vspace*{1mm}
		\State calculate $A_{11} =  \sum_{i \in I^{\text{in},k}} \zeta^l_i \; s_i^{\text{in}} \; \mathbb{I}_{(d_i = 0)}$, 
		\State calculate $A_{12} = \sum_{i \in I^{\text{out},k}} \zeta^l_i \; s_i^{\text{in}} \mathbb{I}_{(d_i = 0)} $, 
		 \State calculate $A_{21} =  \sum_{i \in I^{\text{in},k}} \zeta^l_i \; s_i^{\text{out}}\mathbb{I}_{(d_i = 0)}$ 
		 \State calculate   $A_{22} = \sum_{i \in I^{\text{out},k}} \zeta^l_i \; s_i^{\text{out}} \mathbb{I}_{(d_i = 0)} $
   \vspace*{1mm}
		 \If{$A_{1.} = (0,0)$} 
		  sample $i$ where $\zeta^l_i = 0$,  $s_i^\text{in}>0$  and $d_i = 0$ and set $\zeta^l_i \sim U[0,1]$; recalculate lines 9-12
		 \EndIf
		 \If{$A_{2.} = (0,0)$} 
		  sample $i$ where $\zeta^l_i = 0$, $s_i^\text{out}>0$ and $d_i = 0$ and set $\zeta^l_i \sim U[0,1]$;  recalculate lines 9-12
		 \EndIf
		 \If{$A_{.1} = (0,0)^\top$} 
		 sample $i$ where $\zeta^l_i = 0$, $i \in I^{\text{in},k} $ and $d_i = 0$ and set $\zeta^l_i \sim U[0,1]$;  recalculate lines 9-12
		 \EndIf
		 \If{$A_{.2} = (0,0)^\top$} 
		  sample $i$ where $\zeta^l_i = 0$, $i \in I^{\text{out},k} $ and $d_i = 0$ and set $\zeta^l_i \sim U[0,1]$;  recalculate lines 9-12
		 \EndIf
    \vspace*{1mm}
    \State Calculate the generalized inverse $A^\dagger$, of coefficient matrix $A$
		 \State set $v \gets A^\dagger b$ \Comment{Calculate the rescaling coefficients $v$.} 
		 \State set $ \zeta^l_i \gets v^\text{in} \: \zeta^l_i$ \hspace{.9mm} for $i \in  I^{\text{in},k}$ \hspace{0.9mm}  and $d_i = 0$ \Comment{Rescale shocks of ``in-heavy'' firms.}
		 \State set $ \zeta^l_i \gets v^\text{out} \: \zeta^l_i$ for $i \in  I^{\text{out},k}$  and $d_i = 0$  \Comment{Rescale shocks of ``out-heavy'' firms.}
		 \State set $d_i = 1$ for all $i$ with $\zeta^l_i > 0 $
		 \State set $ \zeta^l_i  \gets \max[0, \min[\zeta^l_i , \; 1]]$
    \vspace*{1mm}
		 \State set $o^\text{in} \gets \big\lvert \sum_{i\in I^k} \zeta^l_i \; s_i^{\text{in}}  \;  - \; \xi^\text{u}_k \: s^{\text{in},k} \big\rvert \quad \text{and} \quad o^\text{out} \gets \big\lvert  \sum_{i\in I^k}\zeta^l_i \; s_i^{\text{out}}  \;  - \; \xi^\text{d}_k \: s^{\text{out},k} \big\rvert $    \Comment{ Update the distance from the targeted shock.}
		\EndWhile
		\State \textbf{end while}
		\State \Return $\zeta^l_i $ 
	\end{algorithmic}
\end{algorithm}

\FloatBarrier

\clearpage

\section{Details on industry-level production losses} \label{SI_section_details_on_sector_losses}

In this section we give an overview of the production losses for all NACE2 industries.
Table \ref{SI_table_sector_losses1} compares theindustry-specific production losses between firm-level production network (FPN) based loss estimates and industry-level production network (IPN) based loss estimates for the NACE2 classes  A01 to F43.  
The first column shows the NACE2 code for which production losses are compared across firm-level and industry-level production losses. 
The second and third column show the aggregation of the initial shock $\zeta$, to the NACE2 level, into the up-stream shock, $\xi^\text{u}$, and the downstream shock, $\xi^\text{d}$, respectively. 
The fourth column shows the FPN based production losses,  $L^k_{\text{firm}}(\psi)$,  for the labor shock, $\psi$ (red `x' symbols in Fig. \ref{boxplot_emp_shock_may_GL_nace2}).
The fifth column shows the average FPN-based production losses,  $\mathbb{E}[L^k_{\text{firm}}(\Psi)]$,  corresponding to the 1,000 synthetic firm-level shock scenarios, $\Psi$ (mean of the boxplots in Fig. \ref{boxplot_emp_shock_may_GL_nace2}). The sixth column shows the IPN-based production losses,  $L^k_{\text{ind.}}(\phi)$,  corresponding to the aggregated industry-level shock scenarios, $\phi$ (blue `+' symbols in Fig. \ref{boxplot_emp_shock_may_GL_nace2}). 
The seventh column shows the mean deviation, $\mathbb{E} \big[ \frac{{L}^k_{\text{ind.}}(\phi)}{{L}^k_{\text{firm}}(\Psi)} - 1 \big] $,  of the industry-level production losses, $L^k_{\text{ind.}}(\phi)$, from the firm-level production losses, $L^k_{\text{firm}}(\Psi)$, across the 1,000 different firm-level shock scenarios $\Psi$. Note that when aggregated, all firm-level initial shocks, $\Psi$, are all identical to the industry-level  shock, $\phi$, that corresponds to the COVID-19 shock $\psi$. 

Table \ref{SI_table_sector_losses2} compares theindustry-specific production losses between firm-level production network (FPN) based loss estimates and industry-level production network (IPN) based loss estimates for NACE2 classes from G45 to U99, incl. the fictional category where the NACE class is not available. The columns are as in Table \ref{SI_table_sector_losses1}.

% latex table generated in R 4.0.3 by xtable 1.8-4 package
% Thu Nov 10 17:22:43 2022
\begin{table}[h] 
	\centering 
	\caption{Comparison of industry-specific production losses between firm-level production network (FPN) based loss estimates and industry-level production network (IPN) based loss estimates for NACE2 classes from A01 to F43.   }
	\begin{tabular}{ccccccc} 
 \hline \hline
		NACE2 & ind. shock, $\xi^\text{u}$ & ind. shock, $\xi^\text{d}$ & FPN-loss, $L^k_{\text{firm}}(\psi)$ & avg. FPN-loss, $L^k_{\text{firm}}(\Psi)$ & IPN-loss, $L_{\text{ind.}}(\phi)$ & avg. IPN/FPN \\
		\hline
A01 & 0.03 & 0.03 & 0.12 & 0.12 & 0.11 & -0.05 \\ 
  A02 & 0.06 & 0.07 & 0.14 & 0.16 & 0.13 & -0.19 \\ 
  A03 & 0.04 & 0.04 & 0.12 & 0.10 & 0.08 & -0.22 \\ 
  B05 & 0.02 & 0.16 & 0.17 & 0.23 & 0.16 & -0.25 \\ 
  B06 & 0.00 & 0.00 & 0.11 & 0.12 & 0.08 & -0.19 \\ 
  B07 & 0.05 & 0.05 & 0.08 & 0.10 & 0.09 & 0.12 \\ 
  B08 & 0.02 & 0.02 & 0.11 & 0.12 & 0.13 & 0.12 \\ 
  B09 & 0.08 & 0.05 & 0.11 & 0.13 & 0.08 & -0.31 \\ 
  C10 & 0.03 & 0.03 & 0.12 & 0.10 & 0.11 & 0.05 \\ 
  C11 & 0.03 & 0.02 & 0.12 & 0.11 & 0.10 & 0.02 \\ 
  C12 & 0.00 & 0.01 & 0.07 & 0.06 & 0.05 & 0.02 \\ 
  C13 & 0.05 & 0.08 & 0.13 & 0.13 & 0.08 & -0.38 \\ 
  C14 & 0.10 & 0.10 & 0.18 & 0.16 & 0.10 & -0.36 \\ 
  C15 & 0.10 & 0.17 & 0.18 & 0.19 & 0.17 & -0.10 \\ 
  C16 & 0.05 & 0.04 & 0.10 & 0.11 & 0.09 & -0.17 \\ 
  C17 & 0.01 & 0.02 & 0.11 & 0.10 & 0.09 & -0.08 \\ 
  C18 & 0.07 & 0.09 & 0.13 & 0.14 & 0.10 & -0.27 \\ 
  C19 & 0.00 & 0.00 & 0.07 & 0.04 & 0.07 & 0.87 \\ 
  C20 & 0.02 & 0.11 & 0.20 & 0.20 & 0.11 & -0.42 \\ 
  C21 & 0.01 & 0.01 & 0.12 & 0.11 & 0.09 & -0.11 \\ 
  C22 & 0.03 & 0.04 & 0.12 & 0.12 & 0.07 & -0.38 \\ 
  C23 & 0.02 & 0.03 & 0.11 & 0.12 & 0.11 & -0.03 \\ 
  C24 & 0.04 & 0.06 & 0.13 & 0.11 & 0.07 & -0.37 \\ 
  C25 & 0.05 & 0.05 & 0.10 & 0.10 & 0.05 & -0.50 \\ 
  C26 & 0.03 & 0.03 & 0.09 & 0.08 & 0.03 & -0.59 \\ 
  C27 & 0.03 & 0.03 & 0.09 & 0.09 & 0.06 & -0.34 \\ 
  C28 & 0.04 & 0.03 & 0.08 & 0.08 & 0.04 & -0.53 \\ 
  C29 & 0.02 & 0.03 & 0.09 & 0.09 & 0.07 & -0.27 \\ 
  C30 & 0.06 & 0.01 & 0.12 & 0.09 & 0.07 & -0.17 \\ 
  C31 & 0.09 & 0.09 & 0.13 & 0.14 & 0.09 & -0.36 \\ 
  C32 & 0.04 & 0.06 & 0.10 & 0.11 & 0.09 & -0.16 \\ 
  C33 & 0.03 & 0.04 & 0.09 & 0.10 & 0.08 & -0.20 \\ 
  D35 & 0.10 & 0.09 & 0.21 & 0.22 & 0.12 & -0.45 \\ 
  E36 & 0.00 & 0.00 & 0.12 & 0.11 & 0.15 & 0.36 \\ 
  E37 & 0.01 & 0.03 & 0.20 & 0.16 & 0.16 & 0.15 \\ 
  E38 & 0.04 & 0.04 & 0.12 & 0.12 & 0.14 & 0.12 \\ 
  E39 & 0.05 & 0.05 & 0.10 & 0.10 & 0.14 & 0.42 \\ 
  F41 & 0.06 & 0.07 & 0.13 & 0.14 & 0.09 & -0.37 \\ 
  F42 & 0.03 & 0.04 & 0.10 & 0.11 & 0.11 & 0.03 \\ 
  F43 & 0.06 & 0.06 & 0.12 & 0.12 & 0.11 & -0.11 \\ 
   \hline
   \hline
	\end{tabular} \label{SI_table_sector_losses1}
\end{table}

\begin{table}[h]
	\centering  
	\caption{Comparison ofindustry-specific production losses between firm-level production network (FPN) based estimates and industry-level production network (IPN) based estimates for NACE2 classes from G45 to U99, incl. the fictional category where the NACE class is not available. }
	\begin{tabular}{ccccccc} 
 \hline \hline
			NACE2 & ind. shock, $\xi^\text{u}$ & ind. shock, $\xi^\text{d}$ & FPN-loss, $L^k_{\text{firm}}(\psi)$ & avg. FPN-loss, $L^k_{\text{firm}}(\Psi)$ & IPN-loss, $L_{\text{ind.}}(\phi)$ & avg. IPN/FPN \\
  \hline
G45 & 0.05 & 0.04 & 0.09 & 0.11 & 0.10 & -0.09 \\ 
  G46 & 0.04 & 0.04 & 0.08 & 0.09 & 0.08 & -0.12 \\ 
  G47 & 0.04 & 0.06 & 0.14 & 0.14 & 0.10 & -0.26 \\ 
  H49 & 0.05 & 0.06 & 0.13 & 0.14 & 0.15 & 0.08 \\ 
  H50 & 0.08 & 0.15 & 0.19 & 0.19 & 0.15 & -0.23 \\ 
  H51 & 0.03 & 0.03 & 0.07 & 0.05 & 0.03 & -0.35 \\ 
  H52 & 0.03 & 0.04 & 0.14 & 0.16 & 0.11 & -0.33 \\ 
  H53 & 0.02 & 0.03 & 0.11 & 0.11 & 0.15 & 0.35 \\ 
  I55 & 0.17 & 0.15 & 0.21 & 0.20 & 0.17 & -0.16 \\ 
  I56 & 0.20 & 0.20 & 0.23 & 0.24 & 0.20 & -0.15 \\ 
  J58 & 0.04 & 0.03 & 0.17 & 0.15 & 0.07 & -0.51 \\ 
  J59 & 0.06 & 0.06 & 0.12 & 0.12 & 0.09 & -0.25 \\ 
  J60 & 0.02 & 0.01 & 0.13 & 0.13 & 0.16 & 0.21 \\ 
  J61 & 0.03 & 0.02 & 0.15 & 0.17 & 0.15 & -0.04 \\ 
  J62 & 0.05 & 0.06 & 0.11 & 0.12 & 0.15 & 0.24 \\ 
  J63 & 0.07 & 0.03 & 0.20 & 0.17 & 0.08 & -0.48 \\ 
  K64 & 0.03 & 0.03 & 0.09 & 0.10 & 0.07 & -0.22 \\ 
  K65 & 0.00 & 0.00 & 0.01 & 0.03 & 0.01 & -0.32 \\ 
  K66 & 0.01 & 0.01 & 0.06 & 0.05 & 0.13 & 1.50 \\ 
  L68 & 0.09 & 0.05 & 0.10 & 0.11 & 0.11 & 0.07 \\ 
  M69 & 0.03 & 0.04 & 0.14 & 0.13 & 0.13 & -0.01 \\ 
  M70 & 0.07 & 0.07 & 0.14 & 0.14 & 0.12 & -0.13 \\ 
  M71 & 0.04 & 0.04 & 0.09 & 0.10 & 0.10 & 0.04 \\ 
  M72 & 0.04 & 0.03 & 0.06 & 0.07 & 0.08 & 0.29 \\ 
  M73 & 0.05 & 0.05 & 0.14 & 0.14 & 0.13 & -0.08 \\ 
  M74 & 0.05 & 0.05 & 0.10 & 0.10 & 0.12 & 0.23 \\ 
  M75 & 0.02 & 0.05 & 0.16 & 0.09 & 0.11 & 0.32 \\ 
  N77 & 0.05 & 0.04 & 0.10 & 0.10 & 0.14 & 0.41 \\ 
  N78 & 0.27 & 0.17 & 0.21 & 0.20 & 0.27 & 0.34 \\ 
  N79 & 0.14 & 0.19 & 0.22 & 0.22 & 0.19 & -0.13 \\ 
  N80 & 0.07 & 0.08 & 0.15 & 0.16 & 0.17 & 0.04 \\ 
  N81 & 0.07 & 0.10 & 0.16 & 0.17 & 0.16 & -0.05 \\ 
  N82 & 0.05 & 0.08 & 0.15 & 0.16 & 0.14 & -0.13 \\ 
  O84 & 0.01 & 0.01 & 0.08 & 0.07 & 0.08 & 0.12 \\ 
  P85 & 0.07 & 0.10 & 0.15 & 0.14 & 0.10 & -0.32 \\ 
  Q86 & 0.10 & 0.06 & 0.10 & 0.10 & 0.10 & -0.02 \\ 
  Q87 & 0.05 & 0.01 & 0.03 & 0.05 & 0.05 & 0.42 \\ 
  Q88 & 0.01 & 0.04 & 0.08 & 0.09 & 0.15 & 0.80 \\ 
  R90 & 0.07 & 0.07 & 0.14 & 0.14 & 0.12 & -0.15 \\ 
  R91 & 0.02 & 0.03 & 0.07 & 0.08 & 0.14 & 0.83 \\ 
  R92 & 0.12 & 0.16 & 0.17 & 0.20 & 0.16 & -0.20 \\ 
  R93 & 0.12 & 0.10 & 0.18 & 0.17 & 0.13 & -0.23 \\ 
  S94 & 0.05 & 0.04 & 0.09 & 0.09 & 0.15 & 0.65 \\ 
  S95 & 0.05 & 0.04 & 0.07 & 0.09 & 0.11 & 0.34 \\ 
  S96 & 0.09 & 0.10 & 0.16 & 0.17 & 0.13 & -0.19 \\ 
  U99 & 0.12 & 0.00 & 0.00 & 0.00 & 0.12 & Inf \\ 
  NA & 0.05 & 0.07 & 0.12 & 0.12 & 0.08 & -0.30 \\ 
   \hline \hline
   \end{tabular} \label{SI_table_sector_losses2}
\end{table}

\clearpage
\FloatBarrier

\begin{figure}[ht]
	\centering
	\includegraphics[width=0.45\columnwidth]{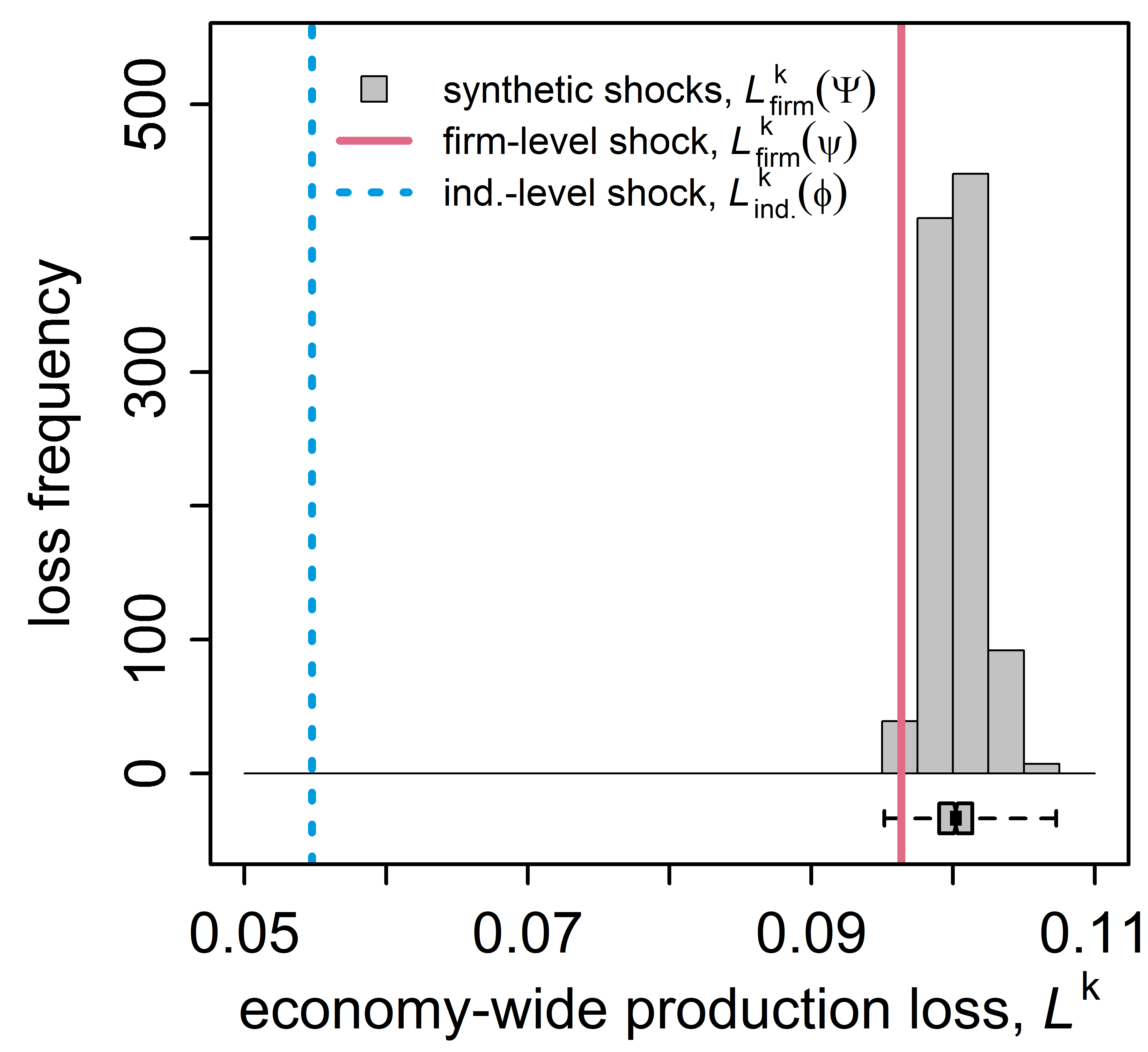} 
	\caption{ Economy-wide production losses, $L$, obtained from an empirically calibrated and 1,000 synthetic COVID-19 shocks propagating \textit{linearly} on the aggregated industry-level production network, IPN, (blue dashed line) and on the firm-level production network, FPN, (red line, histogram). The FPN and IPN correspond to the production network of Hungary in 2019; the firm-level shock, $\psi$, correspond to firms reducing their production level proportional to their reduction in employees between January and May 2020, and are taken from monthly firm-level labor data. The NACE2 level shock, $\phi$, is the aggregation of $\psi$. The 1,000 synthetic shocks, $\Psi$, are sampled such that (when they are aggregated to the NACE2 level) they all have the same size as $\phi$.The empirically calibrated shock, $\psi$, yields a FPN-based loss, ${L}_{\text{firm}}(\psi)$, of 9.6\% (red line). The synthetic shocks yield a distribution of FPN-based production losses, ${L}_{\text{firm}}(\Psi)$, ranging from 9.5\% to 10.7\% of national output (histogram). The median is 10\% (see boxplot). As a reference, the Hungarian GDP declined by 14.2\% in Q2 2020. Note that for the IPN all realizations, $\Psi$, result in the same production loss, ${L}_{\text{ind.}}(\phi) $, of 5.5\%, by construction. The aggregation to the IPN causes a substantial underestimation of the FPN-based production losses. }
 \label{SI_figure_lin_prop_net}
\end{figure}
% # summary of firm-level shocks
%>   summary(emp_shock_may_scens_lin$ESRI[-emp_ind, 1])
%Min. 1st Qu.  Median    Mean 3rd Qu.    Max. 
%0.09514 0.09902 0.10019 0.10024 0.10136 0.10729 
%>   emp_shock_may_scens_lin$ESRI[index_may_median_extension,1]
%ESRI_weight_1 
%0.09634642 
% industry-level propagation: 0.05476606

\section{Results on linear shock propagation}  \label{SI_lin_shocks}

In this section we show how production losses propagate differently on the firm-level and industry-level production network, when all firms and industries have only linear production functions. 
As pointed out in Eq. [\ref{eq_glpf}], each firm $i$ is equipped with a generalized Leontief production function (GLPF), which is defined as
	\begin{equation} 
		x_i = \min\Bigg[
		\min_{k \in \mathcal{I}_i^\text{es}} \Big[ \frac{1}{\alpha_{ik}}\Pi_{ik}\Big], \: 
		\beta_{i} + \frac{1}{\alpha_i} \sum_{k \in  \mathcal{I}_i^\text{ne}} \Pi_{ik} ,  \; \frac{1}{\alpha_{l_i}}l_i, \; \frac{1}{\alpha_{c_i}}c_i \;   \Bigg] \, ,
	\end{equation}
and where $\mathcal{I}_i^\text{es}$  is the set of  essential inputs,  $\mathcal{I}_i^\text{ne}$ is the set of non-essential inputs of firm $i$. 
The linear production function is a special case of the GLPF where all inputs are in the set of non-essential inputs, $\mathcal{I}_i^\text{ne}$. 
We simulate the shocks when for all firms $i$ all inputs, $k \in \mathcal{I}_i^\text{ne}$ belong to, $k\in \{1,2, \dots, m\}$.

We show the estimation errors for network wide production losses from simulating the shock propagation on the IPN, $Z$, instead on the FPN, $W$.
Fig. \ref{SI_figure_lin_prop_net} shows the distribution of network wide production losses, $L_{\text{firm}}(\psi^l)$, in response to the 1,000 synthetic COVID-19 shock scenarios $\Psi$ (defined in the maintext) as histogram and boxplot; loss bins, $L_{\text{firm}}(\psi^l)$, are on the x-axis and frequency of the losses in the respective bins on the y-axis. The variability of losses is economically substantial and ranges from 9.51\% to 10.73\% --- a factor of 1.13. The median and mean losses are 10\% each. 
The variation is substantially smaller than for case with the GLPF shown in Fig. \ref{fig5_total_losses} with losses differing by a factor of up to 1.46 across different shocks.
Note again that the GDP growth in Hungary for Q2 2020 was -14.2\%, indicating a realistic order of magnitude, but a substantial underestimation.
Note again that the initial shocks all have the same monetary size and are identical at the industry-level, i.e. the variation of losses is merely due to the fact that different firms within sectors are initially shocked. 
The distribution is slightly right skewed with a right tail of larger losses. The right tail is substantially smaller than for the GLPF case.
The production loss, $L_{\text{firm}}(\psi) = 9.6\%$, corresponding to the labor shock, $\psi$, (red vertical solid line) lies below the median of the loss distribution. 

% # summary of firm-level shocks
%>   summary(emp_shock_may_scens_lin$ESRI[-emp_ind, 1])
%Min. 1st Qu.  Median    Mean 3rd Qu.    Max. 
%0.09514 0.09902 0.10019 0.10024 0.10136 0.10729 
%>   emp_shock_may_scens_lin$ESRI[index_may_median_extension,1]
%ESRI_weight_1 
%0.09634642 
% industry-level propagation: 0.05476606
%>   # average deviation
%>   estim_errors <- (LIN_nace2_emp_shocks$ESRI[-emp_ind, 1] / emp_shock_may_scens_lin$ESRI[-emp_ind, 1] - 1)
%>   summary(estim_errors)
%Min. 1st Qu.  Median    Mean 3rd Qu.    Max. 
%-0.4896 -0.4597 -0.4534 -0.4535 -0.4469 -0.4244 
%>   quantile(estim_errors, c(0.05, 0.1))
%5%        10% 
%-0.4699551 -0.4656440 
%>   # absolute deviation
%>   mean(LIN_nace2_emp_shocks$ESRI[-emp_ind, 1] - emp_shock_may_scens_lin$ESRI[-emp_ind, 1])
%[1] -0.04547872

The IPN based production losses,  ${L}_{\text{ind.}}(\phi)$, are shown as vertical blue dashed line. As in the main text, firm-level shocks are by construction identical when aggregated to the NACE2 level, each of the 1,000 shock scenarios leads to exactly the same production loss of 5.5\% when propagating on the NACE2 level IPN, $Z$. Interestingly, the IPN estimated production losses, ${L}_{\text{ind.}}(\phi)$, are substantially smaller than the distribution of FPN estimated production losses ${L}_{\text{firm}}(\Psi)$. Therefore, the aggregated network, $Z$ not only can not capture the variation of production losses on the firm-level network, $W$, but  the overall level of shock  propagation is underestimated substantially. 
To quantify the error of estimating the FPN based production loss, ${L}_{\text{firm}}(\Psi)$, with the corresponding IPN based production loss, ${L}_{\text{ind.}}(\phi)$, we calculate the mean absolute error (deviation). 
%The average difference in simulated production losses is -4.5\% ($\mathbb{E} \big[ {L}_{\text{ind.}}(\phi^l) - {L}_{\text{firm}}(\psi^l)$). 
We find that the average estimation error is -45.35\% ($\mathbb{E} \big[ \frac{{L}_{\text{ind.}}(\phi)}{{L}_{\text{firm}}(\Psi)} - 1 \big] $). For the Hungarian production network and the initial shocks, industry-level network shock propagation tends to substantially and systematically underestimate losses from firm-level shock propagation also when production functions are linear.

\begin{figure*}[t]
	\centering
	\includegraphics[width=\textwidth]{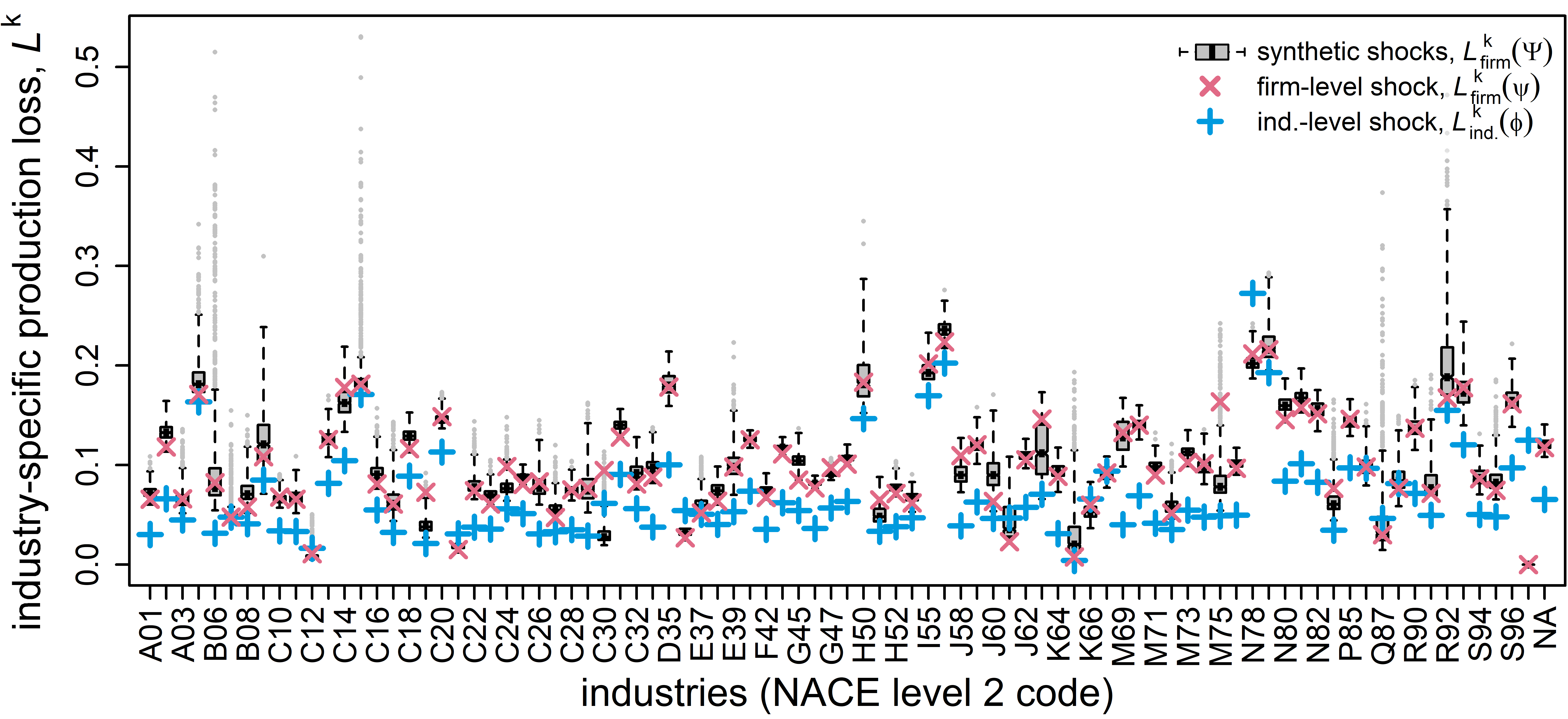} 
	\caption{Comparison of industry-specific production losses, $L^k$, obtained from an empirically calibrated and 1,000 synthetic COVID-19 shocks propagating \textit{linearly} on the aggregated industry-level production network, IPN, (blue `+'es) and on the firm-level production network, FPN, (red `x'es, boxplots). For most industries the FPN-based production losses, ${L}^k_{\text{firm}}(\Psi)$, (boxplots)  vary substantially for few strongly across the synthetic shocks even though shocks have the same size on industry-level. Shock propagation on the industry-level (blue `+'es) can not capture this variation. IPN-based production-losses typically under-estimate the FPN based production losses significantly, on average by about 31.1\%.
 %Comparison of industry-specific production-losses, $L^k$, from the IPN (blue `+'s) and FPN (red `x's) for the same COVID-19 shock scenario and its 1,000 perturbed realizations for the NACE2 industry codes. Loss distributions are given by boxplots. Production losses vary strongly due to slightly different initial shocks. Shock propagation on the industry-level can not capture this variation. IPN-based production-losses typically under-estimate the FPN based production losses, on average by about 31.1\%.
    }
	\label{SI_fig_lin_prop_secs}
\end{figure*}

Fig. \ref{boxplot_emp_shock_may_GL_nace2} shows the distribution of industry  specific production losses, $L^k_{\text{firm}}(\Psi)$, in response to the 1,000 synthetic COVID-19 shock scenarios, $\Psi$, as boxplots. Each boxplot corresponds to an industry, $k$,  with the NACE2 code on the x-axis; the y-axis denotes the losses, $L^k_{\text{firm}}(\Psi)$, of the respective NACE2 codes. The mean overindustry-specific median (mean) losses is 10\% (10.3\%). 
%> summary(mean_sec_shock)
%Min. 1st Qu.  Median    Mean 3rd Qu.    Max. 
%0.00000 0.07208 0.09318 0.10304 0.12944 0.23683 
%> summary(median_sec_shock)
%Min. 1st Qu.  Median    Mean 3rd Qu.    Max. 
%0.00000 0.07044 0.09082 0.10034 0.12847 0.23591
The red `x' symbols represent the production losses, $L^k_{\text{firm}}(\psi)$, corresponding to the original labor shock, $\psi$ and lie  within the boxes.  We clearly see that for many industries remaining production levels vary strongly across initial shocks and the level of variation is very different across industries. The production loss distributions are obviously right skewed --- indicated by extended upper vertical lines (whiskers) --- for all but two industries (H53, N82). Few industries (B05, B06, C15, K65, M75, Q87, and R92) have a substantial amount of outliers (grey dots) that lie outside of 3 times the interquartile range. The minimum and maximum values can differ by factors of up to 9.5 (B06),  7.7 (C12), 5.9 (C30),  5.1 (J61),   41.1 (K65), or 25.8 (Q87). The median (mean) ratios of maximum to minimum loss is 1.27 (1.58). Again, these large deviations do not stem from different sizes of initial shocks, but affecting different firms within industries.  Note that for some sectors the factors, representing the relative variation (maximum loss / minimum loss), are even higher for the case of only linear shock propagation. This is due to the fact that the minimum of the losses are smaller for the linear shock propagation, but the maximum losses are not affected by the non-linearities of the GLPF, i.e. ratios are larger. 
% %max_min_summary[which(max_div_min_shock>5),]
%nace2_labels max_div_min_shock max_sec_shock min_sec_shock
%[1,] "B06"        "9.456"           "0.515"       "0.054"      
%[2,] "C12"        "7.712"           "0.05"        "0.007"      
%[3,] "C30"        "5.858"           "0.113"       "0.019"      
%[4,] "J61"        "5.152"           "0.108"       "0.021"      
%[5,] "K65"        "41.126"          "0.193"       "0.005"      
%[6,] "Q87"        "25.78"           "0.374"       "0.014"       
Fig. \ref{boxplot_emp_shock_may_GL_nace2} shows that the IPN basedindustry-specific production losses,  ${L}^k_{\text{ind.}}(\phi)$, (blue `+' symbols) deviate even stronger from the FPN based losses than for network wide losses. The sectors where IPN based shock propagation underestimates output losses the most are C6 (-62.6\%), C26 (-60.7\%), C29 (-62\%), C33 (-61.5\%), K64 (-66.8\%), K65 (-75.5\%), and M69 (-68.6\%) with negative average relative deviation in parenthesis. Overestimation of losses are highest for sectors, C12 (95.3\%), C21 (70\%), E36 (66.3\%), and 87 (46\%). The average across the mean \textit{absolute} deviation of industries is 31.1\%.
%
%summary(mean_sec_shock)
%   Min. 1st Qu.  Median    Mean 3rd Qu.    Max. 
%0.00000 0.07208 0.09318 0.10304 0.12944 0.23683 
%
%> summary(median_sec_shock)
%   Min. 1st Qu.  Median    Mean 3rd Qu.    Max. 
%0.00000 0.07044 0.09082 0.10034 0.12847 0.23591 
%
%summary(average_deviation_sectors * (average_deviation_sectors < Inf))
%    Min. 1st Qu.  Median    Mean 3rd Qu.    Max.    NA's 
% -0.7553 -0.4948 -0.4131 -0.3109 -0.2807  1.1056       1 
%
%
%> # largest average under estimations
%> average_deviation_sectors[which(average_deviation_sectors < -0.6)]
%       6         26         29         33         64         65         69 
% -0.6256392 -0.6071350 -0.6203416 -0.6153957 -0.6683364 -0.7552520 -0.6855644 
%> # largest average over estimations
%> average_deviation_sectors[which(average_deviation_sectors > 0.4)]
%    12        21        30        36        87        99 
% 0.9526434 0.6973222 1.1055525 0.6628672 0.4582118       Inf 

\FloatBarrier

%\section{Open questions} \label{SI_openquestions}

\end{document}